\documentclass[prx,twocolumn,showpacs,amsmath,amssymb,superscriptaddress, longbibliography]{revtex4-1}

\usepackage[T1]{fontenc}
\usepackage[utf8]{inputenc}
\usepackage{amsmath}
\usepackage{amssymb}
\usepackage{cancel}
\usepackage{graphicx}
\usepackage{esint}
\usepackage{color}
\usepackage{ulem}
\usepackage{booktabs}
\usepackage{makecell}

\usepackage{setspace}
\usepackage{bbding}
\usepackage[table]{xcolor} 
\usepackage{comment}

\usepackage{verbatim}

\usepackage{tikz}
\usetikzlibrary{patterns}
\usetikzlibrary{decorations.pathmorphing}
\tikzset{decoration={snake,amplitude=.4mm,segment length=2mm, post length=0mm,pre length=0mm}}
\tikzset{snake it/.style={decorate, decoration=snake}}

\usetikzlibrary{decorations.pathreplacing,decorations.markings}
\tikzset{
  on each segment/.style={
    decorate,
    decoration={
      show path construction,
      moveto code={},
      lineto code={
        \path [#1]
        (\tikzinputsegmentfirst) -- (\tikzinputsegmentlast);
      },
      curveto code={
        \path [#1] (\tikzinputsegmentfirst)
        .. controls
        (\tikzinputsegmentsupporta) and (\tikzinputsegmentsupportb)
        ..
        (\tikzinputsegmentlast);
      },
      closepath code={
        \path [#1]
        (\tikzinputsegmentfirst) -- (\tikzinputsegmentlast);
      },
    },
  },
  mid arrow/.style={postaction={decorate,decoration={
        markings,
        mark=at position .56 with {\arrow[#1]{stealth}}
      }}},
}

\makeatletter

\usepackage[caption=false]{subfig}
\@ifundefined{showcaptionsetup}{}{\PassOptionsToPackage{caption=false}{subfig}}

\usepackage[english]{babel}
\addto\captionsenglish{}
\renewcommand{\thefigure}{{\@arabic\c@figure}}

\def\@fnsymbol#1{\ensuremath{\ifcase#1\or \dagger\or *\or **\or \ddagger\or
   \mathsection\or \mathparagraph\or \|\or  \dagger\dagger
   \or \ddagger\ddagger \else\@ctrerr\fi}}

\makeatother

\begin{document}

\title{Engineering three dimensional moir\'e flat bands }

\author{Lede Xian}
\altaffiliation{L.X. and A.F contributed equally to this paper.}
\affiliation{Songshan Lake Materials Laboratory, 523808 Dongguan, Guangdong, China}
\affiliation{Max Planck Institute for the Structure and Dynamics of Matter, Center for Free Electron Laser Science, 22761 Hamburg, Germany}

\author{Ammon Fischer}
\altaffiliation{L.X. and A.F contributed equally to this paper.}
\affiliation{Institut f\"ur Theorie der Statistischen Physik, RWTH Aachen University and JARA-Fundamentals of Future Information Technology, 52056 Aachen, Germany}

\author{Martin Claassen}
\affiliation{Department of Physics and Astronomy, University of Pennsylvania, Philadelphia, PA 19104, USA}

\author{Jin Zhang}
\affiliation{Max Planck Institute for the Structure and Dynamics of Matter, Center for Free Electron Laser Science, 22761 Hamburg, Germany}

\author{Angel Rubio}
\altaffiliation{Corresponding author: angel.rubio@mpsd.mpg.de}
\affiliation{Max Planck Institute for the Structure and Dynamics of Matter, Center for Free Electron Laser Science, 22761 Hamburg, Germany}
\affiliation{Center for Computational Quantum Physics, Simons Foundation Flatiron Institute, New York, NY 10010 USA}
\affiliation{Nano-Bio Spectroscopy Group,  Departamento de Fisica de Materiales, Universidad del Pa\'is Vasco, UPV/EHU- 20018 San Sebasti\'an, Spain}

\author{Dante M. Kennes}
\altaffiliation{Corresponding author: dante.kennes@rwth-aachen.de}
\affiliation{Institut f\"ur Theorie der Statistischen Physik, RWTH Aachen University and JARA-Fundamentals of Future Information Technology, 52056 Aachen, Germany}
\affiliation{Max Planck Institute for the Structure and Dynamics of Matter, Center for Free Electron Laser Science, 22761 Hamburg, Germany}

\begin{abstract}

\noindent
\textbf{Abstract:} Twisting two adjacent layers of van der Waals materials with respect to each other can lead to flat two-dimensional electronic bands which enables a wealth of physical phenomena. Here, we generalize this concept of so-called moir\'e flat bands to engineer flat bands in all three spatial dimensions controlled by the twist angle. The basic concept is to stack the material such that the large spatial moir\'e interference patterns are spatially shifted from one twisted layer to the next. We exemplify the general concept by considering graphitic systems, boron nitride and WSe$_2$, but the approach is applicable to any two-dimensional van der Waals material. For hexagonal boron nitride we develop an {\it ab-initio} fitted tight binding model that captures the  corresponding three dimensional  low-energy electronic structure. We outline that interesting three dimensional correlated phases of matter can be induced and controlled following this route, including quantum magnets and unconventional superconducting states. 
\\[1.5em]
\end{abstract}

\maketitle


\section{Introduction}

In the past few years twisting adjacent layers of van der Waals materials has emerged as a versatile route to control the ratio between kinetic, potential and vibrational energy of two-dimensional systems.  Central to the idea of twistronics, the selective suppression of kinetic energy scales permits tuning materials into a regime dominated by electronic interactions, as well as precise control over electronic filling via gating \cite{CoryP,moireqs}. Early experimental and theoretical studies concentrated on graphetic systems of different kinds, such as twisted bilayer graphene \cite{bistritzer2011,cao2018a,cao2018b,Yankowitz18,Kerelsky18}, twisted double bilayer graphene  \cite{liu19,shen19,cao2019electric,tutuc2019,RubioVerdu20}, trilayer rhombohedral graphene on hexagonal boron nitride \cite{chen2019signatures,chen2019evidence} and twisted mono-bilayer graphene \cite{chen2020electrically,shi2020tunable}. More recently,  twisted transition metal dichalcogenides (TMD) moved into the center of attention as another important class of van der Waals materials, such as WSe$_2$ \cite{wang19,an19,pan2020band}, MoS$_2$ \cite{liao2020precise,Naik18,xian20}, and TMD heterostructures \cite{wu17,zhang2019moir,regan2020,tang2020}, allowing access to new regimes,  beyond graphitic systems. With more of these phenomena within experimental reach twisted van der Waals materials are increasingly viewed as potential avenues towards solid-state based platforms of quantum control and quantum materials with properties on demand  \cite{moireqs,Basov2017}. Furthermore, the twist angle allows to control those systems to such a high degree that moir\'e aided metrology outperforms the current gold standard regarding structural questions in van der Waals materials \cite{Halbertal20}. One important question guiding theoretical and experimental research efforts concerns the exploration of the tremendous combinatoric space of chemical compositions of van der Waals materials to shed light onto the basic question of which additional phenomena might be accessible using twistronics. In the quickly expanding cosmos of twisted van der Waals materials, one guiding principle remains the control of the low-energy degrees of freedom, which might realize prototypical models of condensed matter research in a more rigid, clean and penetrable context \cite{moireqs}. In this spirit, and in addition to the directions already outlined above, twisted hexagonal boron nitride was shown to harbor entire families of flat bands \cite{Xian18,zhao2020formation} and twisted two-dimensional magnets, such as CrI$_3$, were identified to realize moir\'e skyrmion lattices and noncollinear twisted magnetic phases \cite{tong2018skyrmions,hejazi2020noncollinear}. Beyond realizing different phenomena in two-dimensions using twist as a control paradigm one might ask whether entirely different dimensionalities can (effectively) be addressed.  Twisted two dimensional monolayers of monochalcogenides (e.g. GeSe) were shown to allow  access to the one-dimensional limit with twist \cite{kennes19,Kariyado19} providing the same unprecedented level of control as in the two-dimensional counterparts. 

However, a practical generalization of twistronics to three dimensions with suppressed kinetic energy scales remains an outstanding challenge (aside from the possibility of adding further synthetic dimensions \cite{Lohse2018,Zilberberg2018}). Here, we address this by showing that stacking van der Waals materials in a predefined fashion allows to engineer three dimensional moir\'e flat bands. With this advance the realization of three dimensional systems controlled by twistronics is no longer elusive, completing the list of systems controlled by twistronics in one, two, and now three dimensions.  Our idea works generically and relies mainly on basic geometric arguments. Importantly, it is achievable within recent experimental advances to fabricate bulk-like artificial twisted materials \cite{liu2020disassembling,kim_AtAtA,pablo_AtAtA} and can be applied to any of the many van der Waals materials, which we will exemplify here for three important materials: graphitic systems, WSe$_2$ and hexagonal boron nitride. The last of these will be examined in more details and we provide a full \textit{ab initio} characterization for its three dimensional twist-dependent band structures. We then consider the effects of correlations on the low-energy bands and find three dimensional magnetic and superconducting states which can be realized as a function of twist angle.

\begin{figure*}
    \centering
    \includegraphics[width=\textwidth]{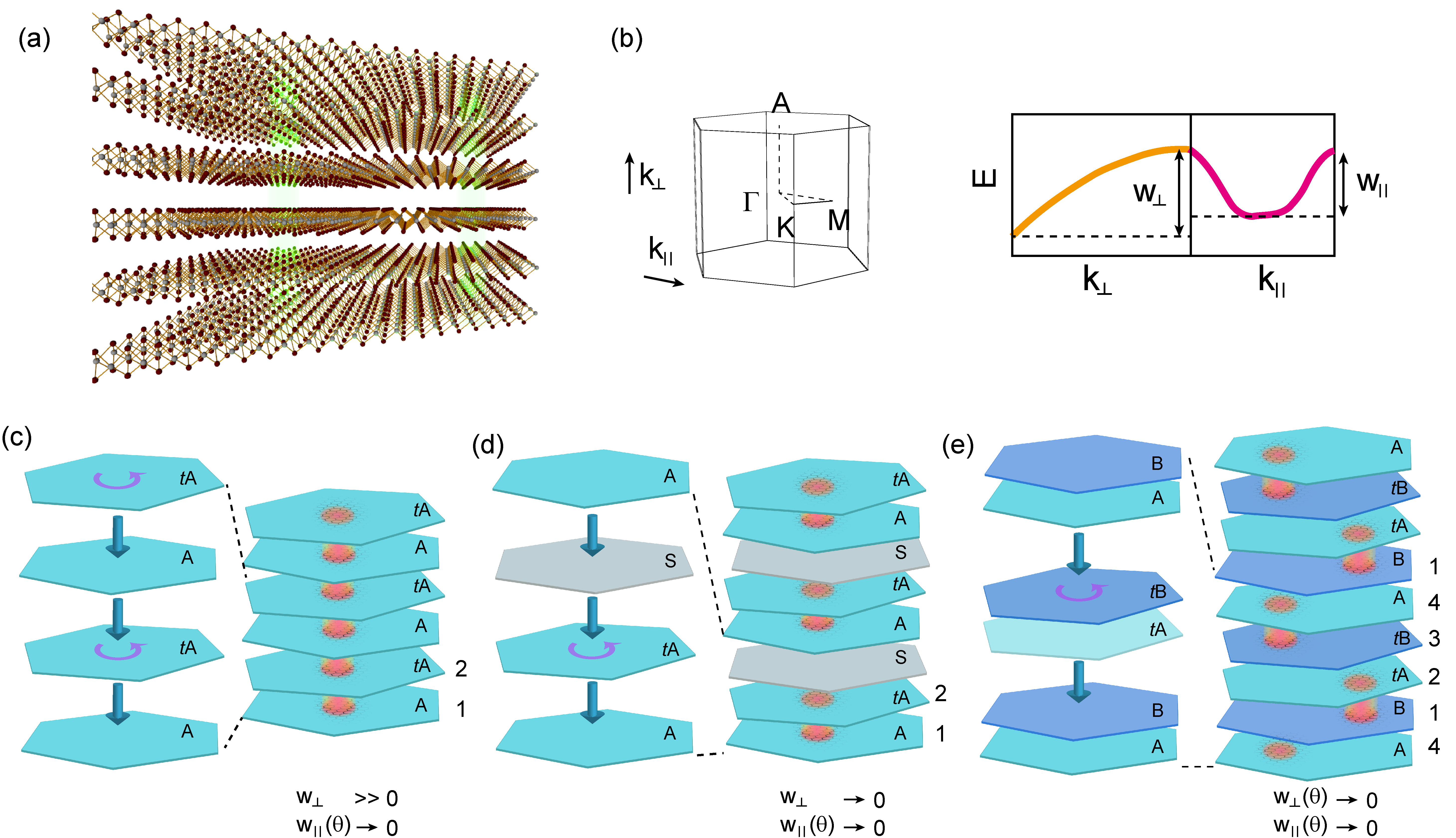}
     \caption{{\bf Stacking ideas to construct 3D flat bands.} Panel (a) exemplifies the general idea of stacking up lattice sites defined by the moir\'e potential in a successive way to approach the three dimensional limit. 
     Panel (b) shows the Brillouin zone  and defines the out-of plane $k_\perp$ and in-plane $k_{\parallel}$ directions (left) as well as a cartoon of the corresponding band widths ${\rm w}_{\perp}$ and ${\rm w}_{\parallel}$ along these directions (right). 
     Panels (c)-(e) show three different configurations to create three dimensional materials out of stacks of two dimensional twisted van der Waals materials. The orange area indicate the moir\'e lattice sites where the charge density localized.  When these moir\'e lattice sites align atop of each other (c) the bands become flat in two of the three directions only. When using spacer layers (d) the band width can be reduced in all three dimensions, but the band width along the stacking direction cannot be controlled by the twist angle. If moir\'e lattice sites by geometric reasons (see main text) are shifted from layer to layer (e), three dimensional flat bands with the flatness in all three dimensions being controlled by the twist angle emerge.  }
    \label{current}
\end{figure*}

\section{Stacking Approaches}
The general idea is shown in Fig.~\ref{current} (a) relying on successively stacking up lattice sites defined by the moir\'e potential. In panel (b) we define the Brillouin zone, the in-plane and out-of-plane directions $k_\perp$ and $k_\parallel$ as well as a cartoon of the corresponding band widths used in the following discussion.  We are looking for ideal 3D flat bands that satisfy the following conditions: (1) the band width is controllable by twist angles at all three dimensions; (2) the system is periodic along all three dimensions such that it remains a well-defined crystal. Therefore, our work is distinct from previous works that study intrinsic 3D flat bands in some solids \cite{kang2020topological,liu2020orbital,zhou2019weyl} with a fixed band width and little tunability. Our approach also is distinctly different to proposals such as the 3D chiral twisted structure \cite{cea2019twists,wu2020three} that render the system a quasicrystal.
To this end we start by considering three different stacking patterns illustrated in Fig.~\ref{current} (c-e). 

Firstly, we consider stacking monolayers of van der Waals materials in an alternating fashion as depicted in panel (c) of Fig.~\ref{current}, meaning that every second layer is aligned perfectly while adjacent layers have a relative twist angle between them.  This can be regarded as stacking twisted bilayers repeatedly. Following this route moir\'e patterns form by the interference between adjacent layers. This type of stacking has been intensively investigated for the study of two dimensional flat bands in twisted trilayer and multilayer graphene \cite{eslam19,carr2020ultraheavy}. Viewed top-down, the moir\'e lattice sites where in-plane charge density localizes, align on top of each other. As a consequence, the electronic bands become flat within the plane; just as is the case for two twisted sheets of van der Waals materials. Conversely, the alignment of moir\'e regions along the out-of-plane direction retains substantial band dispersion in this direction due to significant amount of hybridization between moir\'e sites at adjacent layers (see Fig.~S1 in the SI for an example of 3D twisted boron nitride with such stacking). While this allows to effectively engineer quasi-one-dimensional systems with very low residual coupling along the remaining two spatial directions -- an interesting opportunity of materials engineering in its own rights (for a similar quasi-one-dimensional system it was shown that in-plane confinement albeit imposed by an magnetic field gives rise to a 3D quantum Hall effect \cite{tang2019three})-- it does not allow to realize three dimensional moir\'e flat bands. 

Secondly, one might consider the case in which twisted van der Waals materials with flat bands in their two spatial directions are stacked on top of each other with an insulating buffer layer in between as shown in panel (d) of Fig.~\ref{current}. The properties and thickness of the buffer-layer could then be adjusted such that the hopping from one twisted sheet of van der Waals materials to the next sheet is suppressed substantially. This would lead to flat electronic bands in all three spatial dimensions. However, such an approach has multiple problems. 
First, the flatness of the bands in the out-of-plane direction is mainly determined by the residual coupling between neighboring moir\'e charge localization sites across the insulating layers. This limits available band structures that can be engineered quite substantially compared to the flexible control that twist angle offers with respect to the remaining two directions. Second, with a buffer layer, there is no guarantee to keep the moir\'e sites across the buffer layer well aligned, i.e., the centers of the moir\'e sites of neighboring twisted pairs can relocate to different in-plane positions when stacking up. This could introduce significant amount of disorder along the out-of-plane direction such that the system is no longer a well-defined crystal.


To remedy the shortcomings of the previous two stacking approaches, thirdly, we present an idea using a stacking sequence in which the moir\'e charge localization sites simply due to geometric considerations do not form atop of each other, but are shifted with respect to the out-of-plane directions of the van der Waals materials used. 
Such a configuration can be constructed by expanding the basic stacking unit, e.g., from a twisted bilayer to a twisted double bilayer.  This is visualized in panel (e) of Fig.~\ref{current}. Compared to the first approach, the repeating unit along the out-of-plane direction is changed from layers 1,2 in Fig.~1(a) to layers 1-4 in Fig.~1(c). In this approach, layers 2,3 and layers 4,1 remain at their intrinsic Bernal AB stacking or AA' stacking sequence as in the pristine bulk material and the twisting takes place only between layers 1 and 2 as well as 3 and 4 in the notation of Fig.~\ref{current}(c). Although the in-plane crystal axis of layer 2 is aligned with those of layer 3, the atomic positions of the two layers are translated with respect to each other (as in intrinsic AB stacking), or flipped (as in intrinsic AA' stacking). The same happens for layer 4 and 1. This naturally displaces the moir\'e charge localization sites in these layers with respect to each other, which are now separated by the moir\'e length scale. As the twist angle is decreased and the in-plane distance between moir\'e sites increases, so does the distance between sites on adjacent bilayers. Therefore, the idea of using natural (or intrinsic) bilayer as a stacking unit to construct alternating patterns, allows to engineer robust flat bands in all three dimensions, with the flatness being continuously controlled by the twist. Thus, it satisfy the condition (1) we set for a ideal flat band system. Moreover, as we will show below, such an approach will also generate local stacking regions that resemble the stacking sequence in the pristine bulk crystal, which can act as a low-energy stabilization center to prevent disorder along the out-of-plane direction. Therefore, this approach also meets  condition (2) of a nearly ideal robust flat band crystallographic system.   


\begin{figure*}
    \centering
    \includegraphics[width=\textwidth]{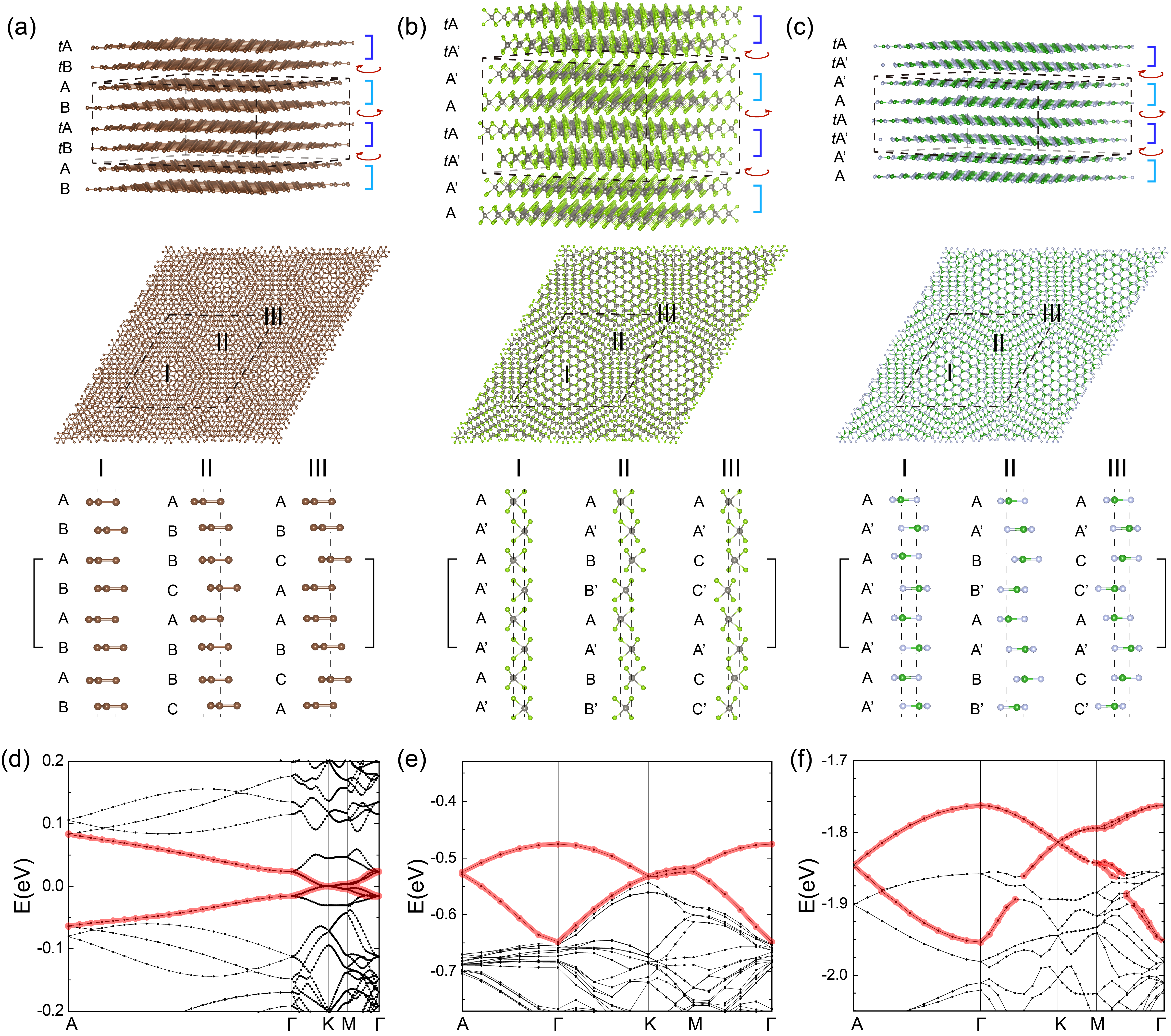}
    \caption{{\bf 3D Flat bands for different materials, Graphene, WSe$_2$ and boron nitride }. Atomic structures of 3D twisted graphene (a), WSe$_2$ (b), boron nitride (c). The top and the middle panels show the perspective and the top views of the structures, respectively. The unit cells are indicated with dashed lines. The bottom panels show the local stacking sequence in the region I, II and III indicated in the middle panels. The repeating units along the out-of-plane direction are indicated with solid brackets. (d-f) The corresponding band structures for graphene at 1.3 degrees (d), for WSe$_2$ at 5.08 degrees (e), and for boron nitride at 5.08 degrees (f). For smaller angles the bands become increasingly flat and detach from other bands. }
    \label{Fig:sys}
\end{figure*}

\section{Flat bands and effective low-energy model}
We put this very general idea to the test by first performing \textit{ab initio} and tight-binding based characterizations of such stacked materials using bilayers of  graphene, WSe$_2$ as well as hexagonal boron nitride. All of these materials were successfully studied in the past for the twisted single bilayer case rendering them ideal starting points to explore the idea we put forward here. The results are summarized for a twist angle of $1.3^\circ$ for graphene and $5.08^\circ$ for both WSe$_2$ and boron nitride in Fig.~\ref{Fig:sys}. We show side and top views of the real space stacking in panels (a)-(c) for graphene, WSe$_2$ and boron nitride, respectively. The unit cell for these bulk twisted systems is formed by a twisted double  bilayer as highlighted by dashed lines in the top row and the solid brackets in the third row. The bottom panels of (a)-(c) show the local stacking sequence in the three representative regions shown in the middle row panels. The stacking sequence in region I is exactly the same as that in the intrinsic untwisted bulk crystal (AB stacking as in graphite, AA' stacking as in 2H WSe$_2$ and bulk boron nitride). This region is expected to pin the in-plane alignment of the layers and naturally prevent accidental layer displacements similar to what is discussed for the case of twisted trilayer graphene \cite{carr2020ultraheavy}.  In panels (d)-(f) we show the respective band structures. Since the angle is not very small for twisted WSe$_2$ and twisted boron nitride, the bands still show substantial dispersion and the set of bands that is flattening, marked in red in Fig.~\ref{Fig:sys}, have not fully separated from other bands yet (a more detailed study of decreasing the twist angle further is given below).The important feature to note is the width of the red set of bands, with respect to variations of both the out-of-plane $k_\perp$ ($A\to\Gamma$ path in the Brillouin zone) and in-plane $k_\parallel$ ($\Gamma\to K\to M \to\Gamma$ path in the Brillouin zone) momentum. Both of these band widths in- and out-of-plane, called $W_{\parallel}$ and $W_{\perp}$ respectively, decrease as the twist angle approaches smaller values (see below and SI). Eventually these bands separate from the rest giving rise to perfectly isolated flat bands, with comparable kinetic energy scales in all three spatial dimensions, whose magnitude can be tuned by the twist, as demonstrated next.

\begin{figure*}
    \centering
    \includegraphics[width=\textwidth]{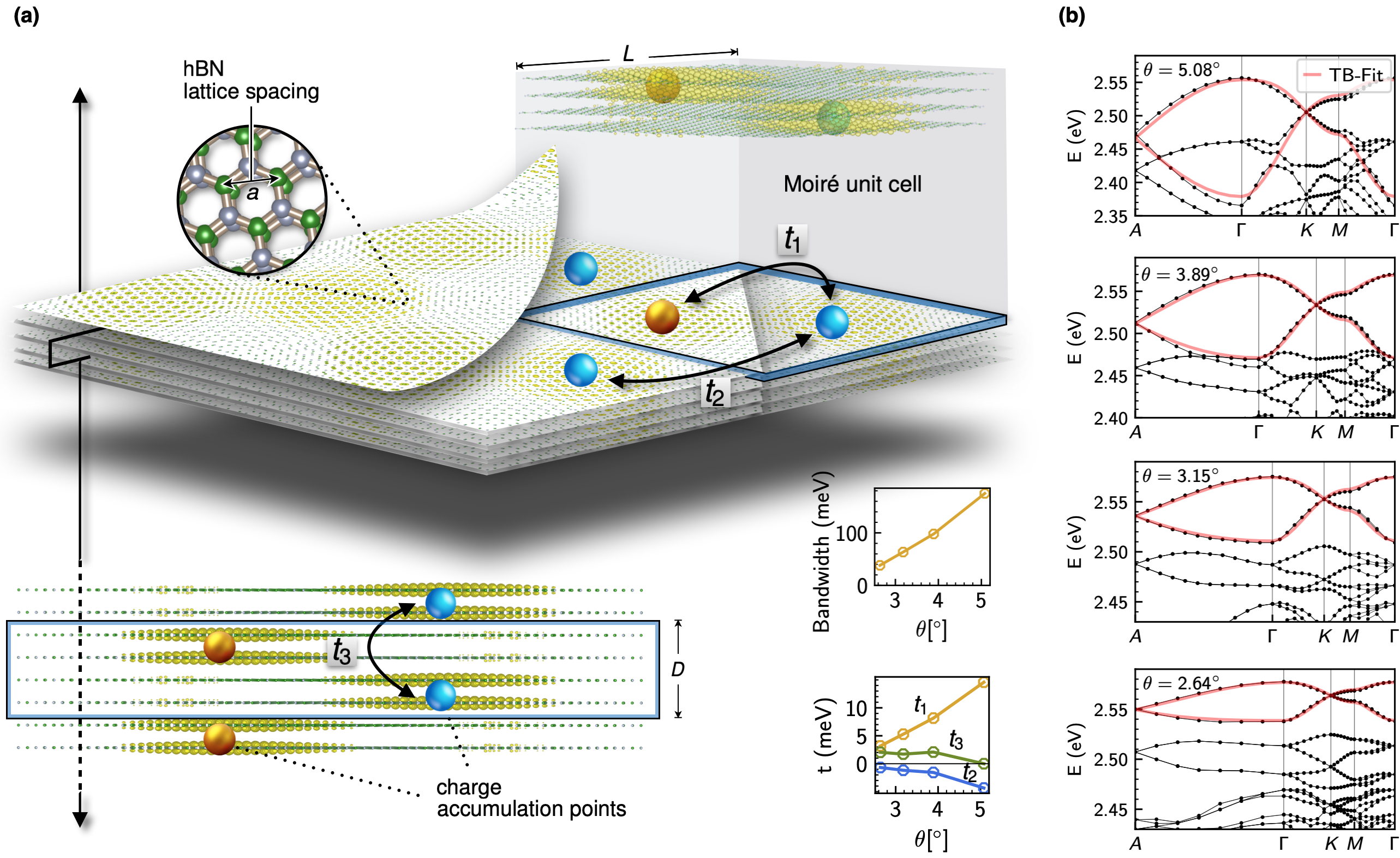}
    \caption{{\bf Low-energy model of three dimensional twisted boron nitride (thBN)}. (a) Stacking pattern and moir\'{e} unit cell of thBN for $\theta = 5.08^{\circ}$. Emerging charge accumulation points (olive green) form an effective lattice that resembles AA-stacked hexagonal multilayers, where one site is shifted by $D/2$ in $z$-direction (blue and gold spheres). (b) Results from our {\it ab initio} fitted tight-binding (TB)  approach (see Supplementary Material S. II for details) for different twist angles $\theta$, taking up to next-nearest neighbor inter- and intralayer hopping terms $t_1$,..,$ t_3$ into account. The band width decreases continuously with the twist angle and, eventually, the low-energy bands detach from the rest of the spectrum.}
    \label{fig:TB_BN}
\end{figure*}

We stress that the idea we present here is general and allows three dimensional flat band engineering also in other materials even beyond the ones discussed explicitly above. However, we are going to illustrate the relevance of this new concept to electronic band engineering taking the specific case of hBN (see Fig.~S2 in the SI for the case of WSe2). The choice of boron nitride is made for convenience (and for being widely used as protective 2D material), as the absence of sharp magic angles, makes it particularly feasible to large scale numerics providing a full-fledged ab-initio characterization of the material's band structure in three dimensions. As the relaxation of twisted boron nitride does not significantly alter the band structure according to the previous work \cite{Xian18}, we fix the atomic structure in the large-scale ab-initio calculations. Our results are summarized in Fig.~\ref{fig:TB_BN}. In panel (a) we show top and side views of the charge localization within the moir\'e unit cell and the position of the B and N atoms. We choose intrinsic AA' bilayers (as in pristine bulk hBN) as the building blocks of our three dimensional structure and therefore there are no ferroelectric domains as recently reported for twisted bilayer systems \cite{yasuda2020stacking,woods2020charge}. Generalizing this constitutes an intriguing avenue of future research.  
In panel (b) we summarize the ab-initio band structure obtained for different twist angles. Importantly, as we approach smaller values of the twist angle, both the in-plane and out-of-plane band width $W_{\parallel}$ and $W_{\perp}$, which are the same with such stacking, decrease and the flat bands detach from the other bands in the spectrum. Strong charge localization marked by blue and golden spheres in panel (a) highlight the emergence of a corresponding three dimensional effective low-energy tight-binding model (see Supplementary Material S. II for details). The model we obtain for the case of boron nitride is an in-plane triangular lattice stacked out of plane, such that the lattice sites of one plane reside in the center of the triangular lattice of the next plane. 

The success of fitting the flat bands within such a low-energy tight-binding model including only short ranged hoppings $t_1,t_2$ and $t_3$ on the moir\'e scale as denoted in panel (a) is demonstrated in panel (b). By fitting the three hopping parameters to the full ab-initio band structure for different angles, almost perfect agreement is achieved (consistent with the earlier study of a single twisted hBN bilayer \cite{Xian18}). The smaller panels left to panel (b) show the extracted values of the hopping as well as the overall bandwidth (in this case $W_\perp=W_\parallel$), demonstrating the success of three dimensional flat band engineering by the twist proposed here. In particular, the hopping parameters $t_{1..3}$ of the low-energy model prove that our initial claim of full twist angle control holds in the case of twisted hBN: the interlayer hopping $t_3$ is nearly independent of the twist angle (small deviations occur for larger twist angles due to mixing of low-energy and valence bands, Fig. 3b), whereas the in-plane hopping $t_2$ and the mixed inter-/intralayer term $t_1$ decrease continuously. Such an effective low-energy model is immensely useful as it can be treated much more efficiently. As a direction of future research one should, building on the above results, set up a continuum theory to further analyze the emergence of three dimensional flat bands.

\begin{figure*}
    \centering
    \includegraphics[width=\textwidth]{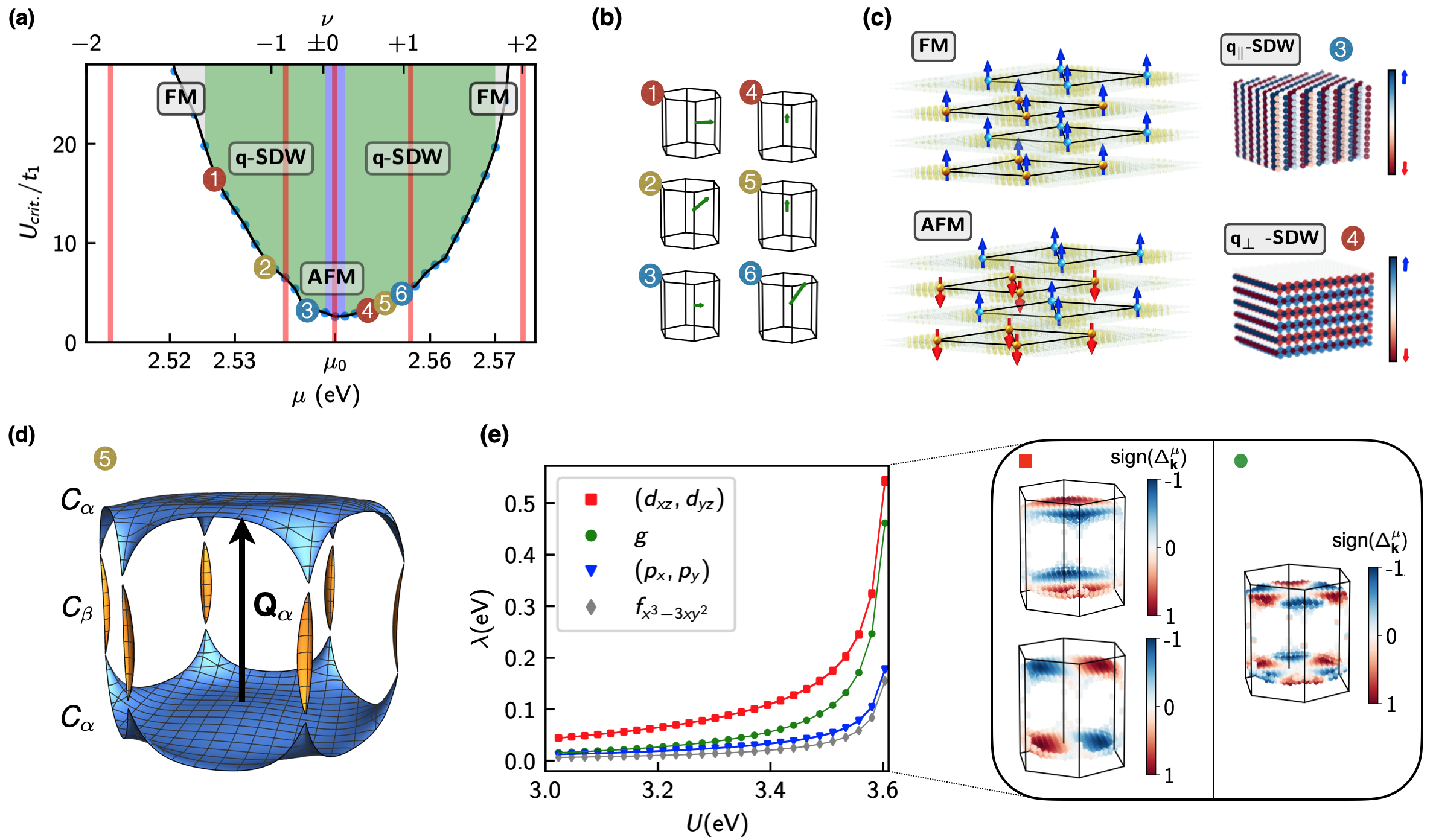}
    \caption{{\bf Correlated phases in three dimensional twisted hexagonal boron nitride for twist angle $\theta = 3.15^{\circ}$.} (a-c) The RPA analysis reveals a variety of magnetic states including antiferromagnetic order (AFM) at charge neutrality, ferromagnetic order (FM) for strong electron/hole doping as well as general spin-density waves $\textbf{q}$-SDW with periodic patterns in all three spatial dimensions. (d) For $\mu =\mu_0+ 5 \, \text{meV}$ the Fermi surface of thBN is almost perfectly nested along the vector $\textbf{Q}_{\alpha}$ resulting in a strong enhancement of particle-particle scattering between these sheets. In particular, the preferred superconducting gap symmetry (e) is two-fold degenerate and of type $(d_{xz}, d_{yz})$. Below $T_c$, the system will minimize its Ginzburg-Landau free energy by assuming the chiral linear combination $(d_{xz}\pm i d_{yz})$ and thus the gap parameter breaks time-reversal symmetry.}
    \label{fig:correlation}
\end{figure*}

\section{Correlated Phases of Matter}
We employ the effective low-energy tight-binding model to outline a putative unexplored phase diagram that could be accessible via the three dimensional twistronics approach. To this end, we consider a local Hubbard interaction $U$ added to the effective flat band model for $\theta = 3.15^{\circ}$ as discussed above. We note that a more realistic model should include longer ranged interactions as well, which should be characterized from first principles. Such a study is unfortunately beyond the scope of the present work and most likely requires a fundamental methodological advance to treat the huge three dimensional unit cell (containing many tens of thousands of atoms at small twist angles).  Here, however, we provide the first step along a characterization of elusive and exciting correlation effects and aim to identify interesting states of matter already present at the level of a Hubbard interaction. To achieve this we first perform a random phase approximation (RPA) study of the system (see Supplementary material S. III) and identify a plethora of magnetic instabilities. A putative magnetic phase diagram is summarized in panel (a) of Fig.~\ref{fig:correlation}. As expected we find ferromagnetic (FM) ordering tendencies as the flat bands are either filled or empty, albeit with a rather large critical $U_{\rm crit}$ driving the transition. Due to the bipartiteness of the lattice (sublattice A and B being charge localization sites marked by blue and golden spheres in Fig.~\ref{fig:TB_BN}(a)) we find antiferromagnetic (AFM) ordering at half filling.  In between these two phases a more general spin density wave with filling dependent ordering vector $q$ emerges. $q$ is illustrated in panel (b) of Fig.~\ref{fig:correlation}, while panel (c) illustrates the magnetic ordering in real space for four different examples depending on the filling: FM, AFM or a spin density wave with a wave vector lying either in the in-plane ($q_\parallel$) or out-of-plane ($q_{\perp}$) direction.  We note that the ordering vectors in general do not align with the crystal axes and therefore, although the underlying mechanisms that determine ordered states (such as Fermi surface nesting or van Hove singularities) are analogous to the two-dimensional case, the phases we find here cannot be described in terms of quasi two-dimensional planes.

For interaction values $U < U_{\text{crit}}$ there is no magnetic ordering and the system is paramagnetic. In this regime, spin and charge fluctuations may provide an effective pairing glue between the electrons leading to the formation of Cooper pairs. To pin down the pairing instability mediated by spin and charge fluctuations, we take the RPA corrected interaction vertex in the fluctuation exchange approximation (see Supplementary Material S. III) and linearize the superconducting gap equation around the critical temperature $T_c$ for slight electron doping $\mu=\mu_0+5$ meV around half filling of the three dimensional flat bands $\mu_0=2.546$ eV. In this scenario, only scattering events between Cooper pairs in the vicinity of the Fermi surface sheets $C_{\alpha, \beta}$ , Fig.~\ref{fig:correlation} (d), contribute notably to the formation of a superconducting state with order parameter $\Delta_{\textbf{k}}^{\mu}$. The fact that (i) the Fermi surface sheets $C_{\alpha}$ are (almost) perfectly connected by the nesting vector $\textbf{Q}_{\alpha}$ at which particle-particle scattering is strongest and (ii) the effective pairing glue contained in the spin-singlet channel is purely repulsive for all scattering events, electron-electron pairing is conditioned on a relative sign change between the pairing form factors $\mu$ connected by the vector $\textbf{Q}_{\alpha}$ i.e. $\Delta_{\textbf{k}}^{\mu} = -\Delta_{\textbf{k} + \textbf{Q}_{\alpha}}^{\mu}$. The linearized gap equation may be written as an eigenvalue problem where the eigenfunctions $\Delta_{\textbf{k}}^{\mu}$ corresponding to the largest eigenvalue $\lambda$ yield the symmetry of the most prominent superconducting state, Fig.~\ref{fig:correlation} (e). Our calculations reveal that the leading gap parameter is of spin-singlet type and is two-fold degenerate with symmetry classification $(d_{xz}, d_{yz})$. The two $d$-wave solutions are characterized by a nodal line along the $k_x$- and $k_y$-direction of the Brillouin zone and thus the system will minimize its Ginzburg-Landau free energy $F$ below $T_c$ by assuming the chiral linear combination $(d_{xz}\pm i d_{yz})$ (see Supplementary Material S. III) which breaks time-reversal symmetry. 

In conclusion our work generalizes the idea of two dimensional twistronics to the three dimensional realm. The main notion relies on cleverly stacking adjacent layers in such a way that the hopping between adjacent charge puddles in all three dimensions gets successively suppressed as the twist angle is lowered. We argue that the proposed stacking method is robust towards small twist angle imperfections and inhomogeneities which might vary within one plane or between adjacent layers. Even more so, since our construction relies purely on geometric arguments even in the presence of imperfections these would simply reflect in slightly inhomogeneous hoppings, like they are present in any (not perfectly clean) crystal and which do not change the overall physics significantly. On the contrary, controlled variations of the twist angle might allow to control the effective disorder and therefore provide a long sought after inroad into tunable strongly disordered systems, from a condensed matter perspective. With this we add the three dimensional realm to the list of low-energy models that can effectively be realized in moir\'e structures. As a side product we also provide an alternative of engineering effectively quasi-one-dimensional structures, by placing the moir\'e sites on top of each other (first scenario in Fig.~\ref{current}). This is not at the center of attention in this work but allows to access similar physics as discussed in the context of the quantum Hall effect in Ref.~\cite{tang2019three}. We already reported on the rich behavior of correlation driven phases in engineered three dimensional flat bands above, but another intriguing avenue of future research should also address the question of three dimensional flat band engineering for purposes of controlling topological properties. This provides a very rich playground that directly opens up by our approach and the future will tell which topological phenomena, such as Weyl physics, and correlated phases beyond the ones discussed here might be tunable by three dimensional twistronics.

\subsection*{Supporting Information}
The Supporting information is available free of charge at http://pubs.acs.org. Further details on ab initio calculations, extended data for low-energy tight-binding models, theoretical details of multi-orbital random-phase approximation (RPA) and linearized gap equation.

\subsection*{Acknowledgments}
This work is supported by the European Research Council (ERC-2015-AdG-694097), Grupos Consolidados (IT1249-19),  and SFB925. AR is supported by the Flatiron Institute, a division of the Simons Foundation. We  acknowledge  funding by the Deutsche Forschungsgemeinschaft (DFG, German Research Foundation) under RTG 1995, within the Priority Program SPP 2244 ``2DMP'', under Germany's Excellence Strategy - Cluster  of  Excellence and Advanced Imaging of Matter (AIM) EXC 2056 - 390715994 and RTG 2247. LX acknowledges the support from Distinguished Junior Fellowship program by the South Bay Interdisciplinary Science Center in the Songshan Lake Materials Laboratory. JZ acknowledges funding received from the European Union Horizon 2020 research and innovation program under Marie Sklodowska-Curie Grant Agreement 886291 (PeSD-NeSL).

\subsection*{Competing interests}
The authors declare no competing interests.

\subsection*{Data and material availability}
All data shown in the manuscript is available upon reasonable request.

\nocite{plimpton1995fast,angeli2018emergent,brenner2002second,kolmogorov2005registry,ouyang2018nanoserpents,bitzek2006structural,tram10,kresse93ab,wang19,Xian18,blochl94,perdew1996generalized,tsmethod09}
\bibliography{reference.bib}

\begin{thebibliography}{60}%
\makeatletter
\providecommand \@ifxundefined [1]{%
 \@ifx{#1\undefined}
}%
\providecommand \@ifnum [1]{%
 \ifnum #1\expandafter \@firstoftwo
 \else \expandafter \@secondoftwo
 \fi
}%
\providecommand \@ifx [1]{%
 \ifx #1\expandafter \@firstoftwo
 \else \expandafter \@secondoftwo
 \fi
}%
\providecommand \natexlab [1]{#1}%
\providecommand \enquote  [1]{``#1''}%
\providecommand \bibnamefont  [1]{#1}%
\providecommand \bibfnamefont [1]{#1}%
\providecommand \citenamefont [1]{#1}%
\providecommand \href@noop [0]{\@secondoftwo}%
\providecommand \href [0]{\begingroup \@sanitize@url \@href}%
\providecommand \@href[1]{\@@startlink{#1}\@@href}%
\providecommand \@@href[1]{\endgroup#1\@@endlink}%
\providecommand \@sanitize@url [0]{\catcode `\\12\catcode `\$12\catcode
  `\&12\catcode `\#12\catcode `\^12\catcode `\_12\catcode `\%12\relax}%
\providecommand \@@startlink[1]{}%
\providecommand \@@endlink[0]{}%
\providecommand \url  [0]{\begingroup\@sanitize@url \@url }%
\providecommand \@url [1]{\endgroup\@href {#1}{\urlprefix }}%
\providecommand \urlprefix  [0]{URL }%
\providecommand \Eprint [0]{\href }%
\providecommand \doibase [0]{http://dx.doi.org/}%
\providecommand \selectlanguage [0]{\@gobble}%
\providecommand \bibinfo  [0]{\@secondoftwo}%
\providecommand \bibfield  [0]{\@secondoftwo}%
\providecommand \translation [1]{[#1]}%
\providecommand \BibitemOpen [0]{}%
\providecommand \bibitemStop [0]{}%
\providecommand \bibitemNoStop [0]{.\EOS\space}%
\providecommand \EOS [0]{\spacefactor3000\relax}%
\providecommand \BibitemShut  [1]{\csname bibitem#1\endcsname}%
\let\auto@bib@innerbib\@empty
\bibitem [{\citenamefont {Balents}\ \emph {et~al.}(2020)\citenamefont
  {Balents}, \citenamefont {Dean}, \citenamefont {Efetov},\ and\ \citenamefont
  {Young}}]{CoryP}%
  \BibitemOpen
  \bibfield  {author} {\bibinfo {author} {\bibfnamefont {Leon}\ \bibnamefont
  {Balents}}, \bibinfo {author} {\bibfnamefont {Cory~R.}\ \bibnamefont {Dean}},
  \bibinfo {author} {\bibfnamefont {Dmitri~K.}\ \bibnamefont {Efetov}}, \ and\
  \bibinfo {author} {\bibfnamefont {Andrea~F.}\ \bibnamefont {Young}},\
  }\bibfield  {title} {\enquote {\bibinfo {title} {Superconductivity and strong
  correlations in moir{\'e} flat bands},}\ }\href {\doibase
  10.1038/s41567-020-0906-9} {\bibfield  {journal} {\bibinfo  {journal} {Nature
  Physics}\ }\textbf {\bibinfo {volume} {16}},\ \bibinfo {pages} {725--733}
  (\bibinfo {year} {2020})}\BibitemShut {NoStop}%
\bibitem [{\citenamefont {Kennes}\ \emph {et~al.}(2021)\citenamefont {Kennes},
  \citenamefont {Claassen}, \citenamefont {Xian}, \citenamefont {Georges},
  \citenamefont {Millis}, \citenamefont {Hone}, \citenamefont {Dean},
  \citenamefont {Basov}, \citenamefont {Pasupathy},\ and\ \citenamefont
  {Rubio}}]{moireqs}%
  \BibitemOpen
  \bibfield  {author} {\bibinfo {author} {\bibfnamefont {Dante~M.}\
  \bibnamefont {Kennes}}, \bibinfo {author} {\bibfnamefont {Martin}\
  \bibnamefont {Claassen}}, \bibinfo {author} {\bibfnamefont {Lede}\
  \bibnamefont {Xian}}, \bibinfo {author} {\bibfnamefont {Antoine}\
  \bibnamefont {Georges}}, \bibinfo {author} {\bibfnamefont {Andrew~J.}\
  \bibnamefont {Millis}}, \bibinfo {author} {\bibfnamefont {James}\
  \bibnamefont {Hone}}, \bibinfo {author} {\bibfnamefont {Cory~R.}\
  \bibnamefont {Dean}}, \bibinfo {author} {\bibfnamefont {D.~N.}\ \bibnamefont
  {Basov}}, \bibinfo {author} {\bibfnamefont {Abhay~N.}\ \bibnamefont
  {Pasupathy}}, \ and\ \bibinfo {author} {\bibfnamefont {Angel}\ \bibnamefont
  {Rubio}},\ }\bibfield  {title} {\enquote {\bibinfo {title} {Moir{\'e}
  heterostructures as a condensed-matter quantum simulator},}\ }\href {\doibase
  10.1038/s41567-020-01154-3} {\bibfield  {journal} {\bibinfo  {journal}
  {Nature Physics}\ }\textbf {\bibinfo {volume} {17}},\ \bibinfo {pages}
  {155--163} (\bibinfo {year} {2021})}\BibitemShut {NoStop}%
\bibitem [{\citenamefont {Bistritzer}\ and\ \citenamefont
  {MacDonald}(2011)}]{bistritzer2011}%
  \BibitemOpen
  \bibfield  {author} {\bibinfo {author} {\bibfnamefont {Rafi}\ \bibnamefont
  {Bistritzer}}\ and\ \bibinfo {author} {\bibfnamefont {Allan~H}\ \bibnamefont
  {MacDonald}},\ }\bibfield  {title} {\enquote {\bibinfo {title} {Moir{\'e}
  bands in twisted double-layer graphene},}\ }\href@noop {} {\bibfield
  {journal} {\bibinfo  {journal} {Proc. Natl. Acad. Sci. U.S.A.}\ }\textbf
  {\bibinfo {volume} {108}},\ \bibinfo {pages} {12233--12237} (\bibinfo {year}
  {2011})}\BibitemShut {NoStop}%
\bibitem [{\citenamefont {Cao}\ \emph {et~al.}(2018{\natexlab{a}})\citenamefont
  {Cao}, \citenamefont {Fatemi}, \citenamefont {Fang}, \citenamefont
  {Watanabe}, \citenamefont {Taniguchi}, \citenamefont {Kaxiras},\ and\
  \citenamefont {Jarillo-Herrero}}]{cao2018a}%
  \BibitemOpen
  \bibfield  {author} {\bibinfo {author} {\bibfnamefont {Yuan}\ \bibnamefont
  {Cao}}, \bibinfo {author} {\bibfnamefont {Valla}\ \bibnamefont {Fatemi}},
  \bibinfo {author} {\bibfnamefont {Shiang}\ \bibnamefont {Fang}}, \bibinfo
  {author} {\bibfnamefont {Kenji}\ \bibnamefont {Watanabe}}, \bibinfo {author}
  {\bibfnamefont {Takashi}\ \bibnamefont {Taniguchi}}, \bibinfo {author}
  {\bibfnamefont {Efthimios}\ \bibnamefont {Kaxiras}}, \ and\ \bibinfo {author}
  {\bibfnamefont {Pablo}\ \bibnamefont {Jarillo-Herrero}},\ }\bibfield  {title}
  {\enquote {\bibinfo {title} {Unconventional superconductivity in magic-angle
  graphene superlattices},}\ }\href@noop {} {\bibfield  {journal} {\bibinfo
  {journal} {Nature}\ }\textbf {\bibinfo {volume} {556}},\ \bibinfo {pages}
  {43} (\bibinfo {year} {2018}{\natexlab{a}})}\BibitemShut {NoStop}%
\bibitem [{\citenamefont {Cao}\ \emph {et~al.}(2018{\natexlab{b}})\citenamefont
  {Cao}, \citenamefont {Fatemi}, \citenamefont {Demir}, \citenamefont {Fang},
  \citenamefont {Tomarken}, \citenamefont {Luo}, \citenamefont
  {Sanchez-Yamagishi}, \citenamefont {Watanabe}, \citenamefont {Taniguchi},
  \citenamefont {Kaxiras}, \citenamefont {Ashoori},\ and\ \citenamefont
  {Jarillo-Herrero}}]{cao2018b}%
  \BibitemOpen
  \bibfield  {author} {\bibinfo {author} {\bibfnamefont {Yuan}\ \bibnamefont
  {Cao}}, \bibinfo {author} {\bibfnamefont {Valla}\ \bibnamefont {Fatemi}},
  \bibinfo {author} {\bibfnamefont {Ahmet}\ \bibnamefont {Demir}}, \bibinfo
  {author} {\bibfnamefont {Shiang}\ \bibnamefont {Fang}}, \bibinfo {author}
  {\bibfnamefont {Spencer~L}\ \bibnamefont {Tomarken}}, \bibinfo {author}
  {\bibfnamefont {Jason~Y}\ \bibnamefont {Luo}}, \bibinfo {author}
  {\bibfnamefont {Javier~D}\ \bibnamefont {Sanchez-Yamagishi}}, \bibinfo
  {author} {\bibfnamefont {Kenji}\ \bibnamefont {Watanabe}}, \bibinfo {author}
  {\bibfnamefont {Takashi}\ \bibnamefont {Taniguchi}}, \bibinfo {author}
  {\bibfnamefont {Efthimios}\ \bibnamefont {Kaxiras}}, \bibinfo {author}
  {\bibfnamefont {Ray~C}\ \bibnamefont {Ashoori}}, \ and\ \bibinfo {author}
  {\bibfnamefont {Pablo}\ \bibnamefont {Jarillo-Herrero}},\ }\bibfield  {title}
  {\enquote {\bibinfo {title} {Correlated insulator behaviour at half-filling
  in magic-angle graphene superlattices},}\ }\href@noop {} {\bibfield
  {journal} {\bibinfo  {journal} {Nature}\ }\textbf {\bibinfo {volume} {556}},\
  \bibinfo {pages} {80} (\bibinfo {year} {2018}{\natexlab{b}})}\BibitemShut
  {NoStop}%
\bibitem [{\citenamefont {Yankowitz}\ \emph {et~al.}(2019)\citenamefont
  {Yankowitz}, \citenamefont {Chen}, \citenamefont {Polshyn}, \citenamefont
  {Zhang}, \citenamefont {Watanabe}, \citenamefont {Taniguchi}, \citenamefont
  {Graf}, \citenamefont {Young},\ and\ \citenamefont {Dean}}]{Yankowitz18}%
  \BibitemOpen
  \bibfield  {author} {\bibinfo {author} {\bibfnamefont {Matthew}\ \bibnamefont
  {Yankowitz}}, \bibinfo {author} {\bibfnamefont {Shaowen}\ \bibnamefont
  {Chen}}, \bibinfo {author} {\bibfnamefont {Hryhoriy}\ \bibnamefont
  {Polshyn}}, \bibinfo {author} {\bibfnamefont {Yuxuan}\ \bibnamefont {Zhang}},
  \bibinfo {author} {\bibfnamefont {K.}~\bibnamefont {Watanabe}}, \bibinfo
  {author} {\bibfnamefont {T.}~\bibnamefont {Taniguchi}}, \bibinfo {author}
  {\bibfnamefont {David}\ \bibnamefont {Graf}}, \bibinfo {author}
  {\bibfnamefont {Andrea~F.}\ \bibnamefont {Young}}, \ and\ \bibinfo {author}
  {\bibfnamefont {Cory~R.}\ \bibnamefont {Dean}},\ }\bibfield  {title}
  {\enquote {\bibinfo {title} {Tuning superconductivity in twisted bilayer
  graphene},}\ }\href {\doibase 10.1126/science.aav1910} {\bibfield  {journal}
  {\bibinfo  {journal} {Science}\ }\textbf {\bibinfo {volume} {363}},\ \bibinfo
  {pages} {1059--1064} (\bibinfo {year} {2019})}\BibitemShut {NoStop}%
\bibitem [{\citenamefont {Kerelsky}\ \emph {et~al.}(2019)\citenamefont
  {Kerelsky}, \citenamefont {McGilly}, \citenamefont {Kennes}, \citenamefont
  {Xian}, \citenamefont {Yankowitz}, \citenamefont {Chen}, \citenamefont
  {Watanabe}, \citenamefont {Taniguchi}, \citenamefont {Hone}, \citenamefont
  {Dean}, \citenamefont {Rubio},\ and\ \citenamefont {Pasupathy}}]{Kerelsky18}%
  \BibitemOpen
  \bibfield  {author} {\bibinfo {author} {\bibfnamefont {Alexander}\
  \bibnamefont {Kerelsky}}, \bibinfo {author} {\bibfnamefont {Leo~J.}\
  \bibnamefont {McGilly}}, \bibinfo {author} {\bibfnamefont {Dante~M.}\
  \bibnamefont {Kennes}}, \bibinfo {author} {\bibfnamefont {Lede}\ \bibnamefont
  {Xian}}, \bibinfo {author} {\bibfnamefont {Matthew}\ \bibnamefont
  {Yankowitz}}, \bibinfo {author} {\bibfnamefont {Shaowen}\ \bibnamefont
  {Chen}}, \bibinfo {author} {\bibfnamefont {K.}~\bibnamefont {Watanabe}},
  \bibinfo {author} {\bibfnamefont {T.}~\bibnamefont {Taniguchi}}, \bibinfo
  {author} {\bibfnamefont {James}\ \bibnamefont {Hone}}, \bibinfo {author}
  {\bibfnamefont {Cory}\ \bibnamefont {Dean}}, \bibinfo {author} {\bibfnamefont
  {Angel}\ \bibnamefont {Rubio}}, \ and\ \bibinfo {author} {\bibfnamefont
  {Abhay~N.}\ \bibnamefont {Pasupathy}},\ }\bibfield  {title} {\enquote
  {\bibinfo {title} {Maximized electron interactions at the magic angle in
  twisted bilayer graphene},}\ }\href {\doibase 10.1038/s41586-019-1431-9}
  {\bibfield  {journal} {\bibinfo  {journal} {Nature}\ }\textbf {\bibinfo
  {volume} {572}},\ \bibinfo {pages} {95--100} (\bibinfo {year}
  {2019})}\BibitemShut {NoStop}%
\bibitem [{\citenamefont {Liu}\ \emph {et~al.}(2020{\natexlab{a}})\citenamefont
  {Liu}, \citenamefont {Hao}, \citenamefont {Khalaf}, \citenamefont {Lee},
  \citenamefont {Ronen}, \citenamefont {Yoo}, \citenamefont {Najafabadi},
  \citenamefont {Watanabe}, \citenamefont {Taniguchi}, \citenamefont
  {Vishwanath} \emph {et~al.}}]{liu19}%
  \BibitemOpen
  \bibfield  {author} {\bibinfo {author} {\bibfnamefont {Xiaomeng}\
  \bibnamefont {Liu}}, \bibinfo {author} {\bibfnamefont {Zeyu}\ \bibnamefont
  {Hao}}, \bibinfo {author} {\bibfnamefont {Eslam}\ \bibnamefont {Khalaf}},
  \bibinfo {author} {\bibfnamefont {Jong~Yeon}\ \bibnamefont {Lee}}, \bibinfo
  {author} {\bibfnamefont {Yuval}\ \bibnamefont {Ronen}}, \bibinfo {author}
  {\bibfnamefont {Hyobin}\ \bibnamefont {Yoo}}, \bibinfo {author}
  {\bibfnamefont {Danial~Haei}\ \bibnamefont {Najafabadi}}, \bibinfo {author}
  {\bibfnamefont {Kenji}\ \bibnamefont {Watanabe}}, \bibinfo {author}
  {\bibfnamefont {Takashi}\ \bibnamefont {Taniguchi}}, \bibinfo {author}
  {\bibfnamefont {Ashvin}\ \bibnamefont {Vishwanath}},  \emph {et~al.},\
  }\bibfield  {title} {\enquote {\bibinfo {title} {Tunable spin-polarized
  correlated states in twisted double bilayer graphene},}\ }\href@noop {}
  {\bibfield  {journal} {\bibinfo  {journal} {Nature}\ }\textbf {\bibinfo
  {volume} {583}},\ \bibinfo {pages} {221--225} (\bibinfo {year}
  {2020}{\natexlab{a}})}\BibitemShut {NoStop}%
\bibitem [{\citenamefont {Shen}\ \emph {et~al.}(2020)\citenamefont {Shen},
  \citenamefont {Chu}, \citenamefont {Wu}, \citenamefont {Li}, \citenamefont
  {Wang}, \citenamefont {Zhao}, \citenamefont {Tang}, \citenamefont {Liu},
  \citenamefont {Tian}, \citenamefont {Watanabe}, \citenamefont {Taniguchi},
  \citenamefont {Yang}, \citenamefont {Meng}, \citenamefont {Shi},
  \citenamefont {Yazyev},\ and\ \citenamefont {Zhang}}]{shen19}%
  \BibitemOpen
  \bibfield  {author} {\bibinfo {author} {\bibfnamefont {Cheng}\ \bibnamefont
  {Shen}}, \bibinfo {author} {\bibfnamefont {Yanbang}\ \bibnamefont {Chu}},
  \bibinfo {author} {\bibfnamefont {QuanSheng}\ \bibnamefont {Wu}}, \bibinfo
  {author} {\bibfnamefont {Na}~\bibnamefont {Li}}, \bibinfo {author}
  {\bibfnamefont {Shuopei}\ \bibnamefont {Wang}}, \bibinfo {author}
  {\bibfnamefont {Yanchong}\ \bibnamefont {Zhao}}, \bibinfo {author}
  {\bibfnamefont {Jian}\ \bibnamefont {Tang}}, \bibinfo {author} {\bibfnamefont
  {Jieying}\ \bibnamefont {Liu}}, \bibinfo {author} {\bibfnamefont {Jinpeng}\
  \bibnamefont {Tian}}, \bibinfo {author} {\bibfnamefont {Kenji}\ \bibnamefont
  {Watanabe}}, \bibinfo {author} {\bibfnamefont {Takashi}\ \bibnamefont
  {Taniguchi}}, \bibinfo {author} {\bibfnamefont {Rong}\ \bibnamefont {Yang}},
  \bibinfo {author} {\bibfnamefont {Zi~Yang}\ \bibnamefont {Meng}}, \bibinfo
  {author} {\bibfnamefont {Dongxia}\ \bibnamefont {Shi}}, \bibinfo {author}
  {\bibfnamefont {Oleg~V.}\ \bibnamefont {Yazyev}}, \ and\ \bibinfo {author}
  {\bibfnamefont {Guangyu}\ \bibnamefont {Zhang}},\ }\bibfield  {title}
  {\enquote {\bibinfo {title} {Correlated states in twisted double bilayer
  graphene},}\ }\href {\doibase 10.1038/s41567-020-0825-9} {\bibfield
  {journal} {\bibinfo  {journal} {Nature Physics}\ }\textbf {\bibinfo {volume}
  {16}},\ \bibinfo {pages} {520--525} (\bibinfo {year} {2020})}\BibitemShut
  {NoStop}%
\bibitem [{\citenamefont {Cao}\ \emph {et~al.}(2020)\citenamefont {Cao},
  \citenamefont {Rodan-Legrain}, \citenamefont {Rubies-Bigorda}, \citenamefont
  {Park}, \citenamefont {Watanabe}, \citenamefont {Taniguchi},\ and\
  \citenamefont {Jarillo-Herrero}}]{cao2019electric}%
  \BibitemOpen
  \bibfield  {author} {\bibinfo {author} {\bibfnamefont {Yuan}\ \bibnamefont
  {Cao}}, \bibinfo {author} {\bibfnamefont {Daniel}\ \bibnamefont
  {Rodan-Legrain}}, \bibinfo {author} {\bibfnamefont {Oriol}\ \bibnamefont
  {Rubies-Bigorda}}, \bibinfo {author} {\bibfnamefont {Jeong~Min}\ \bibnamefont
  {Park}}, \bibinfo {author} {\bibfnamefont {Kenji}\ \bibnamefont {Watanabe}},
  \bibinfo {author} {\bibfnamefont {Takashi}\ \bibnamefont {Taniguchi}}, \ and\
  \bibinfo {author} {\bibfnamefont {Pablo}\ \bibnamefont {Jarillo-Herrero}},\
  }\bibfield  {title} {\enquote {\bibinfo {title} {Author correction: Tunable
  correlated states and spin-polarized phases in twisted bilayer--bilayer
  graphene},}\ }\href {\doibase 10.1038/s41586-020-2393-7} {\bibfield
  {journal} {\bibinfo  {journal} {Nature}\ }\textbf {\bibinfo {volume} {583}},\
  \bibinfo {pages} {E27--E27} (\bibinfo {year} {2020})}\BibitemShut {NoStop}%
\bibitem [{\citenamefont {Burg}\ \emph {et~al.}(2019)\citenamefont {Burg},
  \citenamefont {Zhu}, \citenamefont {Taniguchi}, \citenamefont {Watanabe},
  \citenamefont {MacDonald},\ and\ \citenamefont {Tutuc}}]{tutuc2019}%
  \BibitemOpen
  \bibfield  {author} {\bibinfo {author} {\bibfnamefont {G.~William}\
  \bibnamefont {Burg}}, \bibinfo {author} {\bibfnamefont {Jihang}\ \bibnamefont
  {Zhu}}, \bibinfo {author} {\bibfnamefont {Takashi}\ \bibnamefont
  {Taniguchi}}, \bibinfo {author} {\bibfnamefont {Kenji}\ \bibnamefont
  {Watanabe}}, \bibinfo {author} {\bibfnamefont {Allan~H.}\ \bibnamefont
  {MacDonald}}, \ and\ \bibinfo {author} {\bibfnamefont {Emanuel}\ \bibnamefont
  {Tutuc}},\ }\bibfield  {title} {\enquote {\bibinfo {title} {Correlated
  insulating states in twisted double bilayer graphene},}\ }\href {\doibase
  10.1103/PhysRevLett.123.197702} {\bibfield  {journal} {\bibinfo  {journal}
  {Phys. Rev. Lett.}\ }\textbf {\bibinfo {volume} {123}},\ \bibinfo {pages}
  {197702} (\bibinfo {year} {2019})}\BibitemShut {NoStop}%
\bibitem [{\citenamefont {Rubio-Verdú}\ \emph {et~al.}(2020)\citenamefont
  {Rubio-Verdú}, \citenamefont {Turkel}, \citenamefont {Song}, \citenamefont
  {Klebl}, \citenamefont {Samajdar}, \citenamefont {Scheurer}, \citenamefont
  {Venderbos}, \citenamefont {Watanabe}, \citenamefont {Taniguchi},
  \citenamefont {Ochoa}, \citenamefont {Xian}, \citenamefont {Kennes},
  \citenamefont {Fernandes}, \citenamefont {Ángel Rubio},\ and\ \citenamefont
  {Pasupathy}}]{RubioVerdu20}%
  \BibitemOpen
  \bibfield  {author} {\bibinfo {author} {\bibfnamefont {Carmen}\ \bibnamefont
  {Rubio-Verdú}}, \bibinfo {author} {\bibfnamefont {Simon}\ \bibnamefont
  {Turkel}}, \bibinfo {author} {\bibfnamefont {Larry}\ \bibnamefont {Song}},
  \bibinfo {author} {\bibfnamefont {Lennart}\ \bibnamefont {Klebl}}, \bibinfo
  {author} {\bibfnamefont {Rhine}\ \bibnamefont {Samajdar}}, \bibinfo {author}
  {\bibfnamefont {Mathias~S.}\ \bibnamefont {Scheurer}}, \bibinfo {author}
  {\bibfnamefont {Jörn W.~F.}\ \bibnamefont {Venderbos}}, \bibinfo {author}
  {\bibfnamefont {Kenji}\ \bibnamefont {Watanabe}}, \bibinfo {author}
  {\bibfnamefont {Takashi}\ \bibnamefont {Taniguchi}}, \bibinfo {author}
  {\bibfnamefont {Héctor}\ \bibnamefont {Ochoa}}, \bibinfo {author}
  {\bibfnamefont {Lede}\ \bibnamefont {Xian}}, \bibinfo {author} {\bibfnamefont
  {Dante}\ \bibnamefont {Kennes}}, \bibinfo {author} {\bibfnamefont
  {Rafael~M.}\ \bibnamefont {Fernandes}}, \bibinfo {author} {\bibnamefont
  {Ángel Rubio}}, \ and\ \bibinfo {author} {\bibfnamefont {Abhay~N.}\
  \bibnamefont {Pasupathy}},\ }\href@noop {} {\enquote {\bibinfo {title}
  {Universal moir\'e nematic phase in twisted graphitic systems},}\ } (\bibinfo
  {year} {2020}),\ \bibinfo {note} {https://arxiv.org/abs/2009.11645
  (09/23/2021)}\BibitemShut {NoStop}%
\bibitem [{\citenamefont {Chen}\ \emph
  {et~al.}(2019{\natexlab{a}})\citenamefont {Chen}, \citenamefont {Sharpe},
  \citenamefont {Gallagher}, \citenamefont {Rosen}, \citenamefont {Fox},
  \citenamefont {Jiang}, \citenamefont {Lyu}, \citenamefont {Li}, \citenamefont
  {Watanabe}, \citenamefont {Taniguchi}, \citenamefont {Jung}, \citenamefont
  {Shi}, \citenamefont {Goldhaber-Gordon}, \citenamefont {Zhang},\ and\
  \citenamefont {Wang}}]{chen2019signatures}%
  \BibitemOpen
  \bibfield  {author} {\bibinfo {author} {\bibfnamefont {Guorui}\ \bibnamefont
  {Chen}}, \bibinfo {author} {\bibfnamefont {Aaron~L}\ \bibnamefont {Sharpe}},
  \bibinfo {author} {\bibfnamefont {Patrick}\ \bibnamefont {Gallagher}},
  \bibinfo {author} {\bibfnamefont {Ilan~T}\ \bibnamefont {Rosen}}, \bibinfo
  {author} {\bibfnamefont {Eli~J}\ \bibnamefont {Fox}}, \bibinfo {author}
  {\bibfnamefont {Lili}\ \bibnamefont {Jiang}}, \bibinfo {author}
  {\bibfnamefont {Bosai}\ \bibnamefont {Lyu}}, \bibinfo {author} {\bibfnamefont
  {Hongyuan}\ \bibnamefont {Li}}, \bibinfo {author} {\bibfnamefont {Kenji}\
  \bibnamefont {Watanabe}}, \bibinfo {author} {\bibfnamefont {Takashi}\
  \bibnamefont {Taniguchi}}, \bibinfo {author} {\bibfnamefont {Jeil}\
  \bibnamefont {Jung}}, \bibinfo {author} {\bibfnamefont {Zhiwen}\ \bibnamefont
  {Shi}}, \bibinfo {author} {\bibfnamefont {David}\ \bibnamefont
  {Goldhaber-Gordon}}, \bibinfo {author} {\bibfnamefont {Yuanbo}\ \bibnamefont
  {Zhang}}, \ and\ \bibinfo {author} {\bibfnamefont {Feng}\ \bibnamefont
  {Wang}},\ }\bibfield  {title} {\enquote {\bibinfo {title} {Signatures of
  tunable superconductivity in a trilayer graphene moir{\'e} superlattice},}\
  }\href@noop {} {\bibfield  {journal} {\bibinfo  {journal} {Nature}\ }\textbf
  {\bibinfo {volume} {572}},\ \bibinfo {pages} {215--219} (\bibinfo {year}
  {2019}{\natexlab{a}})}\BibitemShut {NoStop}%
\bibitem [{\citenamefont {Chen}\ \emph
  {et~al.}(2019{\natexlab{b}})\citenamefont {Chen}, \citenamefont {Jiang},
  \citenamefont {Wu}, \citenamefont {Lyu}, \citenamefont {Li}, \citenamefont
  {Chittari}, \citenamefont {Watanabe}, \citenamefont {Taniguchi},
  \citenamefont {Shi}, \citenamefont {Jung}, \citenamefont {Zhang},\ and\
  \citenamefont {Wang}}]{chen2019evidence}%
  \BibitemOpen
  \bibfield  {author} {\bibinfo {author} {\bibfnamefont {Guorui}\ \bibnamefont
  {Chen}}, \bibinfo {author} {\bibfnamefont {Lili}\ \bibnamefont {Jiang}},
  \bibinfo {author} {\bibfnamefont {Shuang}\ \bibnamefont {Wu}}, \bibinfo
  {author} {\bibfnamefont {Bosai}\ \bibnamefont {Lyu}}, \bibinfo {author}
  {\bibfnamefont {Hongyuan}\ \bibnamefont {Li}}, \bibinfo {author}
  {\bibfnamefont {Bheema~Lingam}\ \bibnamefont {Chittari}}, \bibinfo {author}
  {\bibfnamefont {Kenji}\ \bibnamefont {Watanabe}}, \bibinfo {author}
  {\bibfnamefont {Takashi}\ \bibnamefont {Taniguchi}}, \bibinfo {author}
  {\bibfnamefont {Zhiwen}\ \bibnamefont {Shi}}, \bibinfo {author}
  {\bibfnamefont {Jeil}\ \bibnamefont {Jung}}, \bibinfo {author} {\bibfnamefont
  {Yuanbo}\ \bibnamefont {Zhang}}, \ and\ \bibinfo {author} {\bibfnamefont
  {Feng}\ \bibnamefont {Wang}},\ }\bibfield  {title} {\enquote {\bibinfo
  {title} {Evidence of a gate-tunable mott insulator in a trilayer graphene
  moir{\'e} superlattice},}\ }\href@noop {} {\bibfield  {journal} {\bibinfo
  {journal} {Nat. Phys.}\ }\textbf {\bibinfo {volume} {15}},\ \bibinfo {pages}
  {237--241} (\bibinfo {year} {2019}{\natexlab{b}})}\BibitemShut {NoStop}%
\bibitem [{\citenamefont {Chen}\ \emph {et~al.}(2021)\citenamefont {Chen},
  \citenamefont {He}, \citenamefont {Zhang}, \citenamefont {Hsieh},
  \citenamefont {Fei}, \citenamefont {Watanabe}, \citenamefont {Taniguchi},
  \citenamefont {Cobden}, \citenamefont {Xu}, \citenamefont {Dean},\ and\
  \citenamefont {Yankowitz}}]{chen2020electrically}%
  \BibitemOpen
  \bibfield  {author} {\bibinfo {author} {\bibfnamefont {Shaowen}\ \bibnamefont
  {Chen}}, \bibinfo {author} {\bibfnamefont {Minhao}\ \bibnamefont {He}},
  \bibinfo {author} {\bibfnamefont {Ya-Hui}\ \bibnamefont {Zhang}}, \bibinfo
  {author} {\bibfnamefont {Valerie}\ \bibnamefont {Hsieh}}, \bibinfo {author}
  {\bibfnamefont {Zaiyao}\ \bibnamefont {Fei}}, \bibinfo {author}
  {\bibfnamefont {K.}~\bibnamefont {Watanabe}}, \bibinfo {author}
  {\bibfnamefont {T.}~\bibnamefont {Taniguchi}}, \bibinfo {author}
  {\bibfnamefont {David~H.}\ \bibnamefont {Cobden}}, \bibinfo {author}
  {\bibfnamefont {Xiaodong}\ \bibnamefont {Xu}}, \bibinfo {author}
  {\bibfnamefont {Cory~R.}\ \bibnamefont {Dean}}, \ and\ \bibinfo {author}
  {\bibfnamefont {Matthew}\ \bibnamefont {Yankowitz}},\ }\bibfield  {title}
  {\enquote {\bibinfo {title} {Electrically tunable correlated and topological
  states in twisted monolayer--bilayer graphene},}\ }\href {\doibase
  10.1038/s41567-020-01062-6} {\bibfield  {journal} {\bibinfo  {journal}
  {Nature Physics}\ }\textbf {\bibinfo {volume} {17}},\ \bibinfo {pages}
  {374--380} (\bibinfo {year} {2021})}\BibitemShut {NoStop}%
\bibitem [{\citenamefont {Xu}\ \emph {et~al.}(2021)\citenamefont {Xu},
  \citenamefont {Al~Ezzi}, \citenamefont {Balakrishnan}, \citenamefont
  {Garcia-Ruiz}, \citenamefont {Tsim}, \citenamefont {Mullan}, \citenamefont
  {Barrier}, \citenamefont {Xin}, \citenamefont {Piot}, \citenamefont
  {Taniguchi}, \citenamefont {Watanabe}, \citenamefont {Carvalho},
  \citenamefont {Mishchenko}, \citenamefont {Geim}, \citenamefont {Fal'ko},
  \citenamefont {Adam}, \citenamefont {Neto}, \citenamefont {Novoselov},\ and\
  \citenamefont {Shi}}]{shi2020tunable}%
  \BibitemOpen
  \bibfield  {author} {\bibinfo {author} {\bibfnamefont {Shuigang}\
  \bibnamefont {Xu}}, \bibinfo {author} {\bibfnamefont {Mohammed~M.}\
  \bibnamefont {Al~Ezzi}}, \bibinfo {author} {\bibfnamefont {Nilanthy}\
  \bibnamefont {Balakrishnan}}, \bibinfo {author} {\bibfnamefont {Aitor}\
  \bibnamefont {Garcia-Ruiz}}, \bibinfo {author} {\bibfnamefont {Bonnie}\
  \bibnamefont {Tsim}}, \bibinfo {author} {\bibfnamefont {Ciaran}\ \bibnamefont
  {Mullan}}, \bibinfo {author} {\bibfnamefont {Julien}\ \bibnamefont
  {Barrier}}, \bibinfo {author} {\bibfnamefont {Na}~\bibnamefont {Xin}},
  \bibinfo {author} {\bibfnamefont {Benjamin~A.}\ \bibnamefont {Piot}},
  \bibinfo {author} {\bibfnamefont {Takashi}\ \bibnamefont {Taniguchi}},
  \bibinfo {author} {\bibfnamefont {Kenji}\ \bibnamefont {Watanabe}}, \bibinfo
  {author} {\bibfnamefont {Alexandra}\ \bibnamefont {Carvalho}}, \bibinfo
  {author} {\bibfnamefont {Artem}\ \bibnamefont {Mishchenko}}, \bibinfo
  {author} {\bibfnamefont {A.~K.}\ \bibnamefont {Geim}}, \bibinfo {author}
  {\bibfnamefont {Vladimir~I.}\ \bibnamefont {Fal'ko}}, \bibinfo {author}
  {\bibfnamefont {Shaffique}\ \bibnamefont {Adam}}, \bibinfo {author}
  {\bibfnamefont {Antonio Helio~Castro}\ \bibnamefont {Neto}}, \bibinfo
  {author} {\bibfnamefont {Kostya~S.}\ \bibnamefont {Novoselov}}, \ and\
  \bibinfo {author} {\bibfnamefont {Yanmeng}\ \bibnamefont {Shi}},\ }\bibfield
  {title} {\enquote {\bibinfo {title} {Tunable van hove singularities and
  correlated states in twisted monolayer--bilayer graphene},}\ }\href {\doibase
  10.1038/s41567-021-01172-9} {\bibfield  {journal} {\bibinfo  {journal}
  {Nature Physics}\ }\textbf {\bibinfo {volume} {17}},\ \bibinfo {pages}
  {619--626} (\bibinfo {year} {2021})}\BibitemShut {NoStop}%
\bibitem [{\citenamefont {Wang}\ \emph {et~al.}(2020)\citenamefont {Wang},
  \citenamefont {Shih}, \citenamefont {Ghiotto}, \citenamefont {Xian},
  \citenamefont {Rhodes}, \citenamefont {Tan}, \citenamefont {Claassen},
  \citenamefont {Kennes}, \citenamefont {Bai}, \citenamefont {Kim},
  \citenamefont {Watanabe}, \citenamefont {Taniguchi}, \citenamefont {Zhu},
  \citenamefont {Hone}, \citenamefont {Rubio}, \citenamefont {Pasupathy},\ and\
  \citenamefont {Dean}}]{wang19}%
  \BibitemOpen
  \bibfield  {author} {\bibinfo {author} {\bibfnamefont {Lei}\ \bibnamefont
  {Wang}}, \bibinfo {author} {\bibfnamefont {En-Min}\ \bibnamefont {Shih}},
  \bibinfo {author} {\bibfnamefont {Augusto}\ \bibnamefont {Ghiotto}}, \bibinfo
  {author} {\bibfnamefont {Lede}\ \bibnamefont {Xian}}, \bibinfo {author}
  {\bibfnamefont {Daniel~A.}\ \bibnamefont {Rhodes}}, \bibinfo {author}
  {\bibfnamefont {Cheng}\ \bibnamefont {Tan}}, \bibinfo {author} {\bibfnamefont
  {Martin}\ \bibnamefont {Claassen}}, \bibinfo {author} {\bibfnamefont
  {Dante~M.}\ \bibnamefont {Kennes}}, \bibinfo {author} {\bibfnamefont
  {Yusong}\ \bibnamefont {Bai}}, \bibinfo {author} {\bibfnamefont {Bumho}\
  \bibnamefont {Kim}}, \bibinfo {author} {\bibfnamefont {Kenji}\ \bibnamefont
  {Watanabe}}, \bibinfo {author} {\bibfnamefont {Takashi}\ \bibnamefont
  {Taniguchi}}, \bibinfo {author} {\bibfnamefont {Xiaoyang}\ \bibnamefont
  {Zhu}}, \bibinfo {author} {\bibfnamefont {James}\ \bibnamefont {Hone}},
  \bibinfo {author} {\bibfnamefont {Angel}\ \bibnamefont {Rubio}}, \bibinfo
  {author} {\bibfnamefont {Abhay~N.}\ \bibnamefont {Pasupathy}}, \ and\
  \bibinfo {author} {\bibfnamefont {Cory~R.}\ \bibnamefont {Dean}},\ }\bibfield
   {title} {\enquote {\bibinfo {title} {Correlated electronic phases in twisted
  bilayer transition metal dichalcogenides},}\ }\href {\doibase
  10.1038/s41563-020-0708-6} {\bibfield  {journal} {\bibinfo  {journal} {Nature
  Materials}\ }\textbf {\bibinfo {volume} {19}},\ \bibinfo {pages} {861--866}
  (\bibinfo {year} {2020})}\BibitemShut {NoStop}%
\bibitem [{\citenamefont {An}\ \emph {et~al.}(2020)\citenamefont {An},
  \citenamefont {Cai}, \citenamefont {Pei}, \citenamefont {Huang},
  \citenamefont {Wu}, \citenamefont {Zhou}, \citenamefont {Lin}, \citenamefont
  {Ying}, \citenamefont {Ye}, \citenamefont {Feng} \emph {et~al.}}]{an19}%
  \BibitemOpen
  \bibfield  {author} {\bibinfo {author} {\bibfnamefont {Liheng}\ \bibnamefont
  {An}}, \bibinfo {author} {\bibfnamefont {Xiangbin}\ \bibnamefont {Cai}},
  \bibinfo {author} {\bibfnamefont {Ding}\ \bibnamefont {Pei}}, \bibinfo
  {author} {\bibfnamefont {Meizhen}\ \bibnamefont {Huang}}, \bibinfo {author}
  {\bibfnamefont {Zefei}\ \bibnamefont {Wu}}, \bibinfo {author} {\bibfnamefont
  {Zishu}\ \bibnamefont {Zhou}}, \bibinfo {author} {\bibfnamefont {Jiangxiazi}\
  \bibnamefont {Lin}}, \bibinfo {author} {\bibfnamefont {Zhehan}\ \bibnamefont
  {Ying}}, \bibinfo {author} {\bibfnamefont {Ziqing}\ \bibnamefont {Ye}},
  \bibinfo {author} {\bibfnamefont {Xuemeng}\ \bibnamefont {Feng}},  \emph
  {et~al.},\ }\bibfield  {title} {\enquote {\bibinfo {title} {Interaction
  effects and superconductivity signatures in twisted double-bilayer wse2},}\
  }\href@noop {} {\bibfield  {journal} {\bibinfo  {journal} {Nanoscale Horiz.}\
  }\textbf {\bibinfo {volume} {5}},\ \bibinfo {pages} {1309--1316} (\bibinfo
  {year} {2020})}\BibitemShut {NoStop}%
\bibitem [{\citenamefont {Pan}\ \emph {et~al.}(2020)\citenamefont {Pan},
  \citenamefont {Wu},\ and\ \citenamefont {Das~Sarma}}]{pan2020band}%
  \BibitemOpen
  \bibfield  {author} {\bibinfo {author} {\bibfnamefont {Haining}\ \bibnamefont
  {Pan}}, \bibinfo {author} {\bibfnamefont {Fengcheng}\ \bibnamefont {Wu}}, \
  and\ \bibinfo {author} {\bibfnamefont {Sankar}\ \bibnamefont {Das~Sarma}},\
  }\bibfield  {title} {\enquote {\bibinfo {title} {Band topology, hubbard
  model, heisenberg model, and dzyaloshinskii-moriya interaction in twisted
  bilayer ${\mathrm{wse}}_{2}$},}\ }\href {\doibase
  10.1103/PhysRevResearch.2.033087} {\bibfield  {journal} {\bibinfo  {journal}
  {Phys. Rev. Research}\ }\textbf {\bibinfo {volume} {2}},\ \bibinfo {pages}
  {033087} (\bibinfo {year} {2020})}\BibitemShut {NoStop}%
\bibitem [{\citenamefont {Liao}\ \emph {et~al.}(2020)\citenamefont {Liao},
  \citenamefont {Wei}, \citenamefont {Du}, \citenamefont {Wang}, \citenamefont
  {Tang}, \citenamefont {Yu}, \citenamefont {Wu}, \citenamefont {Zhao},
  \citenamefont {Xu}, \citenamefont {Han} \emph {et~al.}}]{liao2020precise}%
  \BibitemOpen
  \bibfield  {author} {\bibinfo {author} {\bibfnamefont {Mengzhou}\
  \bibnamefont {Liao}}, \bibinfo {author} {\bibfnamefont {Zheng}\ \bibnamefont
  {Wei}}, \bibinfo {author} {\bibfnamefont {Luojun}\ \bibnamefont {Du}},
  \bibinfo {author} {\bibfnamefont {Qinqin}\ \bibnamefont {Wang}}, \bibinfo
  {author} {\bibfnamefont {Jian}\ \bibnamefont {Tang}}, \bibinfo {author}
  {\bibfnamefont {Hua}\ \bibnamefont {Yu}}, \bibinfo {author} {\bibfnamefont
  {Fanfan}\ \bibnamefont {Wu}}, \bibinfo {author} {\bibfnamefont {Jiaojiao}\
  \bibnamefont {Zhao}}, \bibinfo {author} {\bibfnamefont {Xiaozhi}\
  \bibnamefont {Xu}}, \bibinfo {author} {\bibfnamefont {Bo}~\bibnamefont
  {Han}},  \emph {et~al.},\ }\bibfield  {title} {\enquote {\bibinfo {title}
  {Precise control of the interlayer twist angle in large scale mos 2
  homostructures},}\ }\href@noop {} {\bibfield  {journal} {\bibinfo  {journal}
  {Nature Communications}\ }\textbf {\bibinfo {volume} {11}},\ \bibinfo {pages}
  {1--8} (\bibinfo {year} {2020})}\BibitemShut {NoStop}%
\bibitem [{\citenamefont {Naik}\ and\ \citenamefont {Jain}(2018)}]{Naik18}%
  \BibitemOpen
  \bibfield  {author} {\bibinfo {author} {\bibfnamefont {Mit~H.}\ \bibnamefont
  {Naik}}\ and\ \bibinfo {author} {\bibfnamefont {Manish}\ \bibnamefont
  {Jain}},\ }\bibfield  {title} {\enquote {\bibinfo {title} {Ultraflatbands and
  shear solitons in moir\'e patterns of twisted bilayer transition metal
  dichalcogenides},}\ }\href {\doibase 10.1103/PhysRevLett.121.266401}
  {\bibfield  {journal} {\bibinfo  {journal} {Phys. Rev. Lett.}\ }\textbf
  {\bibinfo {volume} {121}},\ \bibinfo {pages} {266401} (\bibinfo {year}
  {2018})}\BibitemShut {NoStop}%
\bibitem [{\citenamefont {Xian}\ \emph {et~al.}(2020)\citenamefont {Xian},
  \citenamefont {Claassen}, \citenamefont {Kiese}, \citenamefont {Scherer},
  \citenamefont {Trebst}, \citenamefont {Kennes},\ and\ \citenamefont
  {Rubio}}]{xian20}%
  \BibitemOpen
  \bibfield  {author} {\bibinfo {author} {\bibfnamefont {Lede}\ \bibnamefont
  {Xian}}, \bibinfo {author} {\bibfnamefont {Martin}\ \bibnamefont {Claassen}},
  \bibinfo {author} {\bibfnamefont {Dominik}\ \bibnamefont {Kiese}}, \bibinfo
  {author} {\bibfnamefont {Michael~M.}\ \bibnamefont {Scherer}}, \bibinfo
  {author} {\bibfnamefont {Simon}\ \bibnamefont {Trebst}}, \bibinfo {author}
  {\bibfnamefont {Dante~M.}\ \bibnamefont {Kennes}}, \ and\ \bibinfo {author}
  {\bibfnamefont {Angel}\ \bibnamefont {Rubio}},\ }\href@noop {} {\enquote
  {\bibinfo {title} {Realization of nearly dispersionless bands with strong
  orbital anisotropy from destructive interference in twisted bilayer mos2},}\
  } (\bibinfo {year} {2020}),\ \bibinfo {note}
  {https://arxiv.org/abs/2004.02964 (09/23/2021)}\BibitemShut {NoStop}%
\bibitem [{\citenamefont {Wu}\ \emph {et~al.}(2018)\citenamefont {Wu},
  \citenamefont {Lovorn}, \citenamefont {Tutuc},\ and\ \citenamefont
  {MacDonald}}]{wu17}%
  \BibitemOpen
  \bibfield  {author} {\bibinfo {author} {\bibfnamefont {Fengcheng}\
  \bibnamefont {Wu}}, \bibinfo {author} {\bibfnamefont {Timothy}\ \bibnamefont
  {Lovorn}}, \bibinfo {author} {\bibfnamefont {Emanuel}\ \bibnamefont {Tutuc}},
  \ and\ \bibinfo {author} {\bibfnamefont {A.~H.}\ \bibnamefont {MacDonald}},\
  }\bibfield  {title} {\enquote {\bibinfo {title} {Hubbard model physics in
  transition metal dichalcogenide moir\'e bands},}\ }\href {\doibase
  10.1103/PhysRevLett.121.026402} {\bibfield  {journal} {\bibinfo  {journal}
  {Phys. Rev. Lett.}\ }\textbf {\bibinfo {volume} {121}},\ \bibinfo {pages}
  {026402} (\bibinfo {year} {2018})}\BibitemShut {NoStop}%
\bibitem [{\citenamefont {Zhang}\ \emph {et~al.}(2020)\citenamefont {Zhang},
  \citenamefont {Yuan},\ and\ \citenamefont {Fu}}]{zhang2019moir}%
  \BibitemOpen
  \bibfield  {author} {\bibinfo {author} {\bibfnamefont {Yang}\ \bibnamefont
  {Zhang}}, \bibinfo {author} {\bibfnamefont {Noah F.~Q.}\ \bibnamefont
  {Yuan}}, \ and\ \bibinfo {author} {\bibfnamefont {Liang}\ \bibnamefont
  {Fu}},\ }\bibfield  {title} {\enquote {\bibinfo {title} {Moir\'e quantum
  chemistry: Charge transfer in transition metal dichalcogenide
  superlattices},}\ }\href {\doibase 10.1103/PhysRevB.102.201115} {\bibfield
  {journal} {\bibinfo  {journal} {Phys. Rev. B}\ }\textbf {\bibinfo {volume}
  {102}},\ \bibinfo {pages} {201115} (\bibinfo {year} {2020})}\BibitemShut
  {NoStop}%
\bibitem [{\citenamefont {Regan}\ \emph {et~al.}(2020)\citenamefont {Regan},
  \citenamefont {Wang}, \citenamefont {Jin}, \citenamefont {Utama},
  \citenamefont {Gao}, \citenamefont {Wei}, \citenamefont {Zhao}, \citenamefont
  {Zhao}, \citenamefont {Zhang}, \citenamefont {Yumigeta}, \citenamefont
  {Blei}, \citenamefont {Carlström}, \citenamefont {Watanabe}, \citenamefont
  {Taniguchi}, \citenamefont {Tongay}, \citenamefont {Crommie}, \citenamefont
  {Zettl},\ and\ \citenamefont {Wang}}]{regan2020}%
  \BibitemOpen
  \bibfield  {author} {\bibinfo {author} {\bibfnamefont {Emma~C.}\ \bibnamefont
  {Regan}}, \bibinfo {author} {\bibfnamefont {Danqing}\ \bibnamefont {Wang}},
  \bibinfo {author} {\bibfnamefont {Chenhao}\ \bibnamefont {Jin}}, \bibinfo
  {author} {\bibfnamefont {M.~Iqbal~Bakti}\ \bibnamefont {Utama}}, \bibinfo
  {author} {\bibfnamefont {Beini}\ \bibnamefont {Gao}}, \bibinfo {author}
  {\bibfnamefont {Xin}\ \bibnamefont {Wei}}, \bibinfo {author} {\bibfnamefont
  {Sihan}\ \bibnamefont {Zhao}}, \bibinfo {author} {\bibfnamefont {Wenyu}\
  \bibnamefont {Zhao}}, \bibinfo {author} {\bibfnamefont {Zuocheng}\
  \bibnamefont {Zhang}}, \bibinfo {author} {\bibfnamefont {Kentaro}\
  \bibnamefont {Yumigeta}}, \bibinfo {author} {\bibfnamefont {Mark}\
  \bibnamefont {Blei}}, \bibinfo {author} {\bibfnamefont {Johan~D.}\
  \bibnamefont {Carlström}}, \bibinfo {author} {\bibfnamefont {Kenji}\
  \bibnamefont {Watanabe}}, \bibinfo {author} {\bibfnamefont {Takashi}\
  \bibnamefont {Taniguchi}}, \bibinfo {author} {\bibfnamefont {Sefaattin}\
  \bibnamefont {Tongay}}, \bibinfo {author} {\bibfnamefont {Michael}\
  \bibnamefont {Crommie}}, \bibinfo {author} {\bibfnamefont {Alex}\
  \bibnamefont {Zettl}}, \ and\ \bibinfo {author} {\bibfnamefont {Feng}\
  \bibnamefont {Wang}},\ }\bibfield  {title} {\enquote {\bibinfo {title} {Mott
  and generalized wigner crystal states in wse$_2$/ws$_2$ moir\'e
  superlattices},}\ }\href@noop {} {\bibfield  {journal} {\bibinfo  {journal}
  {Nature}\ }\textbf {\bibinfo {volume} {579}},\ \bibinfo {pages} {359--363}
  (\bibinfo {year} {2020})}\BibitemShut {NoStop}%
\bibitem [{\citenamefont {Tang}\ \emph {et~al.}(2020)\citenamefont {Tang},
  \citenamefont {Li}, \citenamefont {Li}, \citenamefont {Xu}, \citenamefont
  {Liu}, \citenamefont {Barmak}, \citenamefont {Watanabe}, \citenamefont
  {Taniguchi}, \citenamefont {MacDonald}, \citenamefont {Shan},\ and\
  \citenamefont {Mak}}]{tang2020}%
  \BibitemOpen
  \bibfield  {author} {\bibinfo {author} {\bibfnamefont {Yanhao}\ \bibnamefont
  {Tang}}, \bibinfo {author} {\bibfnamefont {Lizhong}\ \bibnamefont {Li}},
  \bibinfo {author} {\bibfnamefont {Tingxin}\ \bibnamefont {Li}}, \bibinfo
  {author} {\bibfnamefont {Yang}\ \bibnamefont {Xu}}, \bibinfo {author}
  {\bibfnamefont {Song}\ \bibnamefont {Liu}}, \bibinfo {author} {\bibfnamefont
  {Katayun}\ \bibnamefont {Barmak}}, \bibinfo {author} {\bibfnamefont {Kenji}\
  \bibnamefont {Watanabe}}, \bibinfo {author} {\bibfnamefont {Takashi}\
  \bibnamefont {Taniguchi}}, \bibinfo {author} {\bibfnamefont {Allan~H.}\
  \bibnamefont {MacDonald}}, \bibinfo {author} {\bibfnamefont {Jie}\
  \bibnamefont {Shan}}, \ and\ \bibinfo {author} {\bibfnamefont {Kin~Fai}\
  \bibnamefont {Mak}},\ }\bibfield  {title} {\enquote {\bibinfo {title}
  {Simulation of hubbard model physics in wse$_2$/ws$_2$ moir\'e
  superlattices},}\ }\href@noop {} {\bibfield  {journal} {\bibinfo  {journal}
  {Nature}\ }\textbf {\bibinfo {volume} {579}},\ \bibinfo {pages} {353--358}
  (\bibinfo {year} {2020})}\BibitemShut {NoStop}%
\bibitem [{\citenamefont {Basov}\ \emph {et~al.}(2017)\citenamefont {Basov},
  \citenamefont {Averitt},\ and\ \citenamefont {Hsieh}}]{Basov2017}%
  \BibitemOpen
  \bibfield  {author} {\bibinfo {author} {\bibfnamefont {D.~N.}\ \bibnamefont
  {Basov}}, \bibinfo {author} {\bibfnamefont {R.~D.}\ \bibnamefont {Averitt}},
  \ and\ \bibinfo {author} {\bibfnamefont {D.}~\bibnamefont {Hsieh}},\
  }\bibfield  {title} {\enquote {\bibinfo {title} {Towards properties on demand
  in quantum materials},}\ }\href {\doibase 10.1038/nmat5017} {\bibfield
  {journal} {\bibinfo  {journal} {Nat. Mater.}\ }\textbf {\bibinfo {volume}
  {16}},\ \bibinfo {pages} {1077--1088} (\bibinfo {year} {2017})}\BibitemShut
  {NoStop}%
\bibitem [{\citenamefont {Halbertal}\ \emph {et~al.}(2021)\citenamefont
  {Halbertal}, \citenamefont {Finney}, \citenamefont {Sunku}, \citenamefont
  {Kerelsky}, \citenamefont {Rubio-Verd{\'u}}, \citenamefont {Shabani},
  \citenamefont {Xian}, \citenamefont {Carr}, \citenamefont {Chen},
  \citenamefont {Zhang}, \citenamefont {Wang}, \citenamefont
  {Gonzalez-Acevedo}, \citenamefont {McLeod}, \citenamefont {Rhodes},
  \citenamefont {Watanabe}, \citenamefont {Taniguchi}, \citenamefont {Kaxiras},
  \citenamefont {Dean}, \citenamefont {Hone}, \citenamefont {Pasupathy},
  \citenamefont {Kennes}, \citenamefont {Rubio},\ and\ \citenamefont
  {Basov}}]{Halbertal20}%
  \BibitemOpen
  \bibfield  {author} {\bibinfo {author} {\bibfnamefont {Dorri}\ \bibnamefont
  {Halbertal}}, \bibinfo {author} {\bibfnamefont {Nathan~R.}\ \bibnamefont
  {Finney}}, \bibinfo {author} {\bibfnamefont {Sai~S.}\ \bibnamefont {Sunku}},
  \bibinfo {author} {\bibfnamefont {Alexander}\ \bibnamefont {Kerelsky}},
  \bibinfo {author} {\bibfnamefont {Carmen}\ \bibnamefont {Rubio-Verd{\'u}}},
  \bibinfo {author} {\bibfnamefont {Sara}\ \bibnamefont {Shabani}}, \bibinfo
  {author} {\bibfnamefont {Lede}\ \bibnamefont {Xian}}, \bibinfo {author}
  {\bibfnamefont {Stephen}\ \bibnamefont {Carr}}, \bibinfo {author}
  {\bibfnamefont {Shaowen}\ \bibnamefont {Chen}}, \bibinfo {author}
  {\bibfnamefont {Charles}\ \bibnamefont {Zhang}}, \bibinfo {author}
  {\bibfnamefont {Lei}\ \bibnamefont {Wang}}, \bibinfo {author} {\bibfnamefont
  {Derick}\ \bibnamefont {Gonzalez-Acevedo}}, \bibinfo {author} {\bibfnamefont
  {Alexander~S.}\ \bibnamefont {McLeod}}, \bibinfo {author} {\bibfnamefont
  {Daniel}\ \bibnamefont {Rhodes}}, \bibinfo {author} {\bibfnamefont {Kenji}\
  \bibnamefont {Watanabe}}, \bibinfo {author} {\bibfnamefont {Takashi}\
  \bibnamefont {Taniguchi}}, \bibinfo {author} {\bibfnamefont {Efthimios}\
  \bibnamefont {Kaxiras}}, \bibinfo {author} {\bibfnamefont {Cory~R.}\
  \bibnamefont {Dean}}, \bibinfo {author} {\bibfnamefont {James~C.}\
  \bibnamefont {Hone}}, \bibinfo {author} {\bibfnamefont {Abhay~N.}\
  \bibnamefont {Pasupathy}}, \bibinfo {author} {\bibfnamefont {Dante~M.}\
  \bibnamefont {Kennes}}, \bibinfo {author} {\bibfnamefont {Angel}\
  \bibnamefont {Rubio}}, \ and\ \bibinfo {author} {\bibfnamefont {D.~N.}\
  \bibnamefont {Basov}},\ }\bibfield  {title} {\enquote {\bibinfo {title}
  {Moir{\'e} metrology of energy landscapes in van der waals
  heterostructures},}\ }\href {\doibase 10.1038/s41467-020-20428-1} {\bibfield
  {journal} {\bibinfo  {journal} {Nature Communications}\ }\textbf {\bibinfo
  {volume} {12}},\ \bibinfo {pages} {242} (\bibinfo {year} {2021})}\BibitemShut
  {NoStop}%
\bibitem [{\citenamefont {Xian}\ \emph {et~al.}(2019)\citenamefont {Xian},
  \citenamefont {Kennes}, \citenamefont {Tancogne-Dejean}, \citenamefont
  {Altarelli},\ and\ \citenamefont {Rubio}}]{Xian18}%
  \BibitemOpen
  \bibfield  {author} {\bibinfo {author} {\bibfnamefont {Lede}\ \bibnamefont
  {Xian}}, \bibinfo {author} {\bibfnamefont {Dante~M.}\ \bibnamefont {Kennes}},
  \bibinfo {author} {\bibfnamefont {Nicolas}\ \bibnamefont {Tancogne-Dejean}},
  \bibinfo {author} {\bibfnamefont {Massimo}\ \bibnamefont {Altarelli}}, \ and\
  \bibinfo {author} {\bibfnamefont {Angel}\ \bibnamefont {Rubio}},\ }\bibfield
  {title} {\enquote {\bibinfo {title} {Multiflat bands and strong correlations
  in twisted bilayer boron nitride: Doping-induced correlated insulator and
  superconductor},}\ }\href@noop {} {\bibfield  {journal} {\bibinfo  {journal}
  {Nano Lett.}\ }\textbf {\bibinfo {volume} {19}},\ \bibinfo {pages}
  {4934--4940} (\bibinfo {year} {2019})}\BibitemShut {NoStop}%
\bibitem [{\citenamefont {Zhao}\ \emph {et~al.}(2020)\citenamefont {Zhao},
  \citenamefont {Yang}, \citenamefont {Zhang},\ and\ \citenamefont
  {Wei}}]{zhao2020formation}%
  \BibitemOpen
  \bibfield  {author} {\bibinfo {author} {\bibfnamefont {Xing-Ju}\ \bibnamefont
  {Zhao}}, \bibinfo {author} {\bibfnamefont {Yang}\ \bibnamefont {Yang}},
  \bibinfo {author} {\bibfnamefont {Dong-Bo}\ \bibnamefont {Zhang}}, \ and\
  \bibinfo {author} {\bibfnamefont {Su-Huai}\ \bibnamefont {Wei}},\ }\bibfield
  {title} {\enquote {\bibinfo {title} {Formation of bloch flat bands in polar
  twisted bilayers without magic angles},}\ }\href@noop {} {\bibfield
  {journal} {\bibinfo  {journal} {Physical Review Letters}\ }\textbf {\bibinfo
  {volume} {124}},\ \bibinfo {pages} {086401} (\bibinfo {year}
  {2020})}\BibitemShut {NoStop}%
\bibitem [{\citenamefont {Tong}\ \emph {et~al.}(2018)\citenamefont {Tong},
  \citenamefont {Liu}, \citenamefont {Xiao},\ and\ \citenamefont
  {Yao}}]{tong2018skyrmions}%
  \BibitemOpen
  \bibfield  {author} {\bibinfo {author} {\bibfnamefont {Qingjun}\ \bibnamefont
  {Tong}}, \bibinfo {author} {\bibfnamefont {Fei}\ \bibnamefont {Liu}},
  \bibinfo {author} {\bibfnamefont {Jiang}\ \bibnamefont {Xiao}}, \ and\
  \bibinfo {author} {\bibfnamefont {Wang}\ \bibnamefont {Yao}},\ }\bibfield
  {title} {\enquote {\bibinfo {title} {Skyrmions in the moir{\'e} of van der
  waals 2d magnets},}\ }\href@noop {} {\bibfield  {journal} {\bibinfo
  {journal} {Nano Lett.}\ }\textbf {\bibinfo {volume} {18}},\ \bibinfo {pages}
  {7194--7199} (\bibinfo {year} {2018})}\BibitemShut {NoStop}%
\bibitem [{\citenamefont {Hejazi}\ \emph {et~al.}(2020)\citenamefont {Hejazi},
  \citenamefont {Luo},\ and\ \citenamefont {Balents}}]{hejazi2020noncollinear}%
  \BibitemOpen
  \bibfield  {author} {\bibinfo {author} {\bibfnamefont {Kasra}\ \bibnamefont
  {Hejazi}}, \bibinfo {author} {\bibfnamefont {Zhu-Xi}\ \bibnamefont {Luo}}, \
  and\ \bibinfo {author} {\bibfnamefont {Leon}\ \bibnamefont {Balents}},\
  }\bibfield  {title} {\enquote {\bibinfo {title} {Noncollinear phases in
  moir{\'e} magnets},}\ }\href@noop {} {\bibfield  {journal} {\bibinfo
  {journal} {Proc. Natl. Acad. Sci. U.S.A.}\ }\textbf {\bibinfo {volume}
  {117}},\ \bibinfo {pages} {10721--10726} (\bibinfo {year}
  {2020})}\BibitemShut {NoStop}%
\bibitem [{\citenamefont {Kennes}\ \emph {et~al.}(2020)\citenamefont {Kennes},
  \citenamefont {Xian}, \citenamefont {Claassen},\ and\ \citenamefont
  {Rubio}}]{kennes19}%
  \BibitemOpen
  \bibfield  {author} {\bibinfo {author} {\bibfnamefont {Dante~M}\ \bibnamefont
  {Kennes}}, \bibinfo {author} {\bibfnamefont {Lede}\ \bibnamefont {Xian}},
  \bibinfo {author} {\bibfnamefont {M}~\bibnamefont {Claassen}}, \ and\
  \bibinfo {author} {\bibfnamefont {A}~\bibnamefont {Rubio}},\ }\bibfield
  {title} {\enquote {\bibinfo {title} {One-dimensional flat bands in twisted
  bilayer germanium selenide},}\ }\href@noop {} {\bibfield  {journal} {\bibinfo
   {journal} {Nat. Commun.}\ }\textbf {\bibinfo {volume} {11}},\ \bibinfo
  {pages} {1--8} (\bibinfo {year} {2020})}\BibitemShut {NoStop}%
\bibitem [{\citenamefont {Kariyado}\ and\ \citenamefont
  {Vishwanath}(2019)}]{Kariyado19}%
  \BibitemOpen
  \bibfield  {author} {\bibinfo {author} {\bibfnamefont {Toshikaze}\
  \bibnamefont {Kariyado}}\ and\ \bibinfo {author} {\bibfnamefont {Ashvin}\
  \bibnamefont {Vishwanath}},\ }\bibfield  {title} {\enquote {\bibinfo {title}
  {Flat band in twisted bilayer bravais lattices},}\ }\href {\doibase
  10.1103/PhysRevResearch.1.033076} {\bibfield  {journal} {\bibinfo  {journal}
  {Phys. Rev. Research}\ }\textbf {\bibinfo {volume} {1}},\ \bibinfo {pages}
  {033076} (\bibinfo {year} {2019})}\BibitemShut {NoStop}%
\bibitem [{\citenamefont {Lohse}\ \emph {et~al.}(2018)\citenamefont {Lohse},
  \citenamefont {Schweizer}, \citenamefont {Price}, \citenamefont
  {Zilberberg},\ and\ \citenamefont {Bloch}}]{Lohse2018}%
  \BibitemOpen
  \bibfield  {author} {\bibinfo {author} {\bibfnamefont {Michael}\ \bibnamefont
  {Lohse}}, \bibinfo {author} {\bibfnamefont {Christian}\ \bibnamefont
  {Schweizer}}, \bibinfo {author} {\bibfnamefont {Hannah~M.}\ \bibnamefont
  {Price}}, \bibinfo {author} {\bibfnamefont {Oded}\ \bibnamefont
  {Zilberberg}}, \ and\ \bibinfo {author} {\bibfnamefont {Immanuel}\
  \bibnamefont {Bloch}},\ }\bibfield  {title} {\enquote {\bibinfo {title}
  {Exploring 4d quantum hall physics with a 2d topological charge pump},}\
  }\href {\doibase 10.1038/nature25000} {\bibfield  {journal} {\bibinfo
  {journal} {Nature}\ }\textbf {\bibinfo {volume} {553}},\ \bibinfo {pages}
  {55--58} (\bibinfo {year} {2018})}\BibitemShut {NoStop}%
\bibitem [{\citenamefont {Zilberberg}\ \emph {et~al.}(2018)\citenamefont
  {Zilberberg}, \citenamefont {Huang}, \citenamefont {Guglielmon},
  \citenamefont {Wang}, \citenamefont {Chen}, \citenamefont {Kraus},\ and\
  \citenamefont {Rechtsman}}]{Zilberberg2018}%
  \BibitemOpen
  \bibfield  {author} {\bibinfo {author} {\bibfnamefont {Oded}\ \bibnamefont
  {Zilberberg}}, \bibinfo {author} {\bibfnamefont {Sheng}\ \bibnamefont
  {Huang}}, \bibinfo {author} {\bibfnamefont {Jonathan}\ \bibnamefont
  {Guglielmon}}, \bibinfo {author} {\bibfnamefont {Mohan}\ \bibnamefont
  {Wang}}, \bibinfo {author} {\bibfnamefont {Kevin~P.}\ \bibnamefont {Chen}},
  \bibinfo {author} {\bibfnamefont {Yaacov~E.}\ \bibnamefont {Kraus}}, \ and\
  \bibinfo {author} {\bibfnamefont {Mikael~C.}\ \bibnamefont {Rechtsman}},\
  }\bibfield  {title} {\enquote {\bibinfo {title} {Photonic topological
  boundary pumping as a probe of 4d quantum hall physics},}\ }\href {\doibase
  10.1038/nature25011} {\bibfield  {journal} {\bibinfo  {journal} {Nature}\
  }\textbf {\bibinfo {volume} {553}},\ \bibinfo {pages} {59--62} (\bibinfo
  {year} {2018})}\BibitemShut {NoStop}%
\bibitem [{\citenamefont {Liu}\ \emph {et~al.}(2020{\natexlab{b}})\citenamefont
  {Liu}, \citenamefont {Wu}, \citenamefont {Bai}, \citenamefont {Chae},
  \citenamefont {Li}, \citenamefont {Wang}, \citenamefont {Hone},\ and\
  \citenamefont {Zhu}}]{liu2020disassembling}%
  \BibitemOpen
  \bibfield  {author} {\bibinfo {author} {\bibfnamefont {Fang}\ \bibnamefont
  {Liu}}, \bibinfo {author} {\bibfnamefont {Wenjing}\ \bibnamefont {Wu}},
  \bibinfo {author} {\bibfnamefont {Yusong}\ \bibnamefont {Bai}}, \bibinfo
  {author} {\bibfnamefont {Sang~Hoon}\ \bibnamefont {Chae}}, \bibinfo {author}
  {\bibfnamefont {Qiuyang}\ \bibnamefont {Li}}, \bibinfo {author}
  {\bibfnamefont {Jue}\ \bibnamefont {Wang}}, \bibinfo {author} {\bibfnamefont
  {James}\ \bibnamefont {Hone}}, \ and\ \bibinfo {author} {\bibfnamefont {X-Y}\
  \bibnamefont {Zhu}},\ }\bibfield  {title} {\enquote {\bibinfo {title}
  {Disassembling 2d van der waals crystals into macroscopic monolayers and
  reassembling into artificial lattices},}\ }\href@noop {} {\bibfield
  {journal} {\bibinfo  {journal} {Science}\ }\textbf {\bibinfo {volume}
  {367}},\ \bibinfo {pages} {903--906} (\bibinfo {year}
  {2020}{\natexlab{b}})}\BibitemShut {NoStop}%
\bibitem [{\citenamefont {Hao}\ \emph {et~al.}(2021)\citenamefont {Hao},
  \citenamefont {Zimmerman}, \citenamefont {Ledwith}, \citenamefont {Khalaf},
  \citenamefont {Najafabadi}, \citenamefont {Watanabe}, \citenamefont
  {Taniguchi}, \citenamefont {Vishwanath},\ and\ \citenamefont
  {Kim}}]{kim_AtAtA}%
  \BibitemOpen
  \bibfield  {author} {\bibinfo {author} {\bibfnamefont {Zeyu}\ \bibnamefont
  {Hao}}, \bibinfo {author} {\bibfnamefont {A.~M.}\ \bibnamefont {Zimmerman}},
  \bibinfo {author} {\bibfnamefont {Patrick}\ \bibnamefont {Ledwith}}, \bibinfo
  {author} {\bibfnamefont {Eslam}\ \bibnamefont {Khalaf}}, \bibinfo {author}
  {\bibfnamefont {Danial~Haie}\ \bibnamefont {Najafabadi}}, \bibinfo {author}
  {\bibfnamefont {Kenji}\ \bibnamefont {Watanabe}}, \bibinfo {author}
  {\bibfnamefont {Takashi}\ \bibnamefont {Taniguchi}}, \bibinfo {author}
  {\bibfnamefont {Ashvin}\ \bibnamefont {Vishwanath}}, \ and\ \bibinfo {author}
  {\bibfnamefont {Philip}\ \bibnamefont {Kim}},\ }\bibfield  {title} {\enquote
  {\bibinfo {title} {Electric field{\textendash}tunable superconductivity in
  alternating-twist magic-angle trilayer graphene},}\ }\href {\doibase
  10.1126/science.abg0399} {\bibfield  {journal} {\bibinfo  {journal}
  {Science}\ }\textbf {\bibinfo {volume} {371}},\ \bibinfo {pages} {1133--1138}
  (\bibinfo {year} {2021})},\ \Eprint
  {http://arxiv.org/abs/https://science.sciencemag.org/content/371/6534/1133.full.pdf}
  {https://science.sciencemag.org/content/371/6534/1133.full.pdf} \BibitemShut
  {NoStop}%
\bibitem [{\citenamefont {Park}\ \emph {et~al.}(2021)\citenamefont {Park},
  \citenamefont {Cao}, \citenamefont {Watanabe}, \citenamefont {Taniguchi},\
  and\ \citenamefont {Jarillo-Herrero}}]{pablo_AtAtA}%
  \BibitemOpen
  \bibfield  {author} {\bibinfo {author} {\bibfnamefont {Jeong~Min}\
  \bibnamefont {Park}}, \bibinfo {author} {\bibfnamefont {Yuan}\ \bibnamefont
  {Cao}}, \bibinfo {author} {\bibfnamefont {Kenji}\ \bibnamefont {Watanabe}},
  \bibinfo {author} {\bibfnamefont {Takashi}\ \bibnamefont {Taniguchi}}, \ and\
  \bibinfo {author} {\bibfnamefont {Pablo}\ \bibnamefont {Jarillo-Herrero}},\
  }\bibfield  {title} {\enquote {\bibinfo {title} {Tunable strongly coupled
  superconductivity in magic-angle twisted trilayer graphene},}\ }\href
  {\doibase 10.1038/s41586-021-03192-0} {\bibfield  {journal} {\bibinfo
  {journal} {Nature}\ }\textbf {\bibinfo {volume} {590}},\ \bibinfo {pages}
  {249--255} (\bibinfo {year} {2021})}\BibitemShut {NoStop}%
\bibitem [{\citenamefont {Kang}\ \emph {et~al.}(2020)\citenamefont {Kang},
  \citenamefont {Fang}, \citenamefont {Ye}, \citenamefont {Po}, \citenamefont
  {Denlinger}, \citenamefont {Jozwiak}, \citenamefont {Bostwick}, \citenamefont
  {Rotenberg}, \citenamefont {Kaxiras}, \citenamefont {Checkelsky},\ and\
  \citenamefont {Comin}}]{kang2020topological}%
  \BibitemOpen
  \bibfield  {author} {\bibinfo {author} {\bibfnamefont {Mingu}\ \bibnamefont
  {Kang}}, \bibinfo {author} {\bibfnamefont {Shiang}\ \bibnamefont {Fang}},
  \bibinfo {author} {\bibfnamefont {Linda}\ \bibnamefont {Ye}}, \bibinfo
  {author} {\bibfnamefont {Hoi~Chun}\ \bibnamefont {Po}}, \bibinfo {author}
  {\bibfnamefont {Jonathan}\ \bibnamefont {Denlinger}}, \bibinfo {author}
  {\bibfnamefont {Chris}\ \bibnamefont {Jozwiak}}, \bibinfo {author}
  {\bibfnamefont {Aaron}\ \bibnamefont {Bostwick}}, \bibinfo {author}
  {\bibfnamefont {Eli}\ \bibnamefont {Rotenberg}}, \bibinfo {author}
  {\bibfnamefont {Efthimios}\ \bibnamefont {Kaxiras}}, \bibinfo {author}
  {\bibfnamefont {Joseph~G.}\ \bibnamefont {Checkelsky}}, \ and\ \bibinfo
  {author} {\bibfnamefont {Riccardo}\ \bibnamefont {Comin}},\ }\bibfield
  {title} {\enquote {\bibinfo {title} {Topological flat bands in frustrated
  kagome lattice cosn},}\ }\href {\doibase 10.1038/s41467-020-17465-1}
  {\bibfield  {journal} {\bibinfo  {journal} {Nature Communications}\ }\textbf
  {\bibinfo {volume} {11}},\ \bibinfo {pages} {4004} (\bibinfo {year}
  {2020})}\BibitemShut {NoStop}%
\bibitem [{\citenamefont {Liu}\ \emph {et~al.}(2020{\natexlab{c}})\citenamefont
  {Liu}, \citenamefont {Li}, \citenamefont {Wang}, \citenamefont {Wang},
  \citenamefont {Wen}, \citenamefont {Jiang}, \citenamefont {Lu}, \citenamefont
  {Yan}, \citenamefont {Huang}, \citenamefont {Shen} \emph
  {et~al.}}]{liu2020orbital}%
  \BibitemOpen
  \bibfield  {author} {\bibinfo {author} {\bibfnamefont {Zhonghao}\
  \bibnamefont {Liu}}, \bibinfo {author} {\bibfnamefont {Man}\ \bibnamefont
  {Li}}, \bibinfo {author} {\bibfnamefont {Qi}~\bibnamefont {Wang}}, \bibinfo
  {author} {\bibfnamefont {Guangwei}\ \bibnamefont {Wang}}, \bibinfo {author}
  {\bibfnamefont {Chenhaoping}\ \bibnamefont {Wen}}, \bibinfo {author}
  {\bibfnamefont {Kun}\ \bibnamefont {Jiang}}, \bibinfo {author} {\bibfnamefont
  {Xiangle}\ \bibnamefont {Lu}}, \bibinfo {author} {\bibfnamefont {Shichao}\
  \bibnamefont {Yan}}, \bibinfo {author} {\bibfnamefont {Yaobo}\ \bibnamefont
  {Huang}}, \bibinfo {author} {\bibfnamefont {Dawei}\ \bibnamefont {Shen}},
  \emph {et~al.},\ }\bibfield  {title} {\enquote {\bibinfo {title}
  {Orbital-selective dirac fermions and extremely flat bands in frustrated
  kagome-lattice metal cosn},}\ }\href@noop {} {\bibfield  {journal} {\bibinfo
  {journal} {Nature communications}\ }\textbf {\bibinfo {volume} {11}},\
  \bibinfo {pages} {1--7} (\bibinfo {year} {2020}{\natexlab{c}})}\BibitemShut
  {NoStop}%
\bibitem [{\citenamefont {Zhou}\ \emph {et~al.}(2019)\citenamefont {Zhou},
  \citenamefont {Jin}, \citenamefont {Huang}, \citenamefont {Wang},\ and\
  \citenamefont {Liu}}]{zhou2019weyl}%
  \BibitemOpen
  \bibfield  {author} {\bibinfo {author} {\bibfnamefont {Yinong}\ \bibnamefont
  {Zhou}}, \bibinfo {author} {\bibfnamefont {Kyung-Hwan}\ \bibnamefont {Jin}},
  \bibinfo {author} {\bibfnamefont {Huaqing}\ \bibnamefont {Huang}}, \bibinfo
  {author} {\bibfnamefont {Zhengfei}\ \bibnamefont {Wang}}, \ and\ \bibinfo
  {author} {\bibfnamefont {Feng}\ \bibnamefont {Liu}},\ }\bibfield  {title}
  {\enquote {\bibinfo {title} {Weyl points created by a three-dimensional flat
  band},}\ }\href@noop {} {\bibfield  {journal} {\bibinfo  {journal} {Physical
  Review B}\ }\textbf {\bibinfo {volume} {99}},\ \bibinfo {pages} {201105}
  (\bibinfo {year} {2019})}\BibitemShut {NoStop}%
\bibitem [{\citenamefont {Cea}\ \emph {et~al.}(2019)\citenamefont {Cea},
  \citenamefont {Walet},\ and\ \citenamefont {Guinea}}]{cea2019twists}%
  \BibitemOpen
  \bibfield  {author} {\bibinfo {author} {\bibfnamefont {Tommaso}\ \bibnamefont
  {Cea}}, \bibinfo {author} {\bibfnamefont {Niels~R}\ \bibnamefont {Walet}}, \
  and\ \bibinfo {author} {\bibfnamefont {Francisco}\ \bibnamefont {Guinea}},\
  }\bibfield  {title} {\enquote {\bibinfo {title} {Twists and the electronic
  structure of graphitic materials},}\ }\href@noop {} {\bibfield  {journal}
  {\bibinfo  {journal} {Nano Letters}\ }\textbf {\bibinfo {volume} {19}},\
  \bibinfo {pages} {8683--8689} (\bibinfo {year} {2019})}\BibitemShut {NoStop}%
\bibitem [{\citenamefont {Wu}\ \emph {et~al.}(2020)\citenamefont {Wu},
  \citenamefont {Zhang},\ and\ \citenamefont {Sarma}}]{wu2020three}%
  \BibitemOpen
  \bibfield  {author} {\bibinfo {author} {\bibfnamefont {Fengcheng}\
  \bibnamefont {Wu}}, \bibinfo {author} {\bibfnamefont {Rui-Xing}\ \bibnamefont
  {Zhang}}, \ and\ \bibinfo {author} {\bibfnamefont {Sankar~Das}\ \bibnamefont
  {Sarma}},\ }\bibfield  {title} {\enquote {\bibinfo {title} {Three-dimensional
  topological twistronics},}\ }\href@noop {} {\bibfield  {journal} {\bibinfo
  {journal} {Phys. Rev. Res.}\ }\textbf {\bibinfo {volume} {2}},\ \bibinfo
  {pages} {022010} (\bibinfo {year} {2020})}\BibitemShut {NoStop}%
\bibitem [{\citenamefont {Khalaf}\ \emph {et~al.}(2019)\citenamefont {Khalaf},
  \citenamefont {Kruchkov}, \citenamefont {Tarnopolsky},\ and\ \citenamefont
  {Vishwanath}}]{eslam19}%
  \BibitemOpen
  \bibfield  {author} {\bibinfo {author} {\bibfnamefont {Eslam}\ \bibnamefont
  {Khalaf}}, \bibinfo {author} {\bibfnamefont {Alex~J.}\ \bibnamefont
  {Kruchkov}}, \bibinfo {author} {\bibfnamefont {Grigory}\ \bibnamefont
  {Tarnopolsky}}, \ and\ \bibinfo {author} {\bibfnamefont {Ashvin}\
  \bibnamefont {Vishwanath}},\ }\bibfield  {title} {\enquote {\bibinfo {title}
  {Magic angle hierarchy in twisted graphene multilayers},}\ }\href {\doibase
  10.1103/PhysRevB.100.085109} {\bibfield  {journal} {\bibinfo  {journal}
  {Phys. Rev. B}\ }\textbf {\bibinfo {volume} {100}},\ \bibinfo {pages}
  {085109} (\bibinfo {year} {2019})}\BibitemShut {NoStop}%
\bibitem [{\citenamefont {Carr}\ \emph {et~al.}(2020)\citenamefont {Carr},
  \citenamefont {Li}, \citenamefont {Zhu}, \citenamefont {Kaxiras},
  \citenamefont {Sachdev},\ and\ \citenamefont
  {Kruchkov}}]{carr2020ultraheavy}%
  \BibitemOpen
  \bibfield  {author} {\bibinfo {author} {\bibfnamefont {Stephen}\ \bibnamefont
  {Carr}}, \bibinfo {author} {\bibfnamefont {Chenyuan}\ \bibnamefont {Li}},
  \bibinfo {author} {\bibfnamefont {Ziyan}\ \bibnamefont {Zhu}}, \bibinfo
  {author} {\bibfnamefont {Efthimios}\ \bibnamefont {Kaxiras}}, \bibinfo
  {author} {\bibfnamefont {Subir}\ \bibnamefont {Sachdev}}, \ and\ \bibinfo
  {author} {\bibfnamefont {Alexander}\ \bibnamefont {Kruchkov}},\ }\bibfield
  {title} {\enquote {\bibinfo {title} {Ultraheavy and ultrarelativistic dirac
  quasiparticles in sandwiched graphenes},}\ }\href {\doibase
  10.1021/acs.nanolett.9b04979} {\bibfield  {journal} {\bibinfo  {journal}
  {Nano Letters}\ }\textbf {\bibinfo {volume} {20}},\ \bibinfo {pages}
  {3030--3038} (\bibinfo {year} {2020})}\BibitemShut {NoStop}%
\bibitem [{\citenamefont {Tang}\ \emph {et~al.}(2019)\citenamefont {Tang},
  \citenamefont {Ren}, \citenamefont {Wang}, \citenamefont {Zhong},
  \citenamefont {Schneeloch}, \citenamefont {Yang}, \citenamefont {Yang},
  \citenamefont {Lee}, \citenamefont {Gu}, \citenamefont {Qiao},\ and\
  \citenamefont {Zhang}}]{tang2019three}%
  \BibitemOpen
  \bibfield  {author} {\bibinfo {author} {\bibfnamefont {Fangdong}\
  \bibnamefont {Tang}}, \bibinfo {author} {\bibfnamefont {Yafei}\ \bibnamefont
  {Ren}}, \bibinfo {author} {\bibfnamefont {Peipei}\ \bibnamefont {Wang}},
  \bibinfo {author} {\bibfnamefont {Ruidan}\ \bibnamefont {Zhong}}, \bibinfo
  {author} {\bibfnamefont {John}\ \bibnamefont {Schneeloch}}, \bibinfo {author}
  {\bibfnamefont {Shengyuan~A}\ \bibnamefont {Yang}}, \bibinfo {author}
  {\bibfnamefont {Kun}\ \bibnamefont {Yang}}, \bibinfo {author} {\bibfnamefont
  {Patrick~A}\ \bibnamefont {Lee}}, \bibinfo {author} {\bibfnamefont {Genda}\
  \bibnamefont {Gu}}, \bibinfo {author} {\bibfnamefont {Zhenhua}\ \bibnamefont
  {Qiao}}, \ and\ \bibinfo {author} {\bibfnamefont {Liyuan}\ \bibnamefont
  {Zhang}},\ }\bibfield  {title} {\enquote {\bibinfo {title} {Three-dimensional
  quantum hall effect and metal--insulator transition in zrte 5},}\ }\href@noop
  {} {\bibfield  {journal} {\bibinfo  {journal} {Nature}\ }\textbf {\bibinfo
  {volume} {569}},\ \bibinfo {pages} {537--541} (\bibinfo {year}
  {2019})}\BibitemShut {NoStop}%
\bibitem [{\citenamefont {Yasuda}\ \emph {et~al.}(2021)\citenamefont {Yasuda},
  \citenamefont {Wang}, \citenamefont {Watanabe}, \citenamefont {Taniguchi},\
  and\ \citenamefont {Jarillo-Herrero}}]{yasuda2020stacking}%
  \BibitemOpen
  \bibfield  {author} {\bibinfo {author} {\bibfnamefont {Kenji}\ \bibnamefont
  {Yasuda}}, \bibinfo {author} {\bibfnamefont {Xirui}\ \bibnamefont {Wang}},
  \bibinfo {author} {\bibfnamefont {Kenji}\ \bibnamefont {Watanabe}}, \bibinfo
  {author} {\bibfnamefont {Takashi}\ \bibnamefont {Taniguchi}}, \ and\ \bibinfo
  {author} {\bibfnamefont {Pablo}\ \bibnamefont {Jarillo-Herrero}},\ }\bibfield
   {title} {\enquote {\bibinfo {title} {Stacking-engineered ferroelectricity in
  bilayer boron nitride},}\ }\href {\doibase 10.1126/science.abd3230}
  {\bibfield  {journal} {\bibinfo  {journal} {Science}\ }\textbf {\bibinfo
  {volume} {372}},\ \bibinfo {pages} {1458--1462} (\bibinfo {year} {2021})},\
  \Eprint
  {http://arxiv.org/abs/https://science.sciencemag.org/content/372/6549/1458.full.pdf}
  {https://science.sciencemag.org/content/372/6549/1458.full.pdf} \BibitemShut
  {NoStop}%
\bibitem [{\citenamefont {Woods}\ \emph {et~al.}(2021)\citenamefont {Woods},
  \citenamefont {Ares}, \citenamefont {Nevison-Andrews}, \citenamefont
  {Holwill}, \citenamefont {Fabregas}, \citenamefont {Guinea}, \citenamefont
  {Geim}, \citenamefont {Novoselov}, \citenamefont {Walet},\ and\ \citenamefont
  {Fumagalli}}]{woods2020charge}%
  \BibitemOpen
  \bibfield  {author} {\bibinfo {author} {\bibfnamefont {C.~R.}\ \bibnamefont
  {Woods}}, \bibinfo {author} {\bibfnamefont {P.}~\bibnamefont {Ares}},
  \bibinfo {author} {\bibfnamefont {H.}~\bibnamefont {Nevison-Andrews}},
  \bibinfo {author} {\bibfnamefont {M.~J.}\ \bibnamefont {Holwill}}, \bibinfo
  {author} {\bibfnamefont {R.}~\bibnamefont {Fabregas}}, \bibinfo {author}
  {\bibfnamefont {F.}~\bibnamefont {Guinea}}, \bibinfo {author} {\bibfnamefont
  {A.~K.}\ \bibnamefont {Geim}}, \bibinfo {author} {\bibfnamefont {K.~S.}\
  \bibnamefont {Novoselov}}, \bibinfo {author} {\bibfnamefont {N.~R.}\
  \bibnamefont {Walet}}, \ and\ \bibinfo {author} {\bibfnamefont
  {L.}~\bibnamefont {Fumagalli}},\ }\bibfield  {title} {\enquote {\bibinfo
  {title} {Charge-polarized interfacial superlattices in marginally twisted
  hexagonal boron nitride},}\ }\href {\doibase 10.1038/s41467-020-20667-2}
  {\bibfield  {journal} {\bibinfo  {journal} {Nature Communications}\ }\textbf
  {\bibinfo {volume} {12}},\ \bibinfo {pages} {347} (\bibinfo {year}
  {2021})}\BibitemShut {NoStop}%
\bibitem [{\citenamefont {Plimpton}(1995)}]{plimpton1995fast}%
  \BibitemOpen
  \bibfield  {author} {\bibinfo {author} {\bibfnamefont {Steve}\ \bibnamefont
  {Plimpton}},\ }\bibfield  {title} {\enquote {\bibinfo {title} {Fast parallel
  algorithms for short-range molecular dynamics},}\ }\href@noop {} {\bibfield
  {journal} {\bibinfo  {journal} {Journal of computational physics}\ }\textbf
  {\bibinfo {volume} {117}},\ \bibinfo {pages} {1--19} (\bibinfo {year}
  {1995})}\BibitemShut {NoStop}%
\bibitem [{\citenamefont {Angeli}\ \emph {et~al.}(2018)\citenamefont {Angeli},
  \citenamefont {Mandelli}, \citenamefont {Valli}, \citenamefont {Amaricci},
  \citenamefont {Capone}, \citenamefont {Tosatti},\ and\ \citenamefont
  {Fabrizio}}]{angeli2018emergent}%
  \BibitemOpen
  \bibfield  {author} {\bibinfo {author} {\bibfnamefont {Mattia}\ \bibnamefont
  {Angeli}}, \bibinfo {author} {\bibfnamefont {D}~\bibnamefont {Mandelli}},
  \bibinfo {author} {\bibfnamefont {ANGELO}\ \bibnamefont {Valli}}, \bibinfo
  {author} {\bibfnamefont {A}~\bibnamefont {Amaricci}}, \bibinfo {author}
  {\bibfnamefont {M}~\bibnamefont {Capone}}, \bibinfo {author} {\bibfnamefont
  {E}~\bibnamefont {Tosatti}}, \ and\ \bibinfo {author} {\bibfnamefont
  {M}~\bibnamefont {Fabrizio}},\ }\bibfield  {title} {\enquote {\bibinfo
  {title} {Emergent d 6 symmetry in fully relaxed magic-angle twisted bilayer
  graphene},}\ }\href@noop {} {\bibfield  {journal} {\bibinfo  {journal}
  {Physical Review B}\ }\textbf {\bibinfo {volume} {98}},\ \bibinfo {pages}
  {235137} (\bibinfo {year} {2018})}\BibitemShut {NoStop}%
\bibitem [{\citenamefont {Brenner}\ \emph {et~al.}(2002)\citenamefont
  {Brenner}, \citenamefont {Shenderova}, \citenamefont {Harrison},
  \citenamefont {Stuart}, \citenamefont {Ni},\ and\ \citenamefont
  {Sinnott}}]{brenner2002second}%
  \BibitemOpen
  \bibfield  {author} {\bibinfo {author} {\bibfnamefont {Donald~W}\
  \bibnamefont {Brenner}}, \bibinfo {author} {\bibfnamefont {Olga~A}\
  \bibnamefont {Shenderova}}, \bibinfo {author} {\bibfnamefont {Judith~A}\
  \bibnamefont {Harrison}}, \bibinfo {author} {\bibfnamefont {Steven~J}\
  \bibnamefont {Stuart}}, \bibinfo {author} {\bibfnamefont {Boris}\
  \bibnamefont {Ni}}, \ and\ \bibinfo {author} {\bibfnamefont {Susan~B}\
  \bibnamefont {Sinnott}},\ }\bibfield  {title} {\enquote {\bibinfo {title} {A
  second-generation reactive empirical bond order (rebo) potential energy
  expression for hydrocarbons},}\ }\href@noop {} {\bibfield  {journal}
  {\bibinfo  {journal} {Journal of Physics: Condensed Matter}\ }\textbf
  {\bibinfo {volume} {14}},\ \bibinfo {pages} {783} (\bibinfo {year}
  {2002})}\BibitemShut {NoStop}%
\bibitem [{\citenamefont {Kolmogorov}\ and\ \citenamefont
  {Crespi}(2005)}]{kolmogorov2005registry}%
  \BibitemOpen
  \bibfield  {author} {\bibinfo {author} {\bibfnamefont {Aleksey~N}\
  \bibnamefont {Kolmogorov}}\ and\ \bibinfo {author} {\bibfnamefont
  {Vincent~H}\ \bibnamefont {Crespi}},\ }\bibfield  {title} {\enquote {\bibinfo
  {title} {Registry-dependent interlayer potential for graphitic systems},}\
  }\href@noop {} {\bibfield  {journal} {\bibinfo  {journal} {Physical Review
  B}\ }\textbf {\bibinfo {volume} {71}},\ \bibinfo {pages} {235415} (\bibinfo
  {year} {2005})}\BibitemShut {NoStop}%
\bibitem [{\citenamefont {Ouyang}\ \emph {et~al.}(2018)\citenamefont {Ouyang},
  \citenamefont {Mandelli}, \citenamefont {Urbakh},\ and\ \citenamefont
  {Hod}}]{ouyang2018nanoserpents}%
  \BibitemOpen
  \bibfield  {author} {\bibinfo {author} {\bibfnamefont {Wengen}\ \bibnamefont
  {Ouyang}}, \bibinfo {author} {\bibfnamefont {Davide}\ \bibnamefont
  {Mandelli}}, \bibinfo {author} {\bibfnamefont {Michael}\ \bibnamefont
  {Urbakh}}, \ and\ \bibinfo {author} {\bibfnamefont {Oded}\ \bibnamefont
  {Hod}},\ }\bibfield  {title} {\enquote {\bibinfo {title} {Nanoserpents:
  Graphene nanoribbon motion on two-dimensional hexagonal materials},}\
  }\href@noop {} {\bibfield  {journal} {\bibinfo  {journal} {Nano letters}\
  }\textbf {\bibinfo {volume} {18}},\ \bibinfo {pages} {6009--6016} (\bibinfo
  {year} {2018})}\BibitemShut {NoStop}%
\bibitem [{\citenamefont {Bitzek}\ \emph {et~al.}(2006)\citenamefont {Bitzek},
  \citenamefont {Koskinen}, \citenamefont {G{\"a}hler}, \citenamefont
  {Moseler},\ and\ \citenamefont {Gumbsch}}]{bitzek2006structural}%
  \BibitemOpen
  \bibfield  {author} {\bibinfo {author} {\bibfnamefont {Erik}\ \bibnamefont
  {Bitzek}}, \bibinfo {author} {\bibfnamefont {Pekka}\ \bibnamefont
  {Koskinen}}, \bibinfo {author} {\bibfnamefont {Franz}\ \bibnamefont
  {G{\"a}hler}}, \bibinfo {author} {\bibfnamefont {Michael}\ \bibnamefont
  {Moseler}}, \ and\ \bibinfo {author} {\bibfnamefont {Peter}\ \bibnamefont
  {Gumbsch}},\ }\bibfield  {title} {\enquote {\bibinfo {title} {Structural
  relaxation made simple},}\ }\href@noop {} {\bibfield  {journal} {\bibinfo
  {journal} {Physical review letters}\ }\textbf {\bibinfo {volume} {97}},\
  \bibinfo {pages} {170201} (\bibinfo {year} {2006})}\BibitemShut {NoStop}%
\bibitem [{\citenamefont {Trambly~de Laissardière}\ \emph
  {et~al.}(2010)\citenamefont {Trambly~de Laissardière}, \citenamefont
  {Mayou},\ and\ \citenamefont {Magaud}}]{tram10}%
  \BibitemOpen
  \bibfield  {author} {\bibinfo {author} {\bibfnamefont {G.}~\bibnamefont
  {Trambly~de Laissardière}}, \bibinfo {author} {\bibfnamefont
  {D.}~\bibnamefont {Mayou}}, \ and\ \bibinfo {author} {\bibfnamefont
  {L.}~\bibnamefont {Magaud}},\ }\bibfield  {title} {\enquote {\bibinfo {title}
  {Localization of dirac electrons in rotated graphene bilayers},}\ }\href@noop
  {} {\bibfield  {journal} {\bibinfo  {journal} {Nano Letters}\ }\textbf
  {\bibinfo {volume} {10}},\ \bibinfo {pages} {804--808} (\bibinfo {year}
  {2010})}\BibitemShut {NoStop}%
\bibitem [{\citenamefont {Kresse}\ and\ \citenamefont
  {Hafner}(1993)}]{kresse93ab}%
  \BibitemOpen
  \bibfield  {author} {\bibinfo {author} {\bibfnamefont {Georg}\ \bibnamefont
  {Kresse}}\ and\ \bibinfo {author} {\bibfnamefont {J{\"u}rgen}\ \bibnamefont
  {Hafner}},\ }\bibfield  {title} {\enquote {\bibinfo {title} {Ab initio
  molecular dynamics for liquid metals},}\ }\href@noop {} {\bibfield  {journal}
  {\bibinfo  {journal} {Phys. Rev. B}\ }\textbf {\bibinfo {volume} {47}},\
  \bibinfo {pages} {558} (\bibinfo {year} {1993})}\BibitemShut {NoStop}%
\bibitem [{\citenamefont {Bl{\"o}chl}(1994)}]{blochl94}%
  \BibitemOpen
  \bibfield  {author} {\bibinfo {author} {\bibfnamefont {Peter~E}\ \bibnamefont
  {Bl{\"o}chl}},\ }\bibfield  {title} {\enquote {\bibinfo {title} {Projector
  augmented-wave method},}\ }\href@noop {} {\bibfield  {journal} {\bibinfo
  {journal} {Phys. Rev. B}\ }\textbf {\bibinfo {volume} {50}},\ \bibinfo
  {pages} {17953} (\bibinfo {year} {1994})}\BibitemShut {NoStop}%
\bibitem [{\citenamefont {Perdew}\ \emph {et~al.}(1996)\citenamefont {Perdew},
  \citenamefont {Burke},\ and\ \citenamefont
  {Ernzerhof}}]{perdew1996generalized}%
  \BibitemOpen
  \bibfield  {author} {\bibinfo {author} {\bibfnamefont {John~P}\ \bibnamefont
  {Perdew}}, \bibinfo {author} {\bibfnamefont {Kieron}\ \bibnamefont {Burke}},
  \ and\ \bibinfo {author} {\bibfnamefont {Matthias}\ \bibnamefont
  {Ernzerhof}},\ }\bibfield  {title} {\enquote {\bibinfo {title} {Generalized
  gradient approximation made simple},}\ }\href@noop {} {\bibfield  {journal}
  {\bibinfo  {journal} {Physical review letters}\ }\textbf {\bibinfo {volume}
  {77}},\ \bibinfo {pages} {3865} (\bibinfo {year} {1996})}\BibitemShut
  {NoStop}%
\bibitem [{\citenamefont {Tkatchenko}\ and\ \citenamefont
  {Scheffler}(2009)}]{tsmethod09}%
  \BibitemOpen
  \bibfield  {author} {\bibinfo {author} {\bibfnamefont {Alexandre}\
  \bibnamefont {Tkatchenko}}\ and\ \bibinfo {author} {\bibfnamefont {Matthias}\
  \bibnamefont {Scheffler}},\ }\bibfield  {title} {\enquote {\bibinfo {title}
  {Accurate molecular van der waals interactions from ground-state electron
  density and free-atom reference data},}\ }\href {\doibase
  10.1103/PhysRevLett.102.073005} {\bibfield  {journal} {\bibinfo  {journal}
  {Phys. Rev. Lett.}\ }\textbf {\bibinfo {volume} {102}},\ \bibinfo {pages}
  {073005} (\bibinfo {year} {2009})}\BibitemShut {NoStop}%
\end{thebibliography}%


\begin{thebibliography}{13}%
\makeatletter
\providecommand \@ifxundefined [1]{%
 \@ifx{#1\undefined}
}%
\providecommand \@ifnum [1]{%
 \ifnum #1\expandafter \@firstoftwo
 \else \expandafter \@secondoftwo
 \fi
}%
\providecommand \@ifx [1]{%
 \ifx #1\expandafter \@firstoftwo
 \else \expandafter \@secondoftwo
 \fi
}%
\providecommand \natexlab [1]{#1}%
\providecommand \enquote  [1]{``#1''}%
\providecommand \bibnamefont  [1]{#1}%
\providecommand \bibfnamefont [1]{#1}%
\providecommand \citenamefont [1]{#1}%
\providecommand \href@noop [0]{\@secondoftwo}%
\providecommand \href [0]{\begingroup \@sanitize@url \@href}%
\providecommand \@href[1]{\@@startlink{#1}\@@href}%
\providecommand \@@href[1]{\endgroup#1\@@endlink}%
\providecommand \@sanitize@url [0]{\catcode `\\12\catcode `\$12\catcode
  `\&12\catcode `\#12\catcode `\^12\catcode `\_12\catcode `\%12\relax}%
\providecommand \@@startlink[1]{}%
\providecommand \@@endlink[0]{}%
\providecommand \url  [0]{\begingroup\@sanitize@url \@url }%
\providecommand \@url [1]{\endgroup\@href {#1}{\urlprefix }}%
\providecommand \urlprefix  [0]{URL }%
\providecommand \Eprint [0]{\href }%
\providecommand \doibase [0]{http://dx.doi.org/}%
\providecommand \selectlanguage [0]{\@gobble}%
\providecommand \bibinfo  [0]{\@secondoftwo}%
\providecommand \bibfield  [0]{\@secondoftwo}%
\providecommand \translation [1]{[#1]}%
\providecommand \BibitemOpen [0]{}%
\providecommand \bibitemStop [0]{}%
\providecommand \bibitemNoStop [0]{.\EOS\space}%
\providecommand \EOS [0]{\spacefactor3000\relax}%
\providecommand \BibitemShut  [1]{\csname bibitem#1\endcsname}%
\let\auto@bib@innerbib\@empty
\bibitem [{\citenamefont {Plimpton}(1995)}]{plimpton1995fast}%
  \BibitemOpen
  \bibfield  {author} {\bibinfo {author} {\bibfnamefont {S.}~\bibnamefont
  {Plimpton}},\ }\href@noop {} {\bibfield  {journal} {\bibinfo  {journal}
  {Journal of computational physics}\ }\textbf {\bibinfo {volume} {117}},\
  \bibinfo {pages} {1} (\bibinfo {year} {1995})}\BibitemShut {NoStop}%
\bibitem [{\citenamefont {Angeli}\ \emph {et~al.}(2018)\citenamefont {Angeli},
  \citenamefont {Mandelli}, \citenamefont {Valli}, \citenamefont {Amaricci},
  \citenamefont {Capone}, \citenamefont {Tosatti},\ and\ \citenamefont
  {Fabrizio}}]{angeli2018emergent}%
  \BibitemOpen
  \bibfield  {author} {\bibinfo {author} {\bibfnamefont {M.}~\bibnamefont
  {Angeli}}, \bibinfo {author} {\bibfnamefont {D.}~\bibnamefont {Mandelli}},
  \bibinfo {author} {\bibfnamefont {A.}~\bibnamefont {Valli}}, \bibinfo
  {author} {\bibfnamefont {A.}~\bibnamefont {Amaricci}}, \bibinfo {author}
  {\bibfnamefont {M.}~\bibnamefont {Capone}}, \bibinfo {author} {\bibfnamefont
  {E.}~\bibnamefont {Tosatti}}, \ and\ \bibinfo {author} {\bibfnamefont
  {M.}~\bibnamefont {Fabrizio}},\ }\href@noop {} {\bibfield  {journal}
  {\bibinfo  {journal} {Physical Review B}\ }\textbf {\bibinfo {volume} {98}},\
  \bibinfo {pages} {235137} (\bibinfo {year} {2018})}\BibitemShut {NoStop}%
\bibitem [{\citenamefont {Brenner}\ \emph {et~al.}(2002)\citenamefont
  {Brenner}, \citenamefont {Shenderova}, \citenamefont {Harrison},
  \citenamefont {Stuart}, \citenamefont {Ni},\ and\ \citenamefont
  {Sinnott}}]{brenner2002second}%
  \BibitemOpen
  \bibfield  {author} {\bibinfo {author} {\bibfnamefont {D.~W.}\ \bibnamefont
  {Brenner}}, \bibinfo {author} {\bibfnamefont {O.~A.}\ \bibnamefont
  {Shenderova}}, \bibinfo {author} {\bibfnamefont {J.~A.}\ \bibnamefont
  {Harrison}}, \bibinfo {author} {\bibfnamefont {S.~J.}\ \bibnamefont
  {Stuart}}, \bibinfo {author} {\bibfnamefont {B.}~\bibnamefont {Ni}}, \ and\
  \bibinfo {author} {\bibfnamefont {S.~B.}\ \bibnamefont {Sinnott}},\
  }\href@noop {} {\bibfield  {journal} {\bibinfo  {journal} {Journal of
  Physics: Condensed Matter}\ }\textbf {\bibinfo {volume} {14}},\ \bibinfo
  {pages} {783} (\bibinfo {year} {2002})}\BibitemShut {NoStop}%
\bibitem [{\citenamefont {Kolmogorov}\ and\ \citenamefont
  {Crespi}(2005)}]{kolmogorov2005registry}%
  \BibitemOpen
  \bibfield  {author} {\bibinfo {author} {\bibfnamefont {A.~N.}\ \bibnamefont
  {Kolmogorov}}\ and\ \bibinfo {author} {\bibfnamefont {V.~H.}\ \bibnamefont
  {Crespi}},\ }\href@noop {} {\bibfield  {journal} {\bibinfo  {journal}
  {Physical Review B}\ }\textbf {\bibinfo {volume} {71}},\ \bibinfo {pages}
  {235415} (\bibinfo {year} {2005})}\BibitemShut {NoStop}%
\bibitem [{\citenamefont {Ouyang}\ \emph {et~al.}(2018)\citenamefont {Ouyang},
  \citenamefont {Mandelli}, \citenamefont {Urbakh},\ and\ \citenamefont
  {Hod}}]{ouyang2018nanoserpents}%
  \BibitemOpen
  \bibfield  {author} {\bibinfo {author} {\bibfnamefont {W.}~\bibnamefont
  {Ouyang}}, \bibinfo {author} {\bibfnamefont {D.}~\bibnamefont {Mandelli}},
  \bibinfo {author} {\bibfnamefont {M.}~\bibnamefont {Urbakh}}, \ and\ \bibinfo
  {author} {\bibfnamefont {O.}~\bibnamefont {Hod}},\ }\href@noop {} {\bibfield
  {journal} {\bibinfo  {journal} {Nano letters}\ }\textbf {\bibinfo {volume}
  {18}},\ \bibinfo {pages} {6009} (\bibinfo {year} {2018})}\BibitemShut
  {NoStop}%
\bibitem [{\citenamefont {Bitzek}\ \emph {et~al.}(2006)\citenamefont {Bitzek},
  \citenamefont {Koskinen}, \citenamefont {G{\"a}hler}, \citenamefont
  {Moseler},\ and\ \citenamefont {Gumbsch}}]{bitzek2006structural}%
  \BibitemOpen
  \bibfield  {author} {\bibinfo {author} {\bibfnamefont {E.}~\bibnamefont
  {Bitzek}}, \bibinfo {author} {\bibfnamefont {P.}~\bibnamefont {Koskinen}},
  \bibinfo {author} {\bibfnamefont {F.}~\bibnamefont {G{\"a}hler}}, \bibinfo
  {author} {\bibfnamefont {M.}~\bibnamefont {Moseler}}, \ and\ \bibinfo
  {author} {\bibfnamefont {P.}~\bibnamefont {Gumbsch}},\ }\href@noop {}
  {\bibfield  {journal} {\bibinfo  {journal} {Physical review letters}\
  }\textbf {\bibinfo {volume} {97}},\ \bibinfo {pages} {170201} (\bibinfo
  {year} {2006})}\BibitemShut {NoStop}%
\bibitem [{\citenamefont {Trambly~de Laissardière}\ \emph
  {et~al.}(2010)\citenamefont {Trambly~de Laissardière}, \citenamefont
  {Mayou},\ and\ \citenamefont {Magaud}}]{tram10}%
  \BibitemOpen
  \bibfield  {author} {\bibinfo {author} {\bibfnamefont {G.}~\bibnamefont
  {Trambly~de Laissardière}}, \bibinfo {author} {\bibfnamefont
  {D.}~\bibnamefont {Mayou}}, \ and\ \bibinfo {author} {\bibfnamefont
  {L.}~\bibnamefont {Magaud}},\ }\href@noop {} {\bibfield  {journal} {\bibinfo
  {journal} {Nano Letters}\ }\textbf {\bibinfo {volume} {10}},\ \bibinfo
  {pages} {804} (\bibinfo {year} {2010})}\BibitemShut {NoStop}%
\bibitem [{\citenamefont {Kresse}\ and\ \citenamefont
  {Hafner}(1993)}]{kresse93ab}%
  \BibitemOpen
  \bibfield  {author} {\bibinfo {author} {\bibfnamefont {G.}~\bibnamefont
  {Kresse}}\ and\ \bibinfo {author} {\bibfnamefont {J.}~\bibnamefont
  {Hafner}},\ }\href@noop {} {\bibfield  {journal} {\bibinfo  {journal} {Phys.
  Rev. B}\ }\textbf {\bibinfo {volume} {47}},\ \bibinfo {pages} {558} (\bibinfo
  {year} {1993})}\BibitemShut {NoStop}%
\bibitem [{\citenamefont {Wang}\ \emph {et~al.}(2020)\citenamefont {Wang},
  \citenamefont {Shih}, \citenamefont {Ghiotto}, \citenamefont {Xian},
  \citenamefont {Rhodes}, \citenamefont {Tan}, \citenamefont {Claassen},
  \citenamefont {Kennes}, \citenamefont {Bai}, \citenamefont {Kim},
  \citenamefont {Watanabe}, \citenamefont {Taniguchi}, \citenamefont {Zhu},
  \citenamefont {Hone}, \citenamefont {Rubio}, \citenamefont {Pasupathy},\ and\
  \citenamefont {Dean}}]{wang19}%
  \BibitemOpen
  \bibfield  {author} {\bibinfo {author} {\bibfnamefont {L.}~\bibnamefont
  {Wang}}, \bibinfo {author} {\bibfnamefont {E.-M.}\ \bibnamefont {Shih}},
  \bibinfo {author} {\bibfnamefont {A.}~\bibnamefont {Ghiotto}}, \bibinfo
  {author} {\bibfnamefont {L.}~\bibnamefont {Xian}}, \bibinfo {author}
  {\bibfnamefont {D.~A.}\ \bibnamefont {Rhodes}}, \bibinfo {author}
  {\bibfnamefont {C.}~\bibnamefont {Tan}}, \bibinfo {author} {\bibfnamefont
  {M.}~\bibnamefont {Claassen}}, \bibinfo {author} {\bibfnamefont {D.~M.}\
  \bibnamefont {Kennes}}, \bibinfo {author} {\bibfnamefont {Y.}~\bibnamefont
  {Bai}}, \bibinfo {author} {\bibfnamefont {B.}~\bibnamefont {Kim}}, \bibinfo
  {author} {\bibfnamefont {K.}~\bibnamefont {Watanabe}}, \bibinfo {author}
  {\bibfnamefont {T.}~\bibnamefont {Taniguchi}}, \bibinfo {author}
  {\bibfnamefont {X.}~\bibnamefont {Zhu}}, \bibinfo {author} {\bibfnamefont
  {J.}~\bibnamefont {Hone}}, \bibinfo {author} {\bibfnamefont {A.}~\bibnamefont
  {Rubio}}, \bibinfo {author} {\bibfnamefont {A.~N.}\ \bibnamefont
  {Pasupathy}}, \ and\ \bibinfo {author} {\bibfnamefont {C.~R.}\ \bibnamefont
  {Dean}},\ }\href {\doibase 10.1038/s41563-020-0708-6} {\bibfield  {journal}
  {\bibinfo  {journal} {Nature Materials}\ }\textbf {\bibinfo {volume} {19}},\
  \bibinfo {pages} {861} (\bibinfo {year} {2020})}\BibitemShut {NoStop}%
\bibitem [{\citenamefont {Xian}\ \emph {et~al.}(2019)\citenamefont {Xian},
  \citenamefont {Kennes}, \citenamefont {Tancogne-Dejean}, \citenamefont
  {Altarelli},\ and\ \citenamefont {Rubio}}]{Xian18}%
  \BibitemOpen
  \bibfield  {author} {\bibinfo {author} {\bibfnamefont {L.}~\bibnamefont
  {Xian}}, \bibinfo {author} {\bibfnamefont {D.~M.}\ \bibnamefont {Kennes}},
  \bibinfo {author} {\bibfnamefont {N.}~\bibnamefont {Tancogne-Dejean}},
  \bibinfo {author} {\bibfnamefont {M.}~\bibnamefont {Altarelli}}, \ and\
  \bibinfo {author} {\bibfnamefont {A.}~\bibnamefont {Rubio}},\ }\href@noop {}
  {\bibfield  {journal} {\bibinfo  {journal} {Nano Lett.}\ }\textbf {\bibinfo
  {volume} {19}},\ \bibinfo {pages} {4934} (\bibinfo {year}
  {2019})}\BibitemShut {NoStop}%
\bibitem [{\citenamefont {Bl{\"o}chl}(1994)}]{blochl94}%
  \BibitemOpen
  \bibfield  {author} {\bibinfo {author} {\bibfnamefont {P.~E.}\ \bibnamefont
  {Bl{\"o}chl}},\ }\href@noop {} {\bibfield  {journal} {\bibinfo  {journal}
  {Phys. Rev. B}\ }\textbf {\bibinfo {volume} {50}},\ \bibinfo {pages} {17953}
  (\bibinfo {year} {1994})}\BibitemShut {NoStop}%
\bibitem [{\citenamefont {Perdew}\ \emph {et~al.}(1996)\citenamefont {Perdew},
  \citenamefont {Burke},\ and\ \citenamefont
  {Ernzerhof}}]{perdew1996generalized}%
  \BibitemOpen
  \bibfield  {author} {\bibinfo {author} {\bibfnamefont {J.~P.}\ \bibnamefont
  {Perdew}}, \bibinfo {author} {\bibfnamefont {K.}~\bibnamefont {Burke}}, \
  and\ \bibinfo {author} {\bibfnamefont {M.}~\bibnamefont {Ernzerhof}},\
  }\href@noop {} {\bibfield  {journal} {\bibinfo  {journal} {Physical review
  letters}\ }\textbf {\bibinfo {volume} {77}},\ \bibinfo {pages} {3865}
  (\bibinfo {year} {1996})}\BibitemShut {NoStop}%
\bibitem [{\citenamefont {Tkatchenko}\ and\ \citenamefont
  {Scheffler}(2009)}]{tsmethod09}%
  \BibitemOpen
  \bibfield  {author} {\bibinfo {author} {\bibfnamefont {A.}~\bibnamefont
  {Tkatchenko}}\ and\ \bibinfo {author} {\bibfnamefont {M.}~\bibnamefont
  {Scheffler}},\ }\href {\doibase 10.1103/PhysRevLett.102.073005} {\bibfield
  {journal} {\bibinfo  {journal} {Phys. Rev. Lett.}\ }\textbf {\bibinfo
  {volume} {102}},\ \bibinfo {pages} {073005} (\bibinfo {year}
  {2009})}\BibitemShut {NoStop}%
\end{thebibliography}%


%

\begin{figure*}
    \centering
    \includegraphics[width=0.4\textwidth]{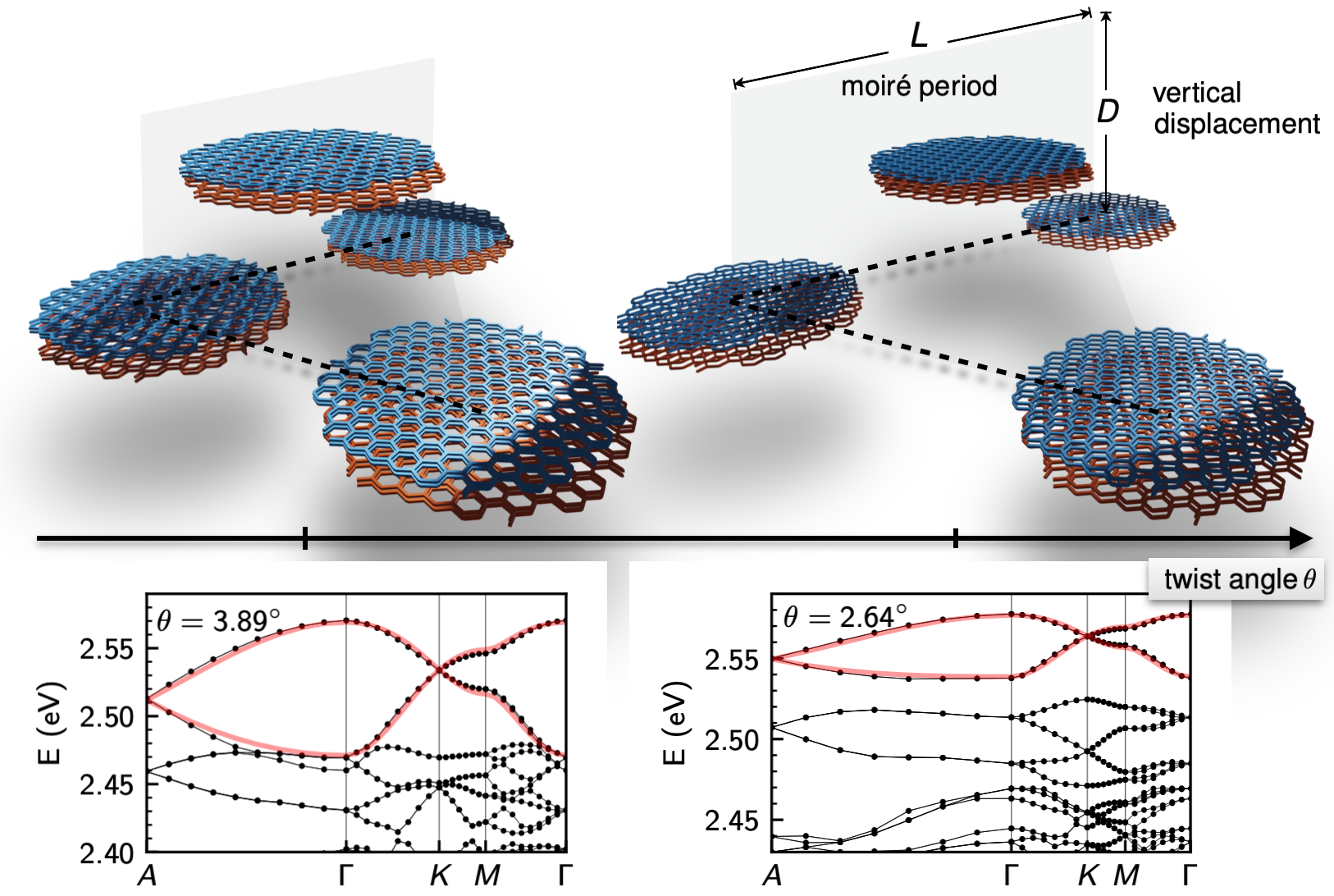}\\[5mm]
    {{\bf TOC Graphic}.}
    \label{fig:toc}
\end{figure*}

\end{document}


\title{Supplemental Material for \\
Engineering Three Dimensional Moir\'e Flat Bands}
\author{Lede Xian}
\altaffiliation{L.X. and A.F contributed equally to this paper.}
\affiliation{Songshan Lake Materials Laboratory, 523808 Dongguan, Guangdong, China}
\affiliation{Max Planck Institute for the Structure and Dynamics of Matter, Center for Free Electron Laser Science, 22761 Hamburg, Germany}

\author{Ammon Fischer}
\altaffiliation{L.X. and A.F contributed equally to this paper.}
\affiliation{Institut f\"ur Theorie der Statistischen Physik, RWTH Aachen University and JARA-Fundamentals of Future Information Technology, 52056 Aachen, Germany}

\author{Martin Claassen}
\affiliation{Department of Physics and Astronomy, University of Pennsylvania, Philadelphia, PA 19104, USA}

\author{Jin Zhang}
\affiliation{Max Planck Institute for the Structure and Dynamics of Matter, Center for Free Electron Laser Science, 22761 Hamburg, Germany}

\author{Angel Rubio}
\altaffiliation{Corresponding author: angel.rubio@mpsd.mpg.de}
\affiliation{Max Planck Institute for the Structure and Dynamics of Matter, Center for Free Electron Laser Science, 22761 Hamburg, Germany}
\affiliation{Center for Computational Quantum Physics, Simons Foundation Flatiron Institute, New York, NY 10010 USA}
\affiliation{Nano-Bio Spectroscopy Group,  Departamento de Fisica de Materiales, Universidad del Pa\'is Vasco, UPV/EHU- 20018 San Sebasti\'an, Spain}

\author{Dante M. Kennes}
\altaffiliation{Corresponding author: dante.kennes@rwth-aachen.de}
\affiliation{Institut f\"ur Theorie der Statistischen Physik, RWTH Aachen University and JARA-Fundamentals of Future Information Technology, 52056 Aachen, Germany}
\affiliation{Max Planck Institute for the Structure and Dynamics of Matter, Center for Free Electron Laser Science, 22761 Hamburg, Germany}

\maketitle
\renewcommand\thesection{S~\Roman{section}}

\section{Computational Details for 3D twisted graphene, WSe2 and Boron Nitride}
For the calculations of 3D twisted graphene, we construct the unit cell with a twisted double bilayer graphene at twist angles close to 0 degree, and impose periodic boundary condition along all three dimensions. As it is not realistic to optimize such a large system with density functional theory (DFT) calculations, we fix the lattice constant along the out-of-plane direction to be 13.415 {\AA}, and set the in-plane lattice constant according to the twist angles such that it corresponds to 2.46 {\AA} for a 1x1 cell. The atomic structure is relaxed using the LAMMPS code \cite{plimpton1995fast} with the same parameters as described in \cite{angeli2018emergent}. The intralayer interactions within each graphene layer are modeled via the second-generation reactive empirical bondorder (REBO) potential \cite{brenner2002second}. The interlayer interactions are modeled via the Kolmogorov-Crespi (KC) potential \cite{kolmogorov2005registry}, using the recent parametrization of \cite{ouyang2018nanoserpents}.  The relaxation is performed using the fast inertial relaxation engine (FIRE) algorithm \cite{bitzek2006structural}. 


We calculate the band structures for 3D twisted graphene using the tight-binding parametrization proposed in Ref. \cite{tram10}
\begin{equation}
H_0 = \sum_{i,j}t(\textbf{r}_i - \textbf{r}_j) c_{i}^{\dagger} c_{j}^{\phantom \dagger}.
\label{tb_tram}
\end{equation}
Here, the operator $c_{i}^{(\dagger)}$ annihilates (creates) an electron in the $p_z$ orbital of the carbon atom at site $\textbf{r}_i$. The $\mathrm p_z$ electrons are coupled via Slater-Koster hopping parameters $t_{ij} = t(\textbf{r}_i - \textbf{r}_j)$ 
\begin{align}
\begin{split}
t(\textbf{d}) &= t_{\parallel}(\textbf{d}) + t_{\bot}(\textbf{d}) \\
&= \left( 1 - n ^ { 2 } \right) \gamma _ {pp\pi } \exp \left[ \lambda _ { 2 } \left( 1 - \frac { \left| \textbf{d} \right| } { c } \right)\right] + n ^ { 2 } \gamma _ {pp\sigma} \exp \left[ \lambda _ { 1 } \left( 1 - \frac { \left| \textbf {d}\right| } { a } \right) \right].
\end{split}
\label{tb_hop}
\end{align}
Due to the internal twist between adjacent graphene sheets, a sufficient description of the interlayer hopping must include contributions from $pp\pi$ bonds $\gamma_{pp\pi} = -2.8$ eV as well as from $pp\sigma$ bonds $\gamma_{pp\pi} = 0.48$ eV \cite{tram10}. To this end, the factor $n = \frac{\textbf{d}\cdot \hat{\textbf{e}}_z}{|\textbf{d}|}$ captures the out-of plane component of the electron transfer integral. Furthermore, $\textbf{e}_z$ is a unit vector which points perpendicular to the graphene sheets, $c=3.364$ {\AA} is the interlayer spacing of graphite, $a=1.42$ {\AA} is the distance between neighboring carbon atoms and ${\lambda }_1=3.15$ and ${\lambda }_2=7.462$ describe the exponential cutoff of the electron hopping.

For the calculations of 3D twisted WSe$_2$ and boron nitride, we perform first principles calculations based on DFT as implemented in the Vienna Ab initio Simulation Package (VASP) \cite{kresse93ab} following similar methods used in previous works \cite{wang19,Xian18}. Plane-wave basis sets are employed with an energy cutoff of 550 eV for WSe$_2$ and 400 eV for boron nitride. The projector augmented wave method (PAW) \cite{blochl94} is used to construct the pseudopotentials felt by the valence electrons. For the calculations of 3D twisted WSe$_2$, the exchange-correlation functionals are treated within the generalized gradient approximation (GGA) \cite{perdew1996generalized}. All the atoms are relaxed until the force on each atom is less than 0.01 eV/{\AA}. Van der Waals interactions are included using the method of Tkatchenko and Scheffler \cite{tsmethod09} during the relaxation. For the calculations of 3D twisted boron nitride, the exchange-correlation functionals are treated within the local density approximation (LDA). As shown in the previous work \cite{Xian18}, the flat bands near the top of the valence band of twisted boron nitride do not change much upon relaxation. Therefore, as the calculations for 3D twisted boron with twist angles down to 2.28 degree are very heavy, we perform these large scale calculations for 3D twisted boron nitride without relaxation.   

\begin{figure}
    \centering
    \includegraphics[width=0.8\textwidth]{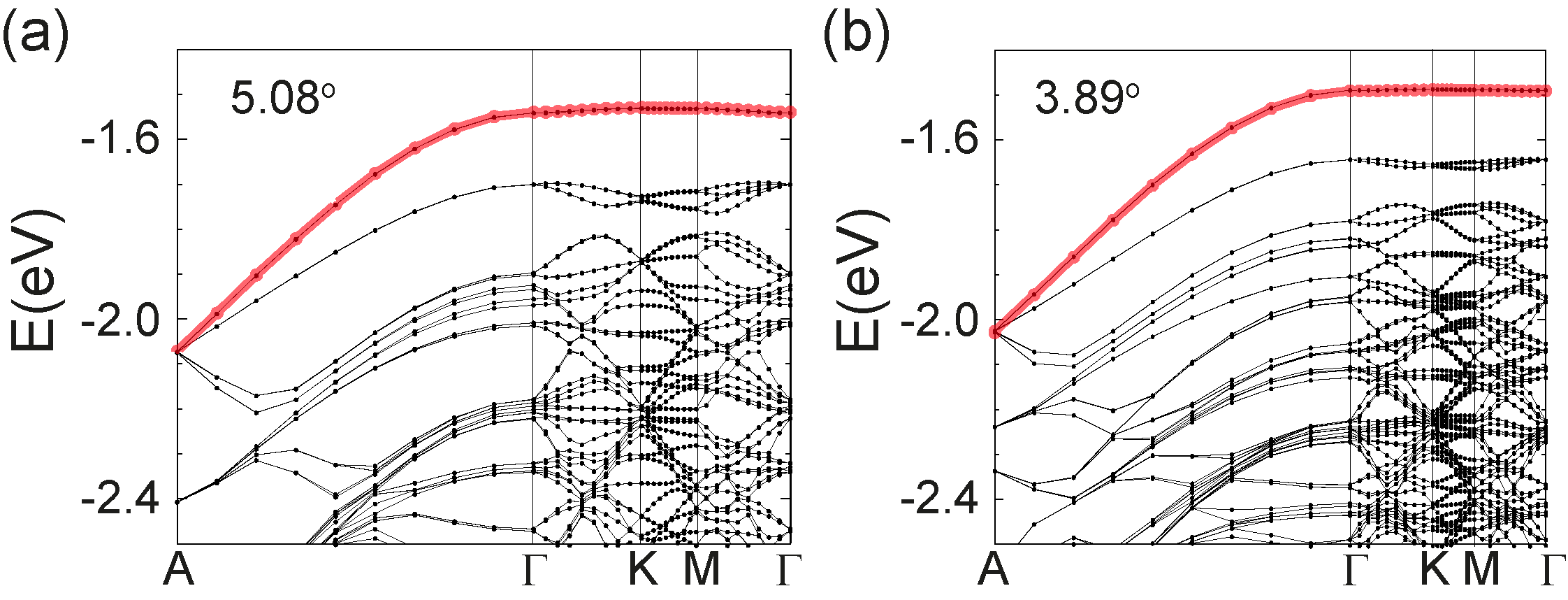}
    \caption{Band structures of 3D twisted boron nitride with type I stacking at 5.08 degrees (a) and 3.89 degrees (b). The band width of the top valence band along the in-plane at k$_z$=0 decreases with twist angles, while the band width along the out-of-plane direction remains highly dispersive. }
    \label{hbnbands}
\end{figure}

\begin{figure}
    \centering
    \includegraphics[width=0.8\textwidth]{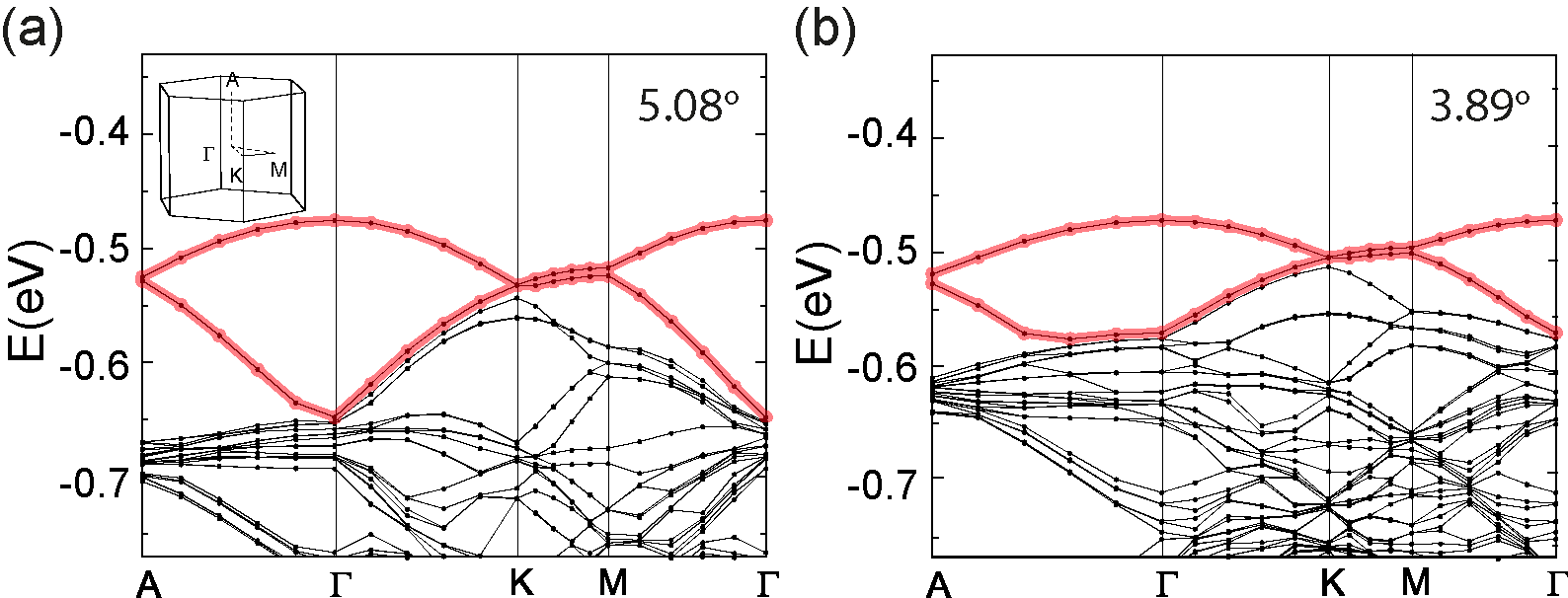}
    \caption{Band structures of 3D twisted WSe$_2$ with type III stacking at 5.08 degrees (a) and 3.89 degrees (b). The band width along both the in-plane and the out-of-plane directions decreases with twist angles. }
    \label{wse2bands}
\end{figure}

\section{Low-energy tight-binding model for twisted boron nitride}

The low-energy physics of twisted hBN (thBN) is captured by an effective three-dimensional tight-binding (TB) model that includes hopping terms between emerging charge localization points at $\textbf{Q}_1 =   (\frac{1}{3}, \frac{1}{3}, \frac{1}{2}  ) $ and $\textbf{Q}_2 = (\frac{2}{3}, \frac{2}{3}, 0 ) $ in the moir\'{e} unit cell. The coordinates are given with respect to the in-plane ($\parallel$) and out-of plane ($\bot$) Bravais lattice vectors $\textbf{L}_1^{\parallel} =  \left (L, 0, 0 \right )$, $\textbf{L}_2^{\parallel} = R(\pi/3) \textbf{L}_1$ and $\textbf{L}_3^{\bot} =  \left (0, 0, D \right )$. The lattice constant $D$ is fixed, while $L$ is twist-angle dependent and it describes the spatial extent of the moir\'{e} pattern, see table \ref{table:constants}. 

The effective structure defined by the charge accumulation points resembles AA-stacked graphene multilayers, where one of the two inequivalent sites, i.e. $\textbf{Q}_1$, is shifted by $D/2$ in z-direction. Hence, in each of the two "effective" planes with $z$-coordinate $0$ and $D/2$ , the charge puddles form a triangular lattice with lattice constant $L$. 

The simplest $SU(2)$ symmetric TB model that can be constructed for this configuration is a single-orbital two-band model that takes up to next-nearest neighbor intra- and interlayer hopping terms between the charge puddles into account
\begin{equation}
H_0 = t_1 \sum_{\langle i,j \rangle} c_{i}^{\dagger} c_{j}^{\phantom \dagger}  
+  t_2 \sum_{\langle i,j \rangle_{\parallel}} c_{i}^{\dagger} c_{j}^{\phantom \dagger} 
+  t_3 \sum_{\langle i,j \rangle_{\bot}} c_{i}^{\dagger} c_{j}^{\phantom \dagger}. 
\label{hop_real}
\end{equation}
Here, $t_1$ denotes the hopping amplitude between neighboring $\textbf{Q}_1$- and $\textbf{Q}_2$-sites , whereas $t_2$ and $t_3$ denote hopping processes between two $\textbf{Q}_1$ ($\textbf{Q}_2$) sites in either the same or different layers. The hopping parameters are determined by fitting the energy eigenvalues of $H_0$ to the flat bands of the \textit{ab-initio} band structure of thBN. The single-particle spectrum for the periodic system is then modeled by the following Bloch Hamiltonian
\begin{equation}
H_0 = \sum_{\textbf{k}} h_{\textbf{k}} = \sum_{\textbf{k}}
\begin{pmatrix}
h_0(\textbf{k}) & h_1(\textbf{k}) \\
h_1^*(\textbf{k}) & h_0(\textbf{k})
\end{pmatrix},
\end{equation}
which is labeled in the order of the two charge localization points $\textbf{Q}_1$,$\textbf{Q}_2$. The matrix elements are obtained by a Fourier transform of the real-space hopping matrix Eq.~\eqref{hop_real} to (Bloch) momentum space
\begin{equation}
\begin{split}
h_0(\textbf{k}) =\, &2t_2 \left [ \text{cos}(  \textbf{k}\cdot ( \textbf{L}_1-\textbf{L}_2) ) +  \text{cos}(\textbf{k}\cdot\textbf{L}_1) + \text{cos}(\textbf{k}\cdot\textbf{L}_2)\right ] + 2t_3 \text{cos}(\textbf{k}\cdot \textbf{L}_3),  \\
h_1(\textbf{k}) = \, &t_1 \left [1 + e^{-i \textbf{k}\cdot \textbf{L}_1 }  + e^{-i \textbf{k}\cdot \textbf{L}_2 }  \right] \left [ 1+ e^{-i \textbf{k}\cdot \textbf{L}_3} \right ]. 
\end{split}
\end{equation}
The matrix $h_{\textbf{k}}$ can then be diagonalized in orbital space for each momentum $\textbf{k}$ to obtain the bandstructure $\epsilon_b(\textbf{k})$ and orbital-to-band transformation $u_{\textbf{r}}^{b}(\textbf{k})$, $b = 1 .. N$: 
\begin{align}
H_0 &= \sum_{\textbf{k}, b} \epsilon_b(\textbf{k}) \gamma_{\textbf{k}, b}^{\dagger} \gamma_{\textbf{k}, b}^{\phantom \dagger} \qquad \text{with} \ \   \gamma_{\textbf{k}, b} = u_{\textbf{r}}^{b}(\textbf{k}) c_{\textbf{k}, \textbf{r}}.
\label{o2b}
\end{align}

\renewcommand{\arraystretch}{1.3}
\setlength\tabcolsep{16pt}
\begin{table*}[htbp]
\begin{tabular}{rccccc}

\toprule
\begin{tabular}[c]{@{}c@{}}twist angle \\ $\theta$\end{tabular} & \multicolumn{3}{c}{\begin{tabular}[c]{@{}c@{}}Hopping parameters\\  (meV)\end{tabular}} & \multicolumn{2}{c}{\begin{tabular}[c]{@{}c@{}}Lattice constants\\ (\AA)\end{tabular}} \\ \cline{2-6} 
                                                                & $t_1$                        & $t_2$                       & $t_3$                       & L                                                   & D                                                   \\ \midrule
                                                                
 5.08$^{\circ}$                                                  & 14.59                        & -4.35                       & 0.00                        & 28.31                                                    &   12.92                                                  \\
3.89$^{\circ}$                                                  & 8.16                         & -1.56                       & 2.08                        & 37.00                                               &  12.92                                                   \\
3.15$^{\circ}$                                                  & 5.29                         & -1.17                       & 1.72                        & 45.70                                               & 12.92                                               \\
2.64$^{\circ}$                                                  & 3.18                         & -0.68                       & 2.04                        &  63.10                                                   &  12.92                                                   \\ \bottomrule
\end{tabular}
\caption{Hopping parameters of the effective $SU(2)$-symmetric tight-binding model for different twist angles $\theta$ according to Fig.~3 in the main text. The structure constants $D$ and $L$ (see Fig.~\ref{fig:TB_BN}(a)) describe the spatial extent of the moir\'{e} cell in in-plane and out-of-plane direction, respectively. }
\label{table:constants}
\end{table*}

\section{Fluctuation exchange approximation in multi-orbital systems}

\subsection{3D multi-orbital susceptibility}
We define the free Matsubara Green's function in orbital-momentum (frequency) space as
\begin{equation}
g_{\textbf{r}, \textbf{r}^{\prime}} (i\omega, \textbf{k}) = (i\omega - (H_0(\textbf{k}))_{\textbf{r}, \textbf{r}^{\prime}}  )^{-1} = \sum_{b} u_{\textbf{r}}^{b}(\textbf{k}) g^b (i \omega, \textbf{k}) u_{\mathbf{r^{\prime}}}^{b*}(\textbf{k}) =
 	\begin{tikzpicture}[baseline = -2.5]
	
		\path [draw=black,postaction={on each segment={mid arrow=black}}, line width=0.8pt]
		(0,0) -- (3,0) 	node[above=2, xshift = -3 cm]{$\textbf{r}$}
					node[above=2, xshift = +0.1 cm]{$\mathbf{r^{\prime}}$} 
					node[below=2, xshift = -1.45 cm]{$(b, \textbf{k})$}
		;
	\end{tikzpicture}
\label{green}
\end{equation}
where $u^b_{\textbf{r}}$ are the orbital-to-band transformations that render the unperturbed Hamiltonian $H_0$ and the free Green's function $g^b(i \omega, \textbf{k}) = (i \omega -e^b(\textbf{k}))^{-1}$ diagonal. The orbital indices $\textbf{r} = \{ \textbf{Q}_1, \textbf{Q}_2\}$ are restricted to the same unit cell and the momenta $\textbf{k}$ lie in the first Brillouin zone. To this end, we define the free polarization function $\hat{\chi}_0(q) = \chi_{0_{\textbf{r}, \textbf{r}^{\prime}}}(q) $ as
\begin{equation}
\chi_{0_{\textbf{r}, \textbf{r}^{\prime}}}(q) = \chi_{0_{\textbf{r}, \textbf{r}^{\prime}}}(\textbf{q}, i\omega) = \frac{1}{N \beta} \sum_{\textbf{k}, \omega^{\prime}} g_{\textbf{r}, \textbf{r}^{\prime}}(i\omega^{\prime}, \textbf{k})g_{\textbf{r}^{\prime}, \textbf{r}}\left ( i(\omega^{\prime}+ \omega \right ), \textbf{k}+ \textbf{q}).
\label{chi0}
\end{equation}
The Matsubara summation occuring in Eq. \eqref{chi0} can be evaluated analytically giving the well-known Lindhard function for multi-orbital systems
\begin{equation}
\label{chi0_mats}
\chi_{0_{\textbf{r}, \textbf{r}^{\prime}}}(\textbf{q}, i\omega) =   \frac{1}{N} \sum_{\textbf{k}, b, b^{\prime}} \frac{n_F(\epsilon_{b^{\prime }}(\textbf{k})) - n_F(\epsilon_b(\textbf{k}+\textbf{q})) }{i \omega + \epsilon_{b^{\prime}}(\textbf{k}) - \epsilon_b(\textbf{k}+\textbf{q}) } u_{\textbf{r}}^{b^{\prime}}(\textbf{k})  u_{\textbf{r}^{\prime}}^{b^{\prime}*}(\textbf{k}) u_{\textbf{r}}^{b*}(\textbf{k}+\textbf{q}) u_{\textbf{r}^{\prime}}^{b}(\textbf{k}+\textbf{q}),
\end{equation}
where $n_F(\epsilon) = (1+e^{\beta \epsilon})^{-1}$ is the Fermi function.

\subsection{Random-phase approximation for multi-orbital systems}

To study correlated states of matter in thBN that arise due to the presence of electron-electron interaction, we employ a repulsive Hubbard term for electrons with opposite spin $\sigma \in \{-1,1\}$ with $\overline{\sigma} = - \sigma$ residing on site $\textbf{r}$ in moir\'{e} supercell $\textbf{R}$
\begin{equation}
V = \frac{1}{2}\sum_{\textbf{R}, \textbf{r}_i, \sigma} U n_{\textbf{R}, \textbf{r}_i, \sigma} n_{\textbf{R}, \textbf{r}_i, \overline{\sigma}} = \frac{1}{2N} \sum_{\textbf{k}, \textbf{k}^{\prime}, \textbf{q}} \sum_{\textbf{r}, \sigma}  U c_{\textbf{k}, \sigma}^{{\dagger}_{\textbf{r}}} c_{\textbf{k}^{\prime}, \overline{\sigma}}^{{\dagger}_{\textbf{r}}} c_{\textbf{k}^{\prime}-\textbf{q}, \overline{\sigma}}^{\phantom{\dagger}_{\textbf{r}}} c_{\textbf{k}+\textbf{q}, \sigma}^{\phantom{\dagger}_{\textbf{r}}}
\label{hubbard}
\end{equation}
Here, the occupation number operator is defined as $n_{\textbf{R}, \textbf{r}_i, \sigma} = c^{\dagger}_{\textbf{R}, \textbf{r}_i, \sigma} c_{\textbf{R}, \textbf{r}_i, \sigma}$. We calculate the renormalized interactions within the random-phase approximation (RPA) to analyze the electronic instabilities mediated by spin-fluctuation exchange between electrons to high order in the bare coupling $U$. Admittedly, this approach is biased as it does not capture the interwind fluctuations in different two-particle scattering channels, which would require the use of unbiased fRG techniques. 

\begin{equation}
\centering
 	\begin{tikzpicture}[baseline = 10]
	
		\path [draw=black,postaction={on each segment={mid arrow=black}}, line width=0.8pt]
		(1,0) -- (0,0)	node[below=0, xshift = -0.1 cm]{\small $\textbf{r}_1,k_1, \downarrow$}
		(2,0) -- (1,0)	node[below=0, xshift =  1.2 cm]{\small $\textbf{r}_2, k_2, \downarrow$}
		(0,1) -- (1,1)	node[above=0, xshift = -1.0 cm]{\small $\textbf{r}_1,k_1+q, \uparrow$}
		(1,1) -- (2,1)	node[above=0, xshift = +0.2 cm]{\small $\textbf{r}_2,k_2+q, \uparrow$} 
		;
		
		\draw[pattern = north west lines, pattern color = black ]
		(0.8, 0) rectangle ++(0.4, 1) node[above=-0.7cm, xshift = 2.5 cm]{$=$}
		;
		
		\begin{scope}[shift={(5,0)}]		
		\path [draw=black,postaction={on each segment={mid arrow=black}}, line width=0.8pt]
		(1,0) -- (0,0)	node[below=0, xshift = -0.1 cm]{\small $\textbf{r}_1,k_1, \downarrow$}
		(2,0) -- (1,0)	node[below=0, xshift =  1.2 cm]{\small $\textbf{r}_1, k_2, \downarrow$}
		(0,1) -- (1,1)	node[above=0, xshift = -1.0 cm]{\small $\textbf{r}_1,k_1+q, \uparrow$}
		(1,1) -- (2,1)	node[above=0, xshift = +0.2 cm]{\small $\textbf{r}_1,k_2+q, \uparrow$} 
		;
		
		\path [draw=black,snake it]
		(1,1) -- (1,0)	node[above = 8, xshift = 0.3 cm]{\small $U$} node[above=+0.3cm, xshift = 1.8 cm]{$+$} 
		;
		\end{scope}
		
		\begin{scope}[shift={(8.8,0)}]
		\path [draw=black,postaction={on each segment={mid arrow=black}}, line width=0.8pt]
		(1,0) -- (0,0)	node[below=0, xshift = 1.25 cm]{\small $\textbf{r}^{\phantom \prime}$}
		(3,0) -- (1,0)	node[below=0, xshift = 1.0 cm]{\small$k \downarrow$}
		(0,1) -- (1,1)	node[above=0, xshift = 0.2 cm]{\small$\textbf{r}$}
		(1,1) -- (3,1)	node[above=0, xshift = -1.0 cm]{\small$k$+$q$ $\uparrow$}
		(4,0) -- (3,0)	node[below=0, xshift = -0.1 cm]{\small$\textbf{r}^{\prime}$}
		(3,1) -- (4,1)	node[above=0, xshift = -1.1 cm]{\small$\textbf{r}^{\prime}$}
		;
		
		\path [draw=black,snake it]
		(1,1) -- (1,0)	node[above = 8, xshift = 0.3 cm]{\small $U$}
		;
		
		\draw[pattern = north west lines, pattern color = black] 
		(2.8, 0) rectangle ++(0.4, 1)
		;
		
		\end{scope}

	\end{tikzpicture}
\label{RPA_magnetism_diagrams}	
\end{equation}

The renormalized interaction in RPA approximation Eq. \eqref{RPA_magnetism_diagrams}  is then given by $\hat{V}_{\text{RPA}}(q) = U/[1+U \hat{\chi_0}(q)]$. Magnetic instabilities can be classified according to a generalized Stoner criterion: The effective (RPA) interaction diverges, when the smallest eigenvalue $\lambda_0$ of $\hat{\chi}_0(\textbf{q}, i\omega)$ reaches $-1/U$, marking the onset of magnetic order for all interaction strengths  $U \geq U_{\text{crit.}} = -1/\lambda_0$. The corresponding eigenvector $v^{(0)}(\textbf{q}, i\omega)$ is expected to dominate the spatial structure of orbital magnetisation.

\subsection{Pairing Symmetry}
We may write the general particle-particle scattering vertex between electrons with opposite momenta $(\textbf{k}_1, - \textbf{k}_1) \to (\textbf{k}_2, - \textbf{k}_2)$ as

\begin{equation}
 V = \frac{1}{2N} \sum_{s, s^{\prime}} \sum_{\textbf{r}_1, ..., \textbf{r}_4} \sum_{\textbf{k}_1, \textbf{k}_2} \Gamma_{\textbf{k}_1,-\textbf{k}_1 \to \textbf{k}_2, -\textbf{k}_2}^{\textbf{r}_1 \textbf{r}_2 \to \textbf{r}_3 \textbf{r}_4} c_{\textbf{k}_2 s}^{\dagger_{\textbf{r}_3}} c_{-\textbf{k}_2 s^{\prime}}^{\dagger_{\textbf{r}_4}}  c_{-\textbf{k}_1 s^{\prime}}^{ \phantom \dagger_{\textbf{r}_2}} c_{\textbf{k}_1 s}^{\phantom \dagger_{\textbf{r}_1}}  = 
 	\begin{tikzpicture}[baseline = 12]
	
		\path [draw=black,postaction={on each segment={mid arrow=black}}, line width=0.8pt]
		(0,0) -- (1,0)	node[below=0, xshift = -0.8 cm]{\small $\textbf{r}_2,-k_1, s^{\prime}$}
		(1,0) -- (2,0)	node[below=0, xshift =  -0.1 cm]{\small $\textbf{r}_4, -k_2,s^{\prime}$}
		(0,1) -- (1,1)	node[above=0, xshift = -0.8 cm]{\small $\textbf{r}_1,k_1, s$}
		(1,1) -- (2,1)	node[above=0, xshift = -0.1 cm]{\small $\textbf{r}_3,k_2, s$} 
		;
		
		\draw[pattern = north west lines, pattern color = black ]
		(0.8, 0) rectangle ++(0.4, 1)	node[above=-0.9cm, xshift = 1.3 cm]{$\Gamma_{\textbf{k}_1,-\textbf{k}_1 \to \textbf{k}_2, -\textbf{k}_2}^{\textbf{r}_1 \textbf{r}_2 \to \textbf{r}_3 \textbf{r}_4}$}
		;
	\end{tikzpicture}
\label{V}
\end{equation} 

For interaction values $U < U_{\text{crit}}$ the magnetic instabilities prescribed by the RPA analysis might not be strong enough to actually occur. In this paramagnetic regime, spin and charge fluctuations contained in the transverse and longitudinal spin channel can give rise to an effective interaction between electrons that may lead to the formation of Cooper pairs. The diagrams can be separated into spin-singlet and spin-triplet contributions, depending on whether pairing same/opposite spins, i.e. $s \neq s^{\prime}$ (singlet) or $s = s^{\prime}$ (triplet). In general, we may separate the dependence of the gap parameter on momentum, spatial and spin degrees of freedom
\begin{equation}
\Delta_{\textbf{k} s_1 s_2}^{\phantom \dagger_{\textbf{r}_1 \textbf{r}_2}} = f(\textbf{k}, \textbf{r}_1, \textbf{r}_2) \chi(s_1, s_2).
\label{gap_general_sym}
\end{equation}
Since for spin singlet gaps the spin function $ \chi(s_1, s_2)$ is antisymmetric under exchange of indices, i.e.  $\chi(s_1, s_2) = - \chi(s_2, s_1)$, the spatial and momentum dependence must be symmetric in order to fulfill the Pauli principle. For spin triplet gaps we hence require $ f(\textbf{k}, \textbf{r}_1, \textbf{r}_2) =  - f(-\textbf{k},\textbf{r}_2, \textbf{r}_1)$. Since the system is assumed to be paramagnetic, pairing same/opposite spins yields the same result after explicitly symmetrizing/anti-symmetrizing the interaction vertex in orbital-momentum space.

Restricting the pairing to Kramer's degenerate pairs $(\textbf{k}_1, \uparrow)$ and $(-\textbf{k}_1, \downarrow)$, the particle-particle scattering vertex in FLEX approximation is given by transverse ($t$) and longitudinal ($l$) spin fluctuations. For simplicity, we will use the abbreviation $\Gamma_{\textbf{k}_1,-\textbf{k}_1 \to \textbf{k}_2, -\textbf{k}_2}^{\textbf{r}_1 \textbf{r}_2 \to \textbf{r}_3 \textbf{r}_4} = \Gamma^{\textbf{r}_1, \textbf{r}_2}_{{\textbf{k}_1, \textbf{k}_2}}$ in the following. The diagrams contributing to these spin channels are shown below.

\begin{figure}
    \captionsetup{justification = centerlast}
    \subfloat[Diagrams contributing to the transverse spin-fluctuation mediated pairing interaction $\Gamma^{t^{\textbf{r}_1, \textbf{r}_2}}_{2_{\textbf{k}_1, \textbf{k}_2}}$. The momentum transfer occurring in the polarization function in RPA is given by $q_t = k_1+k_2$ due to momentum conservation.]{
	\begin{tikzpicture}[scale = 0.95]
	
		\path [draw=black,postaction={on each segment={mid arrow=black}}, line width=0.8pt]
		(0,0) -- (1,0)	node[below=0, xshift = -1.2 cm]{\small $\textbf{r}_2,-k_1, \downarrow$}
		(1,0) -- (2,0)	node[below=0, xshift =  0.3 cm]{\small $\textbf{r}_1, -k_2, \downarrow$}
		(0,1) -- (1,1)	node[above=0, xshift = -1.2 cm]{\small $\textbf{r}_1,k_1+q, \uparrow$}
		(1,1) -- (2,1)	node[above=0, xshift = 0.3 cm]{\small $\textbf{r}_2,k_2+q, \uparrow$} 
		;
		
		\draw[pattern = north west lines, pattern color = black ]
		(0.8, 0) rectangle ++(0.4, 1)	node[above=-0.7cm, xshift = 2.0cm]{$=$} 
								node[above=-0.9cm, xshift = 0.6 cm]{$\Gamma^{t^{\textbf{r}_1, \textbf{r}_2}}_{2_{\textbf{k}_1, \textbf{k}_2}}$}
		;
		
		\begin{scope}[shift={(4,0)}]
		\path [draw=black,postaction={on each segment={mid arrow=black}}, line width=0.8pt]
		(1,0) -- (0,0)	node[below=-4, xshift = 1.25 cm]{$\textbf{r}_1^{\phantom \prime}$}
		(3,0) -- (1,0)	node[below=0, xshift = 1.0 cm]{$k$}
		(0,1) -- (1,1)	node[above=0, xshift = 0.2 cm]{$\textbf{r}_1$}
		(1,1) -- (3,1)	node[above=0, xshift = -1.0 cm]{$k+q_t$}
		(4,0) -- (3,0)	node[below=-4, xshift = -0.2 cm]{$\textbf{r}_2^{\phantom \prime}$}
		(3,1) -- (4,1)	node[above=0, xshift = -1.2 cm]{$\textbf{r}_2$}
		
		(2,-0.75) -- (4,-1.5)	
		(0,-1.5) -- (2,-0.75)	
		;
		
		\path [draw=black, line width=0.8pt]
		(0,0) -- (2,-0.75)
		(2,-0.75)--(4,0)
		;
		
		\path [draw=black,snake it]
		(1,1) -- (1,0)	node[above = 8, xshift = 0.3 cm]{\small $U$}
		;
		
		\path [draw=black,snake it]
		(3,1) -- (3,0)	node[above = 8, xshift = 0.3 cm]{\small $U$}
					node[above=+0.28cm, xshift = 1.65 cm]{$+$}
		;
		\end{scope}
		
		\begin{scope}[shift={(9.0,0)}]
		\path [draw=black,postaction={on each segment={mid arrow=black}}, line width=0.8pt]
		(1,0) -- (0,0)	
		(2,0) -- (1,0)	node[below=0, xshift =0.5 cm]{$k$}
		(3,0) -- (2.,0)	node[below=0, xshift =0.5 cm]{$k^{\prime}$}
		(0,1) -- (1,1)	node[above=0, xshift = 0.5 cm]{$k$+$q_t$}
		(1,1) -- (2,1)	node[above=0, xshift = 0.5 cm]{$k^{\prime}$+$q_t$}
		(2,1) -- (3,1)	
		(4,0) -- (3,0)	
		(3,1) -- (4,1)	
		
		(2,-0.75) -- (4,-1.5)	
		(0,-1.5) -- (2,-0.75)	
		;
		
		\path [draw=black, line width=0.8pt]
		(0,0) -- (2,-0.75)
		(2,-0.75)--(4,0)
		;

		\path [draw=black,snake it]
		(1,1) -- (1,0)	node[above = 8, xshift = 0.3 cm]{\small $U$}
		(2,1) -- (2,0)	node[above = 8, xshift = 0.3 cm]{\small $U$}
		(3,1) -- (3,0)	node[above = 8, xshift = 0.3 cm]{\small $U$}
					node[above=+0.3cm, xshift = 1.8 cm]{$+$}
					node[above=+0.3cm, xshift = 3 cm]{...}
		;
		
		
		\end{scope}

	\end{tikzpicture}

}\quad

\subfloat[Diagrams contributing to the longitudinal spin-fluctuation mediated pairing interaction $\Gamma^{l^{\textbf{r}_1, \textbf{r}_2}}_{2_{\textbf{k}_1, \textbf{k}_2}}$. The momentum transfer occurring in the polarization function in RPA is given by $q_l = k_1-k_2$ due to momentum conservation. Only an even number of particle-hole bubbles is allowed in the diagrammatic expansion in order to preserve the spin in the upper and lower leg of the pairing interaction. The diagrams that are resummed in the longitudinal channel are connected to the particle-hole susceptibility describing screening effects of the bare Coulomb interaction.]{
    \centering
  	\begin{tikzpicture}[scale = 0.95]
	
		\path [draw=black,postaction={on each segment={mid arrow=black}}, line width=0.8pt]
		(0,0) -- (1,0)	node[below=0, xshift = -1.2 cm]{\small $\textbf{r}_2,-k_1, \downarrow$}
		(1,0) -- (2,0)	node[below=0, xshift =  0.3 cm]{\small $\textbf{r}_2, -k_2, \downarrow$}
		(0,1) -- (1,1)	node[above=0, xshift = -1.2 cm]{\small $\textbf{r}_1,k_1+q, \uparrow$}
		(1,1) -- (2,1)	node[above=0, xshift = 0.3 cm]{\small $\textbf{r}_1,k_2+q, \uparrow$} 
		;
		
		\draw[pattern = north west lines, pattern color = black ]
		(0.8, 0) rectangle ++(0.4, 1) node[above=-0.7cm, xshift = 1.5 cm]{$=$} node[above=-0.9cm, xshift = 0.6 cm]{$\Gamma^{l^{\textbf{r}_1, \textbf{r}_2}}_{2_{\textbf{k}_1, \textbf{k}_2}}$}
		;
		
		\begin{scope}[shift={(5,0)}]
		\path [draw=black,postaction={on each segment={mid arrow=black}}, line width=0.8pt]
		(0,0) -- (1,0)	node[below=0, xshift = -1.2 cm]{\small $\textbf{r}_2,-k_1, \downarrow$}
		(1,0) -- (2,0)	node[below=0, xshift =  0.3 cm]{\small $\textbf{r}_2, -k_2, \downarrow$}
		(0,1) -- (1,1)	node[above=0, xshift = -1.2 cm]{\small $\textbf{r}_1,k_1+q, \uparrow$}
		(1,1) -- (2,1)	node[above=0, xshift = 0.3 cm]{\small $\textbf{r}_1,k_2+q, \uparrow$} 
		;
		
		\path [draw=black,snake it]
		(1,1) -- (1,0)	node[above = 8, xshift = 0.3 cm]{\small $U$} node[above=+0.3cm, xshift =3 cm]{$+$} 
		;
		\end{scope}
		
		\begin{scope}[shift={(9.5,-1)}]
		\path [draw=black, postaction={on each segment={mid arrow=black}}, line width=0.8pt]
		(0,0) -- (1,0)	node[below=0, xshift = -0.8 cm]{\small $\textbf{r}_2,-k_1, \downarrow$}
		(3,0) -- (4,0)	node[below=0, xshift =  -0.1 cm]{\small $\textbf{r}_2, -k_2, \downarrow$}
		(0,3) -- (1,3)	node[above=0, xshift = -0.8 cm]{\small $\textbf{r}_1,k_1+q, \uparrow$}
		(3,3) -- (4,3)	node[above=0, xshift = -0.1 cm]{\small $\textbf{r}_1,k_2+q, \uparrow$} 
		;
		\path [draw=black, line width=0.8pt]
		(1,0)--(3,0)
		(1,3)--(3,3)
		;
		
		\draw (1.5,2) ellipse (0.5cm and 0.3cm);
		\draw[->] (1.5,2.3) -- (1.55, 2.3) node[above=-1.5, xshift = 0 cm]{\footnotesize $k$+$q_l$$\downarrow$};
		\draw[->] (1.5,1.7) -- (1.45, 1.7)	node[below=0.5, xshift = 0.05 cm]{\footnotesize $k$$\downarrow$}
								node[above=4, xshift = 0.75 cm]{\small $\textbf{r}$}
								node[above=2, xshift = -0.7 cm]{\small $\textbf{r}_1$};
		
		\draw (2.5,1) ellipse (0.5cm and 0.3cm);
		\draw[->] (2.5,1.3) -- (2.55, 1.3)	node[above=-1.5, xshift = 0.1 cm]{\footnotesize $k^{\prime}$+$q_l$$\uparrow$};
		\draw[->] (2.5,0.7) -- (2.45, 0.7)	node[below=0.5, xshift = 0.05 cm]{\footnotesize $k^{\prime}$$\uparrow$}
								node[above=0, xshift = 0.75 cm]{\small $\textbf{r}_{2}$}
								node[above=2, xshift = -0.7 cm]{\small $\textbf{r}$};

		\path [draw=black,snake it]
		(1.,3) -- (1.,2)	 
		(2.,2) -- (2,1)	 
		(3,0) -- (3,1)	 node[above=+0.3cm, xshift = 1.8 cm]{$+$}
					node[above=+0.3cm, xshift = 3 cm]{...}
					
		;

		\end{scope}

	\end{tikzpicture}

}%
\label{longitudinal_transverse}
\end{figure}

The effective spin-mediated interaction in the opposite spin channel thus becomes
\begin{equation}
\Gamma_{\textbf{k}_1,-\textbf{k}_1 \to \textbf{k}_2, -\textbf{k}_2}^{{\textbf{r}_1 \textbf{r}_2 \to \textbf{r}_3 \textbf{r}_4}}  = \delta_{\textbf{r}_1, \textbf{r}_3} \delta_{\textbf{r}_2, \textbf{r}_4}  \left [ \hat{U}  +  \frac{U^3\hat{\chi}_0^2(q_l)}{1-U^2\hat{\chi}_0^2(q_l)} \right ]  +  \delta_{\textbf{r}_1, \textbf{r}_4} \delta_{\textbf{r}_2, \textbf{r}_3} \left [ - \frac{U^2 \hat{\chi}_0(q_t)}{1+U\hat{\chi}_0(q_t)} \right ]
\label{v_eff_translation}
\end{equation}

The spin-dependence of the susceptibilities occuring in the diagrammatic expansion above can be neglected due to the emergent $SU(2)$ symmetry in the paramagnetic phase. To obtain the effective interaction in the singlet ($s$) and triplet ($t$) channel, we symmetrize/anti-symmetrize the interaction vertex, i.e.

\begin{equation}
\Gamma^{\text{s/t}}  = \frac{1}{2}
 	\begin{tikzpicture}[baseline = 12]
	
		\path [draw=black,postaction={on each segment={mid arrow=black}}, line width=0.8pt]
		(0,0) -- (1,0)	node[below=0, xshift = -0.8 cm]{\small $\textbf{r}_2,-k_1, s^{\prime}$}
		(1,0) -- (2,0)	node[below=0, xshift =  -0.1 cm]{\small $\textbf{r}_4, -k_2,s^{\prime}$}
		(0,1) -- (1,1)	node[above=0, xshift = -0.8 cm]{\small $\textbf{r}_1,k_1, s$}
		(1,1) -- (2,1)	node[above=0, xshift = -0.1 cm]{\small $\textbf{r}_3,k_2, s$} 
		;
		
		\draw[pattern = north west lines, pattern color = black ]
		(0.8, 0) rectangle ++(0.4, 1)	
		;
	\end{tikzpicture}
+ \sigma
	 \begin{tikzpicture}[baseline = 12]
	
		\path [draw=black,postaction={on each segment={mid arrow=black}}, line width=0.8pt]
		(0,0) -- (1,0)	node[below=0, xshift = -0.8 cm]{\small $\textbf{r}_2,-k_1, s^{\prime}$}
		(1,0) -- (2,0)	node[below=0, xshift =  -0.1 cm]{\small $\textbf{r}_3,k_2, s$} 
		(0,1) -- (1,1)	node[above=0, xshift = -0.8 cm]{\small $\textbf{r}_1,k_1, s$}
		(1,1) -- (2,1)	node[above=0, xshift = -0.1 cm]{\small $\textbf{r}_4, -k_2,s^{\prime}$}
		;
		
		\draw[pattern = north west lines, pattern color = black ]
		(0.8, 0) rectangle ++(0.4, 1)	
		;
	\end{tikzpicture}
\label{V_sym}
\end{equation}

\subsection{Linearized Gap Equation}
Assuming that spin- and charge fluctuation provide the superconducting glue in the system, we confine our considerations to the vicinity of the Fermi surface and only treat scattering processes of a Cooper pair from state ($\textbf{k}, - \textbf{k}$) on fermi surface $C_{b}$ to the state ($\textbf{k}^{\prime}, - \textbf{k}^{\prime}$) on fermi surface $C_{b^{\prime}}$. To this end, we project the pairing vertex Eq. \eqref{v_eff_translation} from orbital to band space and only take intra-band scattering into account
\begin{equation}
\Gamma_{s/t}^{b b^{\prime}}(\textbf{k}, \textbf{k}^{\prime}) = \text{Re} \left [\sum_{\textbf{r}_1, \textbf{r}_2, \textbf{r}_3, \textbf{r}_4}  \Gamma^{s/t}  u_{\textbf{r}_1}^{b^*}(\textbf{k}) u_{\textbf{r}_2}^{b^*}(-\textbf{k}) u_{\textbf{r}_3}^{b^{\prime}}(\textbf{k}^{\prime}) u_{\textbf{r}_4}^{b^{\prime}}(-\textbf{k}^{\prime}) \right ].
\label{proj_v}
\end{equation}
The momenta $\textbf{k}$ and $\textbf{k}^{\prime}$ are restricted to the various fermi surface sheets $\{ C \}$, such that $\textbf{k} \in C_{b}$ and $\textbf{k}^{\prime} \in C_{b^{\prime}}$ with $b$ and $b^{\prime}$ being the band indices of the fermi sheets. Neglecting the frequency dependence of $\Gamma$, we can proceed further by considering only the real part of the pairing interaction. We then solve the linearized gap equation in order to obtain strength and pairing symmetry of the superconducting order parameter, which takes the form of a generalized eigenvalue problem
\begin{equation}
- \frac{1}{V_{\text{BZ}}} \sum_{b^{\prime}} \oint_{\text{FS}_{b^{\prime}}} \frac{\Gamma_{s/t}^{b b^{\prime}}(\textbf{k}, \textbf{k}^{\prime})}{v_F^{b^{\prime}}(\textbf{k}^{\prime})} \Delta_{b^{\prime}}(\textbf{k}^{\prime}) = \lambda \Delta_{b}(\textbf{k}).
\label{lin_gap}
\end{equation}
Here, $v_F^{b}(\textbf{k}) = | \nabla \epsilon_{b}(\textbf{k})|$ is the Fermi velocity at $\textbf{k}^{\prime}$ in band $b$. The largest eigenvalue $\lambda>0$ for a given interaction kernel $\Gamma_{s/t}^{b b^{\prime}}(\textbf{k}, \textbf{k}^{\prime})$, will lead to the highest transition temperature $T_c$ and the corresponding eigenfunction $\Delta_{b}(\textbf{k})$ determines the symmetry of the gap. The effective lattice model obtained from the charge accumulation points has point group $D_{3h}$. The symmetry of the gap can thus by classified according to the irreducible representations of $D_{3h}$ that are listed in Table \ref{table:sym}. 

The linearized gap equation \eqref{lin_gap} only accounts for the leading pairing symmetry at the transition temperature $T_c$ of the superconducting phase. In the case of degenerate eigenvalues (e.g. $d$-wave instabilities $\{d_{xz}, d_{yz}\}$) belonging to a two-dimensional irreducible representation, an arbitrary linear combination might be favored for $T<T_c$. In order to find the linear combination that is preferred by the system below the transition temperature, we compute the free energy of the system
\begin{equation}
\begin{split}
F = E-TS = &\frac{1}{N} \sum_{\textbf{k}, {b}} \left [ E_{b}(\textbf{k}) n_F(E_{b}(\textbf{k}))  - \frac{|\Delta_{b}(\textbf{k})|}{E_{b}(\textbf{k})} \tanh \left (\frac{E_{b}(\textbf{k})}{2T} \right)\right]  \\ 
&+ \frac{T}{N} \sum_{\textbf{k}, {b}}  \left[ n_F(E_{b}(\textbf{k})) \ln (n_F(E_{b}(\textbf{k}))) + n_F(-E_{b}(\textbf{k})) \ln (n_F(-E_{b}(\textbf{k}))) \right ].
\end{split}
\label{free_energy}
\end{equation}
Here, $E_{b}(\textbf{k})$ is the energy of the Bogoliubov quasi-particles resulting from diagonalization of the BdG Hamiltonian
\begin{equation}
H_{\text{BdG}} = \sum_{\textbf{k}, b } \psi^{\dagger}_{b \textbf{k}}  
\begin{pmatrix} \epsilon_{b}(\textbf{k})- \mu & \Delta_{b}(\textbf{k}) \\ \Delta^{\dagger}_{b}(\textbf{k}) & -\epsilon_{b}(-\textbf{k})+ \mu  \\\end{pmatrix}
\psi^{\phantom \dagger}_{b \textbf{k}} 
= \sum_{\textbf{k}, b } \psi^{\dagger}_{b \textbf{k}}  \left [ \delta_{b}(\textbf{k}) \cdot \mathbf{\tau} \right ] \psi^{\phantom \dagger}_{b \textbf{k}}, 
\label{bdg}
\end{equation}
where $ \delta_{b}(\textbf{k}) = (\Re[\Delta_{b}(\textbf{k})], \Im[\Delta_{b}(\textbf{k})] , \epsilon_{b}(\textbf{k})- \mu  )^T$ and $\mathbf{\tau}$ are the Pauli matrices. In the expression of the free energy Eq. \eqref{free_energy}, we only account for states at the Fermi surface as contributions from $\textbf{k}$ points far away from the Fermi surface are negligible $\epsilon_{b}(\textbf{k}) \gg |\Delta_{b}(\textbf{k})|$.

\begin{figure}
    \centering
    \includegraphics[width=0.6\textwidth]{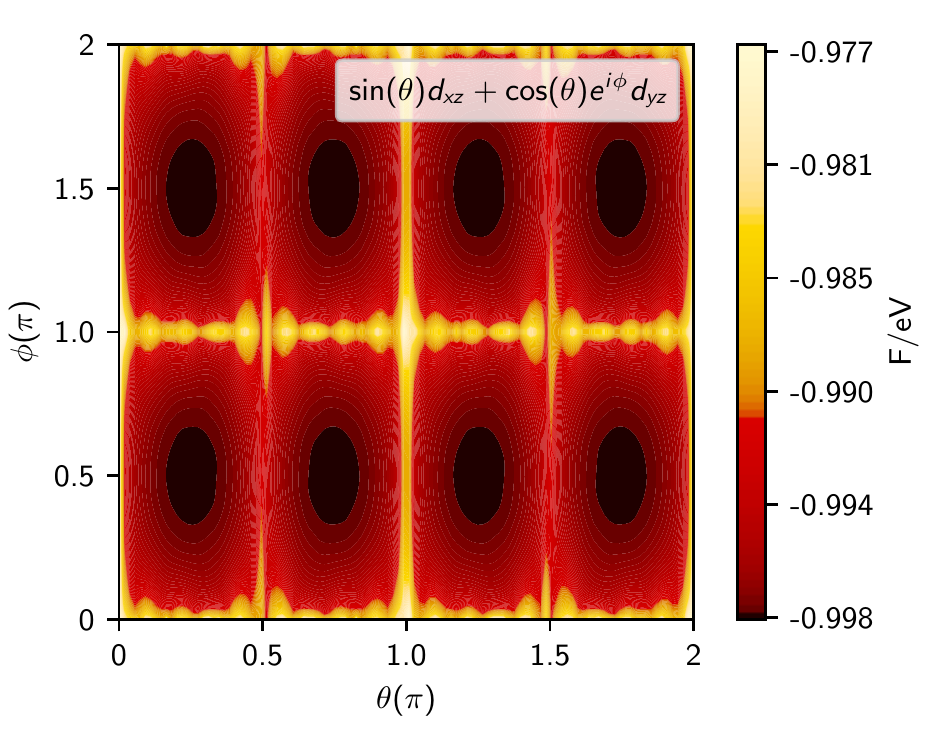}
    \caption{Free energy of the linear combination $\Delta_{b}(\textbf{k}) = \sin (\theta) d_{xz}(\textbf{k}) + \cos (\theta) e^{i\phi} d_{yz}(\textbf{k})$ corresponding to the leading pairing symmetry at $\mu \approx\mu_0 +  5 \, \text{meV}$. The system minimizes its free energy by choosing the linear combination $d_{xz}(\textbf{k}) \pm i d_{yz}(\textbf{k})$.
    }
    \label{fig:free_energy}
\end{figure}

At the filling $\mu \approx \mu_0 +  5 \, \text{meV}$ studied in the manuscript, the leading pairing symmetry is the $d$-wave which belongs to a two-dimensional irreducible representation. To minimize the free energy of the system we make the ansatz
\begin{equation}
\Delta_{b}(\textbf{k}) = \sin (\theta) d_{xz}(\textbf{k}) + \cos (\theta) e^{i\phi} d_{yz}(\textbf{k}), 
\end{equation}
where the form factors are are given by $d_{xz}(\textbf{k}) = \text{sin}(k_x)\text{sin}(k_z)$ and $d_{yz}(\textbf{k}) = \text{sin}(k_y)\text{sin}(k_z)$. The free parameters $\theta$ and $\phi$ are extracted by minizing the free energy of the system Eq. \eqref{free_energy}. In Fig. \ref{fig:free_energy} we show that the linear combination $\Delta^{b}_\textbf{k} \propto \left [d_{xz}(\textbf{k}) \pm i d_{yz}(\textbf{k}) \right ] = \left [ \text{sin}(k_x)\text{sin}(k_z) \pm i \text{sin}(k_y)\text{sin}(k_z) \right  ]$ is generally preferred for the given filling.

\begin{table}[h]
\begin{tabular}{cc}
\hline
singlet                     & triplet                           \\ \hline
$s$                         & $p_z$                             \\
$(d_{x^2-y^2}, d_{xy})\cdot p_z$     & $(d_{x^2-y^2}, d_{xy})$ \\
$(d_{xz}, d_{yz})$      & $(p_x, p_y)$                      \\
$f_{x(x^2-3y^2)}\cdot p_z$  & $f_{x(x^2-3y^2)}$                 \\
$f_{y(y^2-3x^2)} \cdot p_z$ & $f_{y(y^2-3x^2)}$                 \\ \hline
\end{tabular}
\caption{Pairing symmetries for the effective lattice model of thBN separated into contributions to spin singlet and triplet channel.}
\label{table:sym}
\end{table}
\bibliography{reference.bib}